\documentclass[fleqn,usenatbib]{mnras}

\usepackage{txfonts}

\usepackage[T1]{fontenc}

\DeclareRobustCommand{\VAN}[3]{#2}
\let\VANthebibliography\thebibliography
\def\thebibliography{\DeclareRobustCommand{\VAN}[3]{##3}\VANthebibliography}

\usepackage{savesym}
\savesymbol{iint}
\savesymbol{iiint}
\savesymbol{iiiint}
\savesymbol{idotsint}
\usepackage{newtxtext,newtxmath}
\usepackage{graphicx}	
\usepackage{longtable}
\usepackage[dvipsnames]{xcolor}
\usepackage{listings}
\usepackage{gensymb}
\usepackage{algorithm}
\usepackage{algpseudocode}
\usepackage{siunitx}
\title[GC candidates around the M\,81 triplet of galaxies]{J-PLUS: A catalogue of globular cluster candidates around the M\,81/M\,82/NGC\,3077 triplet of galaxies}

\author[Ana L. Chies-Santos et al.]{Ana L. Chies-Santos,$^{1,2}$\thanks{E-mail: ana.chies@ufrgs.br}
Rafael S. de Souza,$^{2}$\thanks{E-mail: drsouza@shao.ac.cn}
Juan P. Caso, $^{3,4}$\thanks{E-mail: jpcaso@fcaglp.unlp.edu.ar}
Ana I. Ennis, $^{3,4}$
Camila P. E. de Souza,$^{5}$
\newauthor
Renan S. Barbosa,$^{6}$ 
Peng Chen,$^{2,7}$ 
A. Javier Cenarro,$^{8}$ 
Alessandro Ederoclite,$^{8}$ 
David Crist\'obal-Hornillos,$^{8}$ 
\newauthor
Carlos Hern\'andez-Monteagudo,$^{9,10}$
Carlos L\'opez-Sanjuan,$^{8}$ 
Antonio Mar\'in-Franch,$^{8}$ 
Mariano Moles,$^{8}$ 
\newauthor
Jes\'us Varela,$^{8}$ 
H\'ector V\'azquez Rami\'o,$^{8}$ 
Renato Dupke,$^{11}$ 
Laerte Sodr\'e Jr.,$^{12}$ 
Raul E. Angulo$^{13}$ 
\\
$^{1}$Instituto de Física, Universidade Federal do Rio Grande do Sul (UFRGS), Av. Bento Gonçalves, 9500, Porto Alegre, RS, Brazil\\
$^{2}$Shanghai Astronomical Observatory, Chinese Academy of Sciences, 80 Nandan Rd., Shanghai 200030, China\\
$^{3}$Facultad de Ciencias Astron\'omicas y Geof\'isicas de la Universidad Nacional de La Plata, and Instituto de Astrof\'isica de La Plata\\ (CCT La Plata -- CONICET, UNLP), Paseo del Bosque S/N, B1900FWA La Plata, Argentina\\
$^{4}$Consejo Nacional de Investigaciones Cient\'ificas y T\'ecnicas, Godoy Cruz 2290, C1425FQB, Ciudad Aut\'onoma de Buenos Aires, Argentina\\
$^{5}$Department of Statistical and Actuarial Sciences, University of Western Ontario, London, ON, Canada \\
$^{6}$Instituto Tecnológico de Aeronáutica, Departamento de Ciência e Tecnologia Aeroespacial, Praça Marechal Eduardo Gomes 50,\\ São José dos Campos, SP, 12228-900, Brazil\\
$^{7}$ Shanghai Institute of Technology, 100 Haiquan Rd., Shanghai 201418, China\\
$^{8}$ Centro de Estudios de F\'isica del Cosmos de Arag\'on, Unidad Asociada al CSIC, Plaza San Juan 1, 44001 Teruel, Spain \\
$^{9}$ Instituto de Astrof\'isica de Canarias, Calle V\'ia L\'actea SN, ES38205 La Laguna, Spain\\
$^{10}$ Departamento de Astrof\'isica, Universidad de La Laguna, ES38205, La Laguna, Spain\\
$^{11}$ Observat\'orio Nacional - MCTI (ON), Rua Gal. Jos\'e Cristino 77, S\~ao Crist\'ov\~ao, 20921-400, Rio de Janeiro, Brazil\\
$^{12}$ Universidade de S\~ao Paulo (USP), Instituto de Astronomia, Geof\'isica e Ci\^encias Atmosf\'ericas, R. do Mat\~ao 1226, S\~ao Paulo, SP, 05508-090, Brazil\\
$^{13}$Ikerbasque, Basque Foundation for Science, E-48013 Bilbao, Spain
}

\date{Accepted 2022 July 12. Received 2021, July 8; in original form 2022 February 23}

\pubyear{2022}

\begin{document}
\label{firstpage}
\pagerange{\pageref{firstpage}--\pageref{lastpage}}
\maketitle

\begin{abstract}
Globular clusters (GCs) are proxies of the formation assemblies of their host galaxies. However, few studies exist targeting GC systems of spiral galaxies up to several effective radii. Through 12-band Javalambre Photometric Local Universe Survey (J-PLUS) imaging,  we study the point sources around the M\,81/M\,82/NGC\,3077 triplet in search of new GC candidates. We develop a tailored classification scheme to search for GC candidates based on their similarity to known GCs via a principal components analysis (PCA) projection. Our method accounts for missing data and photometric errors. We report 642 new GC candidates in a region of 3.5 deg$^2$ around the triplet, ranked according to their Gaia astrometric proper motions when available. We find tantalising evidence for an overdensity of GC candidate sources forming a bridge connecting M\,81 and M\,82. Finally, the spatial distribution of the GC candidates $(g-i)$ colours is consistent with halo/intra-cluster GCs, i.e. it gets bluer as they get further from the closest galaxy in the field. We further employ a regression-tree based model to estimate the metallicity distribution of the GC candidates based on their J-PLUS bands. The metallicity distribution of the sample candidates is broad and displays a bump towards the metal-rich end. Our list increases the population of GC candidates around the triplet by 3-fold, stresses the usefulness of multi-band surveys in finding these objects, and provides a testbed for further studies analysing their spatial distribution around nearby (spirals) galaxies.
\end{abstract}




\section{Introduction}

Understanding the assembly history of the baryonic content in galaxies is pivotal for studying the cosmic growth of large-scale structures.
Globular clusters (GC) are found around galaxies spanning an extensive range of masses, from dwarfs to giants (\citealt{brodie06}; \citealt{beasley20}) and are discrete bright beacons that help shed light on the evolution of their host galaxies up to distances of hundreds of Mpc \citep{Harris2017, AM17}.
In addition to their high intrinsic brightness, another property makes them of vital interest to galaxy evolution studies. 
Having mean ages older than $\sim$10\,Gyrs (\citealt{strader05}; \citealt{chies11}) GCs act as fossil tracers of galaxy evolution and its environment. 

The properties of GC systems are intrinsically related to the accretion history of their host galaxies not only through the physical processes ruling their origin (\citealt{kruijssen19},  \citealt{choksi19}) but also due to subsequent assembly episodes that shape their current properties through the contribution of accreted populations (e.g. \citealt{forbes11}, \citealt{caso17}, \citealt{longo18}, \citealt{villaume20}, \citealt{fensch2020}).
GCs are not only found in the bodies of their host galaxies but also free floating in galaxy clusters, not necessarily bound to a host galaxy (\citealt{Blakeslee1999}, \citealt{bassino03}; \citealt{west95}; \citealt{lee2010}; \citealt{AM17}, \citealt{harris2020}).
The Virgo, the Fornax, the Coma and the Abell\,1689 galaxy clusters all appear to have rich populations of intracluster GCs. Moreover, the Milky Way satellite dwarf galaxies Large and Small Magellanic Clouds (LMC/SMC) have a bridge population of GCs (e.g. \citealt{bica2015}). 
Further out in the Local Group (LG), a rich population of stream GCs has been uncovered by the Pandas Survey in M\,31 (e.g. \citealt{huxor2014}) as well as a population of intragroup GCs, not associated with any particular galaxy from the LG (\citealt{zinn2015}).
Several studies point to the relevance of the environment in the build-up and later evolution of GC systems \citep{bortoli22}, including stripping \citep{bassino06} and potential signs of supra-galactic formation processes \citep{forte19}. This is supported by the constant GC-to-halo mass relation, described in both observational \citep[e.g.][]{Hudson2014,Harris2017} and numerical studies \citep[e.g.][]{elbadry19, reina2021, doppel2021}, and their common use in the literature as dynamical tracers of the galaxy halo \citep[e.g.][]{schuberth12,alabi17}. 

In addition to the GC systems associated with the Milky Way, galaxies in the LG and a few other exceptions (e.g. \citealt{gonzalez2016, gonzalez2019}), GC system studies have traditionally targeted early-type galaxies (ETGs). In ETGs, GCs are found in large numbers and are more easily detectable against a smoother galactic background. Age-metallicity distribution of GCs for Milky Way-type simulations at the present-day Universe show a remarkable variety of distributions, which arises due to differences in the formation and assembly histories of the host galaxies (\citealt{kruijssen19}, \citealt{choksi2018}, \citealt{li2019}). 
Although the halo to total mass relation seems to rule the richness of a GC system, observational studies point to second-order differences based on the morphological type, with late-type galaxies appearing less efficient per unit mass in forming GCs, and having metal-rich GC fractions slightly higher than early-types \citep{harris15}. The extension of the GC system and the properties of those GCs located in the distant halo might also be relevant to describe the evolutionary history of the system \citep[e.g.][]{marchi19}. In this sense, the brightest galaxies in the LG arise as a natural reference for spiral galaxies. For instance, \citet{laevens14} found a Galactic GC at a distance of $\sim 145$\,kpc, and \citet{her19} used RRLyrae to measure precise spatial distances to 13 Galactic GCs, spanning up to 90\,kpc, and a fraction of unknown distant GCs are yet to be found  \citep{webb21}. Going further out from our spiral neighbour Andromeda \citet{ditul15} uncovered distant GCs associated with M31. They surveyed a large portion of the LG, resulting in 17 candidates associated with M31 GCs with projected distances of 137\,kpc. This same work found five intragroup GC candidates not associated with any particular galaxy.
Hence, wide-field studies of GC systems associated with nearby spirals are critical to a comprehensive picture of GCs and their role in galaxy evolution.

The region around the M\,81/M\,82/NGC\,3077 triplet (from now one refereed as the triplet) has been the subject of several campaigns targeting its stellar cluster systems. However, they target regions close to the respective host galaxies. A detailed picture of the GC population in the vicinity of this interacting system is still unknown.
M\,81 is a spiral located at 3.6\,Mpc \citep{tully2013}, being the dominant galaxy of a group conformed by $\sim30$ members. The foreground extinction in the direction of M\,81 is $\sim$ 0.16 in the r-band (\citealt{Schlafly_Finkbeiner2011}). It is classified as a spiral with a classical bulge by \citet{fisher2008}, and its bulge mass is huge given its stellar halo mass, occupying an unusual region in the bulge mass-stellar halo mass diagram in the sample of \citet{bell2017}. It presents a stellar mass of $\sim3-8.5\times10^{10}\,{\rm M_{\odot}}$ (\citealt{kk2014} and \citealt{oehm2017}) and a dark matter halo mass of $\sim1\times10^{12}\,{\rm M_{\odot}}$, (\citealt{oehm2017}). Its halo shows a flat colour profile, indicating negligible halo population variations as a function of galactocentric distances \citep{monachesi2013}.
 \cite{perelmuter1995} present the kinematics and metallicity of 25 GCs in M\,81 from 82 bright spectra of GC candidates and computes relative strengths of H$\delta$, CaI $\lambda$4227, and FeI $\lambda$4045 absorption lines to distinguish stellar images of M\,81 globular clusters from stars in the Milky Way.
 \cite{nantais2010_spec} obtain spectra for 74 GCs in M\,81, finding a mean GC metallicity of $\sim$-1.06, higher than either M31 or the Milky Way. The authors report a similar rotation pattern among blue and red GC subpopulations to the Milky Way ones. Clusters at small projected radii and metal-rich clusters rotate firmly, while clusters at large projected radii and metal-poor clusters show weaker evidence of rotation.
 \cite{nantais2010_phot_ext} present a catalogue of extended objects in the vicinity of M\,81 based on a set of 24 Hubble Space Telescope (HST) Advanced Camera for Surveys (ACS) I-band images. They find a total of 233 good GC candidates, 92 candidate H{\sc ii} regions, OB associations, or diffuse open clusters.
 \cite{nantais2011_photGCs} study over 400 GC candidates from HST/ACS photometry. The blue and red GC candidates and the metal-rich and metal-poor spectroscopically confirmed clusters are similar in half-light radius. The total population of confirmed and "good" candidates shows an increase in half-light radius as a function of galactocentric distance. 
 More recently, \cite{ma2017} derives structural parameters of two old and massive GCs in the halo of M\,81 - GC1, GC2 through the Galaxy Evolution Explorer (GALEX), the Beijing-Arizona-Taiwan-Connecticut (BATC), the Two Micron All-Sky Survey (2MASS) and HST/Wide Field Camera 3 (WFC3) imaging. The effective radius versus $M_V$ diagram shows that GC2 is an ultra-compact dwarf (UCD).

M\,82, located at 3.5\,Mpc (\citealt{tully2013}), has a baryonic and halo mass of $\sim1\times10^{10}\,{\rm M_{\odot}}$ and $\sim5\times10^{11}\,{\rm M_{\odot}}$ respectively,  (\citealt{oehm2017}). It is the textbook example of starburst galaxy, with a star formation rate $SFR=13M_{\odot}/yr$ \citep{adebahr2017}. The foreground extinction in the direction of M\,82 is estimated to be $\sim$ 0.32 in the r-band but drops to 0.17 halfway towards M81 (\citealt{Schlafly_Finkbeiner2011}).
Through Subaru Faint Object Camera and Spectrograph (FOCAS) imaging and spectroscopy \cite{saito2005} identify two {\it bona fide} GCs that should have formed at the epoch of M\,82's formation. They also identify a few young star clusters in M\,82, likely produced during the tidal-interaction episode with M\,81.
\cite{lim2013} finds over 1000 star clusters through  UBVIYJH imaging. The colours of halo clusters are similar to GCs in the Milky Way, and their ages are estimated to be older than 1 Gyr.
\cite{cuevas21} extracts structural parameters for a sample of 99 intermediate-age super star clusters (SSCs) in the disc of M\,82 and carry out a survival analysis using a semi-analytical cluster evolution code.
NGC\,3077 is an irregular galaxy located at 3.8\,Mpc (\citealt{tully2013}) with a baryonic and halo mass of $\sim2\times10^{10}\,{\rm M_{\odot}}$ and $\sim5\times10^{11}\,{\rm M_{\odot}}$ respectively (\citealt{oehm2017}). Its foreground extinction is of $\sim$0.14 in the r-band (\citealt{Schlafly_Finkbeiner2011}). 
\cite{davidge2004} investigates the near-infrared photometric properties of NGC\,3077 
through the Canada-France-Hawaii Telescope
and \cite{harris2004} studies the star clusters candidates of NGC\,3077 through HST/ACS broad (F300W, F547M, and F814W) and narrow-band (F487N and F656N) filters. They estimate the age and mass of each star cluster, which provides constraints on the recent star formation histories of the host galaxy.

The Javalambre Photometric Local Universe Survey (J-PLUS) is currently undertaking observations of thousands square degrees
of the sky visible from Observatorio Astrof\'{\i}sico de Javalambre (OAJ, Teruel, Spain; \citealt{oaj}) with the panoramic camera T80Cam \citep{t80cam} at the Javalambre Auxiliary Survey Telescope (JAST80) using a set of 12 broad, intermediate and narrow-band optical filters (\citealt{cen19}, see also \citealt{mendes19} for the southern counterpart, S-PLUS).
The wide-field (1.4 deg $\times$ 1.4 deg) capabilities of T80Cam allow the study of nearby systems out to great galactocentric distances (see \citealt{brito2022}, \citealt{buzzo2022}).
Our goal is to make optimum use of the J-PLUS photometric bands to identify extragalactic GCs candidates in a region of 3.5 deg$^2$ around the M\,81 triplet. Most studies focus on two main criteria to identify GCs, ranging from simple cuts in colour space to fits of their spectral energy distributions (SEDs). Our study employs an approach that naturally utilises the full information provided by the J-PLUS SEDs while accounting for missing data and photometric errors applying a principled statistical learning technique.

We organise the paper as follows. In Sect.~\ref{sec:data} we present the J-PLUS data details and outline the photometric procedures adopted to extract a list of point sources in the analysed pointings. In Sect.~\ref{sec:met} we present the heuristic GC search methodology procedure adopted and the training sample literature data used that allowed us to derive a list of candidate GCs around the triplet. In Sect.~\ref{sec:res} we show the analysis and results, and finally, in Sect.~\ref{sec:conc} we present a summary and the concluding remarks.

\section{Data}
\label{sec:data}

Here we outline the J-PLUS data used in the present study and explain the procedures we adopt to extract photometry when J-PLUS pipeline magnitudes are unavailable. We supplement our analysis with literature catalogues (see \autoref{apend:lit}). Ancillary spectroscopic data from Sloan Digital Sky Survey (SDSS) and Gaia Early Data Release 3 (EDR3) are presented in Sect~\ref{sec:anc} and used later on. Appendices \ref{apend:flux_ex} and \ref{apend:extinction} discuss the GAIA flux excess and extinction.

\subsection{J-PLUS}
\label{sec:jplusdata}
\begin{figure}
\centering
 \includegraphics[width=1\columnwidth]{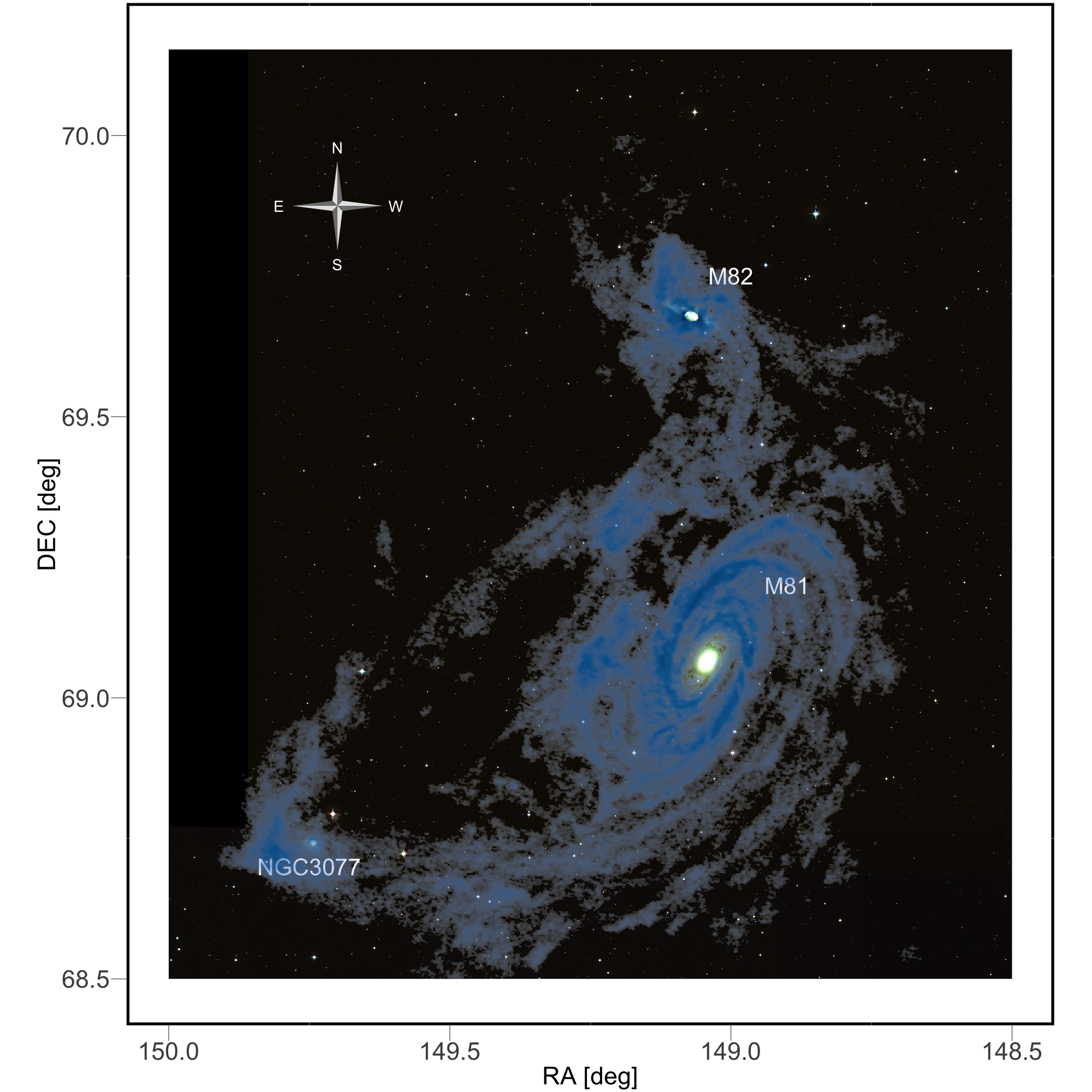}
 \caption{A $gri$ colour composite J-PLUS image exemplifying the region around the triplet. The map between the $gri$ bands and  RGB colours are made using a $\rm asinh$ stretch \citep[see e.g.][]{Lupton2004} Overplotted in tones of blue is the H\,{\sc i} data from \citet{deb18}. Besides the evidence of interaction in H\,{\sc i} \citep{deb18}, the disturbed appearance of the diffuse optical light in the galaxies is also apparent \citep{oka15,sme20}.}
 \label{fig:triplet}
\end{figure}

The J-PLUS dataset consists of the processed images in the 12 available broad  ({\it u, g, r, i and z}) and narrow ({\it J0378, J0395, J0410, J0430, J0515, J0660, J0861}) filters of the J-PLUS survey \citep{cen19} for three pointings from the J-PLUS second data release (DR2), downloaded from the J-PLUS collaboration website\footnote{http://www.j-plus.es/datareleases/data\_release\_dr2}. These pointings cover the central region of the M\,81 group, which contains the two brightest galaxies in the group, M\,81 and M\,82, and they extend to the South, including other less massive members like NGC\,3077.
The field of view of each pointing is $\sim 2.1$\,deg$^2$, with a pixel scale of 0.55\,arcsec. In this work, the analysis is limited to the region spanning $148.4\, \rm deg < {\rm RA} < 151.1\, \rm deg$ and ${\rm DE > 68\,deg}$ (see Figure\,\ref{fig:triplet}), which matches the H\,{\sc i} emission described in \citet{deb18}, and spans a total of $\sim 3.5$ deg$^2$.

\subsubsection{Photometry and preliminary catalogue}
\label{sec:phot}

The J-PLUS DR2 catalogue offers a list of detected sources for each field with their corresponding estimated magnitudes. The detection of sources is good enough at large galactocentric distances. However, the completeness of the J-PLUS catalogue significantly decreases in the vicinity of galaxies, making it mandatory to pre-process the images to recover objects in such regions. Here, we require aperture corrected magnitudes and a homogeneous treatment of the data, hence the photometry was rerun for all 12 J-PLUS filters across the 3 pointings analysed.

Based on bright point sources from the images, the seeing typically spans $1-1.5$\,arcsec, with some degradation towards the blue side of the spectral range. There are examples of extended clusters in the literature, whose nature have been largely discussed \citep[e.g.][]{brodie11,bru12,norris19}. However the mean effective radius for old GCs is $\sim 3$\,pc \citep[e.g. \citealt{har96}, 2010 Edition;][]{pen08}, i.e. $\sim 0.2$\,arcsec at the distance of M\,81, which is close to the limit of detection for several devoted algorithms \citep[e.g. ISHAPE,][]{lar99}. Therefore, we assume that GCs can be treated as point sources in the following.

We run a median filter of size 100\,px on each J-PLUS filter, and then subtracted it from the original image. This procedure slightly increases the noise of the images, but the removal of the extended emission of the galaxies largely improves the source detection. The initial catalogues are build using SE{\sc xtractor} version 2.19.5 \citep{ber96} in dual mode with the $r$ filter acting as a reference image. We consider every group of three connected pixels with a number of counts above $2 \sigma$ the sky level as a positive detection. The analysis threshold is $2 \sigma$ above sky level, except for the filters bluer than $g$, for which we reduced the threshold to $1 \sigma$. Aperture photometry is performed in the 12 filters, assuming in each case an integer aperture diameter close to 3 times the FWHM of point sources. Several bright and isolated stars are used to build the point spread function (PSF) and to calculate the aperture corrections in different sections of the fields, focused on the region analysed in this work. Such aperture corrections span $0.14-0.30$\,mag. Finally, zero points for each field and filter are calculated from the crossmatch of the sources with the J-PLUS photometric catalogue. To this end, a second aperture photometry run is carried out, with a standard diameter of 5.45\,px ($\sim 3$\,arcsec), to facilitate the comparison with the photometric catalogue from the J-PLUS Data Release\,2. In all cases, more than 100 bright sources are used, and the scatter ranges from $0.01$ to $0.04$\,mag. At this point, the photometric catalogue that spans the 3 J-PLUS fields described above contains 17,800 sources.

\subsubsection{Selection of point sources}
\label{sec:ps}

A representative value for the FWHM is derived for the redder broad bands ($r$, $i$, and $z$ filters), as the average of the values measured from the two filters presenting lower seeing. This leads to a similar FWHM distribution for sources from the three fields, with a sudden peak at 2\,px, and a smooth slope towards larger values. A preliminary catalogue of point-sources is built from sources presenting ${\rm FWHM}<6$\,px and a stellarity index from SE{\sc xtractor} (\citealt{sex}) in the $r$ filter larger than $0.5$. Such criteria have proven to be useful to discard the majority of the extended sources but relaxed enough to be fulfilled by sources embedded in the disk of M\,81. Additionally, since we aim to perform a multi-band analysis, we require sources to have as many photometric data points as possible. The completeness drops towards the blue direction of the spectral range, and it is particularly low in the $u$ band. For this reason, we choose to exclude sources with photometry available in less than 11 filters. This results in a final sample of 7\,200 point-like sources. The 90-percentile of the photometric errors in the $r$ filter, assumed as the reference one, increases from 0.01 at $r=17$\,mag to 0.1 at $r=20.5$\,mag.

\subsubsection{Consistency check for the filtering procedure}
To be certain that the filtering procedure does not affect the photometry of point-like sources, we run a test in a region of $20\times20$ arcmin$^2$, centred on M\,81. The previously generated PSF is used to add 250 artificial stars to the original images, and the procedure is repeated 40 times until we achieve a final sample of 10,000 artificial stars. Following this, we applied the filtering and repeated the SE{\sc xtractor} photometry in the same manner. The results show that the filtering does not affect the photometry, with a typical scatter after removing outliers of 0.12\,mag for artificial stars brighter than $r= 20$\,mag. Moreover, no trends related to the distance to the galaxy centre are found.

\subsection{Published GCs}
\label{sec:confGCs}
\begin{figure}
\centering
 \includegraphics[width=\linewidth]{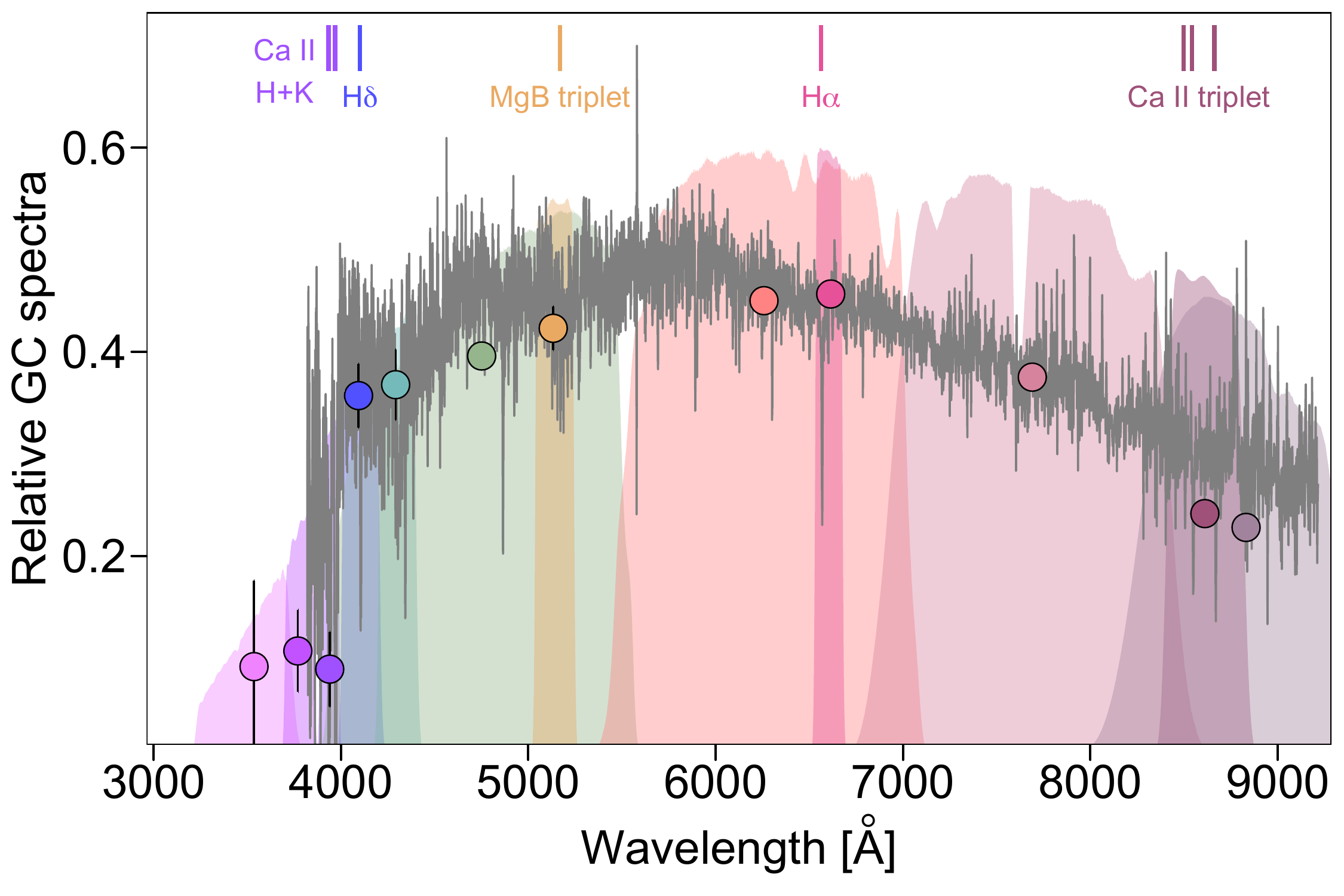}
 \caption{J-PLUS photo-spectrum 
 and SDSS spectrum of intragroup GC-2 from \citet{jan12} on top of the J-PLUS filter curves. The location of key spectral features of the J-PLUS filters is highlighted at the top of the figure.}
 \label{fig:gc_spectra}
\end{figure}

A number of observational studies, focused on the GCs from M\,81 and M\,82, have produced both photometric \citep{nantais2011_photGCs,lim2013} and spectroscopic catalogues \citep{perelmuter1995,saito2005,nantais2010_spec} of the GC systems. Even a couple of intragroup GCs in the region between M\,81 and M\,82 have already been reported \citep{jan12,ma2017}. 
As an example, we show the J-PLUS SED of GC-2 from \cite{jan12}, in \autoref{fig:gc_spectra}, along side its SDSS spectrum.

\begin{figure}
\centering
\includegraphics[width=\columnwidth]{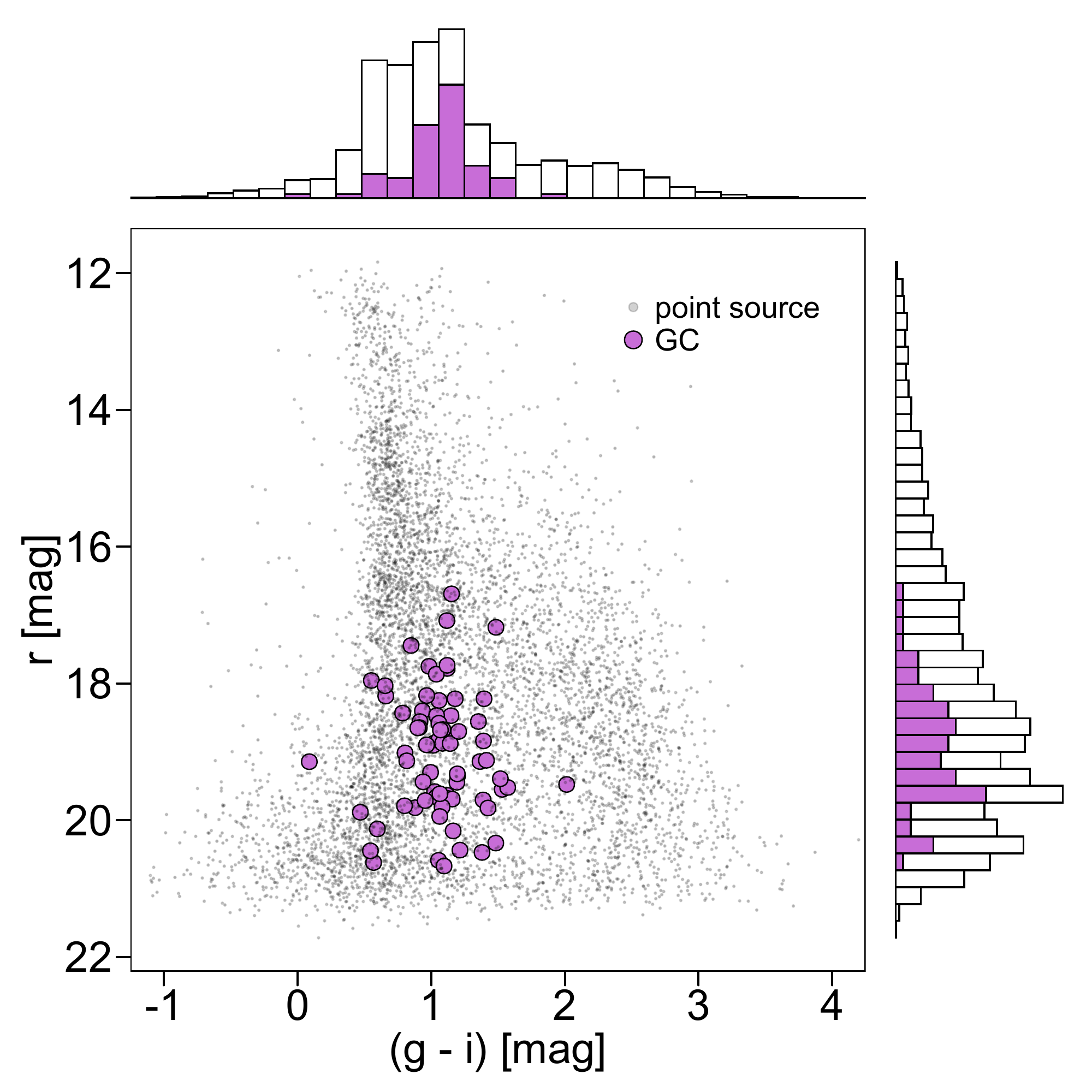}
 \caption{Colour magnitude diagram for the catalogue of point sources (grey dots), with the 73 spectroscopically confirmed GCs used in our statistical analysis highlighted with violet filled circles. The sided panels present histograms, re-scaled to a maximum unit height per group, of the colour and brightness distribution for the confirmed GCs and the general sample of point sources in violet and grey colours.}
 \label{fig:dcm1}
\end{figure}

We gather a sample of 105 GCs that have spectroscopy available. In the majority of cases, objects are marginally resolved through observations, but for some, the classification as GCs was based on the relative strengths of spectral lines \citep{perelmuter1995}. This spectroscopic sample of literature confirmed GCs is listed in Table\,\ref{tab:litGCs} (see Appendix~\ref{apend:lit}).
The crossmatch between these 105 objects and the 7,200 source catalogue described in Section\,\ref{sec:ps} leads to 73 objects, whose locus on the colour-magnitude diagram is highlighted in Figure\,\ref{fig:dcm1}. Marginal distributions are shown on the side panels re-scaled to have a maximum height of one per group for clarity. All subsequent marginals shown in the paper will follow a similar normalisation.

Another 21 objects are detected in such catalogue. However, photometry is only available for a subset of the filters. In \autoref{sec:met} we use the 73 previously known GCs with J-PLUS photometry in at least 11 filters as our training set.

\subsection{Ancillary data}
\label{sec:anc}
We cross-match  our 7,200 point-source catalog with  SDSS\,DR16\footnote{http://skyserver.sdss.org/dr16/en/home.aspx} \citep{ahu20}, to include radial velocity measurements and find a total of 53 SDSS sources.
Based on the NASA/IPAC Extragalactic Database (NED)\footnote{The NASA/IPAC Extragalactic Database (NED) is operated by the Jet Propulsion Laboratory, California Institute of Technology, under contract with the National Aeronautics and Space Administration.}, the heliocentric velocities of the galaxies of the triplet are in the range between $-40$ and $270 \,km\,s^{-1}$.
From a visual inspection of such SDSS spectra, we find 27 objects with ${\rm V_R \lesssim  250\,km\,s^{-1}}$, which is consistent with the systemic velocity of the group. Most of these present spectral features and continuum slopes consistent with bluer (A, F) or redder (M) spectral stellar types, with seven of them classified by the SDSS as G or K stars. For instance, the previously classified intragroup GC-2 \citep{jan12} is classified as a G2 star in SDSS16 (\autoref{fig:gc_spectra}).
On the other hand, the 26 objects with ${\rm V_R \gtrsim  250\,km\,s^{-1}}$ seem to be more consistent with background galaxies, displaying bluer broad-band colours than typical GCs, and a similar brightness range. 

\begin{figure}
\centering
 \includegraphics[width=\columnwidth]{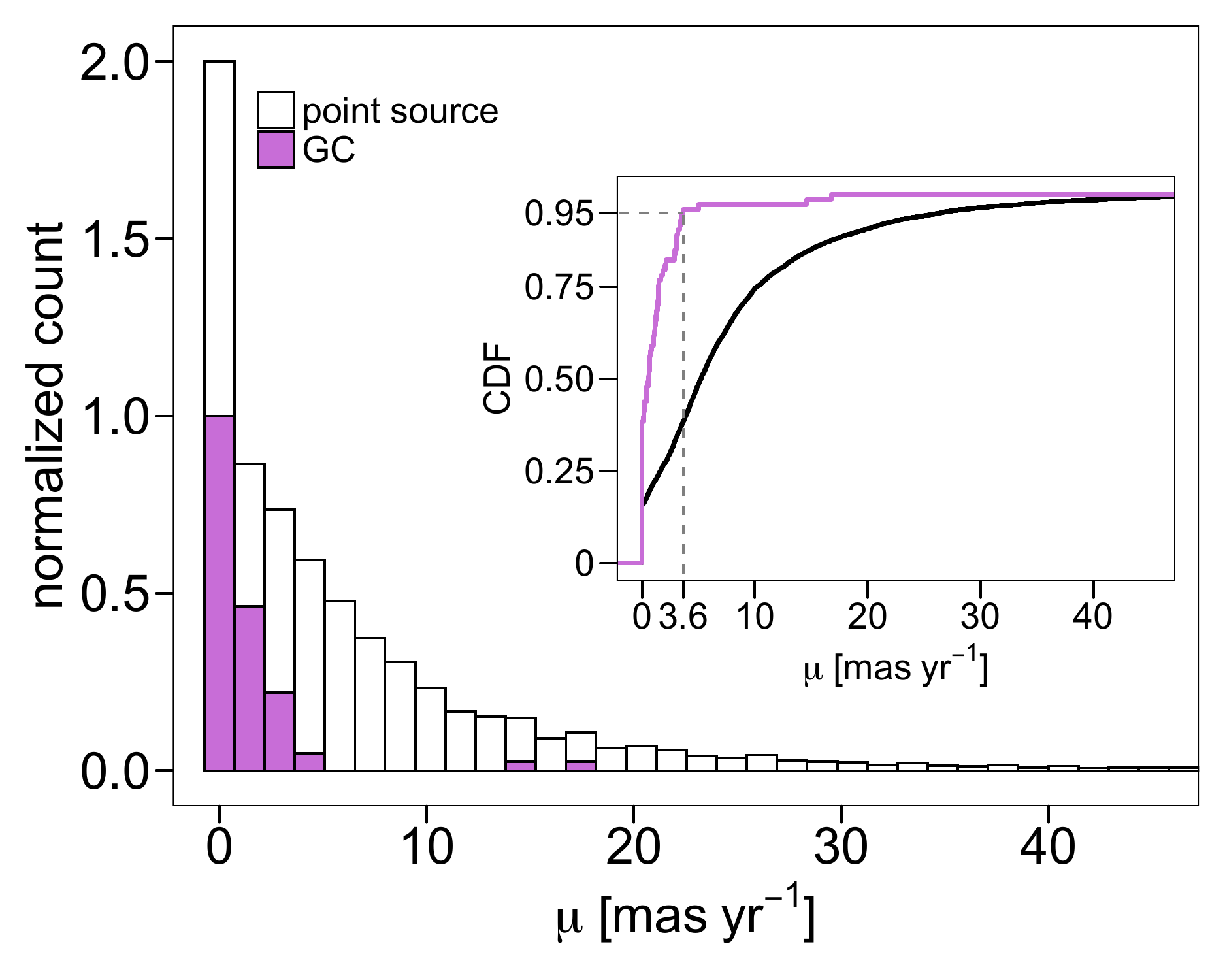}
 \caption{The  distribution of proper motions ($\mu$) for the general sample of point sources as the open histogram and the spectroscopically confirmed GCs as the filled violet histogram. The indent shows the cumulative density function (CDF) of the distributions of spectroscopically confirmed GCs (violet line) and point sources (black line).} 
 \label{fig:pm_hist}
\end{figure}

We further cross-match our J-PLUS point source catalogue with Gaia\,EDR3 \citep{gaia20} to add information on proper motion, resulting in $\sim$ 6,000 positive detections (Section\,\ref{sec:ps}). 
\autoref{fig:pm_hist} displays the proper motion distribution for the point sources and confirmed GCs and the respective empirical cumulative distribution function as inset. From 73 confirmed GCs in our sample, 45 of them have proper motion measurements, and only 4 have values of $\mu > 3.6\,{\rm mas\,yr^{-1}}$, representing only 5 per cent of the sample. In contrast, the point sources have a much broader distribution, with more than 60 per cent of the sample presenting values above this threshold. Additionally, from the 27  objects in our sample with ${\rm V_R \lesssim  250\,km\,s^{-1}}$, 25 present proper motion measurements, and 16 have $\mu > 3.6\,{\rm mas\,yr^{-1}}$. This represents more than 60 per cent of the sample, including the majority of those with spectral types from G to K. While this cannot provide a hard cut for GC detection, this is valuable information to discriminate the most plausible candidates. 

\section{Methodology}
\label{sec:met}

The set of confirmed GCs account for a very small fraction ($\sim$ 1 per cent) of the catalogue of point sources used on this project. Therefore, the selection of GC candidates requires a more crafted statistical procedure than an off-the-shelf machine learning heuristics. However, a data-driven approach offers a few advantages. For instance, by using a set of actual GCs as our training sample, our method will avoid classifying as GCs objects with unusual properties, even if these objects can be reproduced by an unrealistic configuration of parameters from a template fitting based approach. Besides, it allows us to automatically exploit the bulk of the information available in the 12 J-PLUS filters instead of relying on a few {\it ad-hoc} combinations. Other data issues are also considered, including missing data and errors-in-measurements.  

\subsection{Missing data: Multiple Copula Imputation}
\begin{figure}
\centering
 \includegraphics[width=\columnwidth]{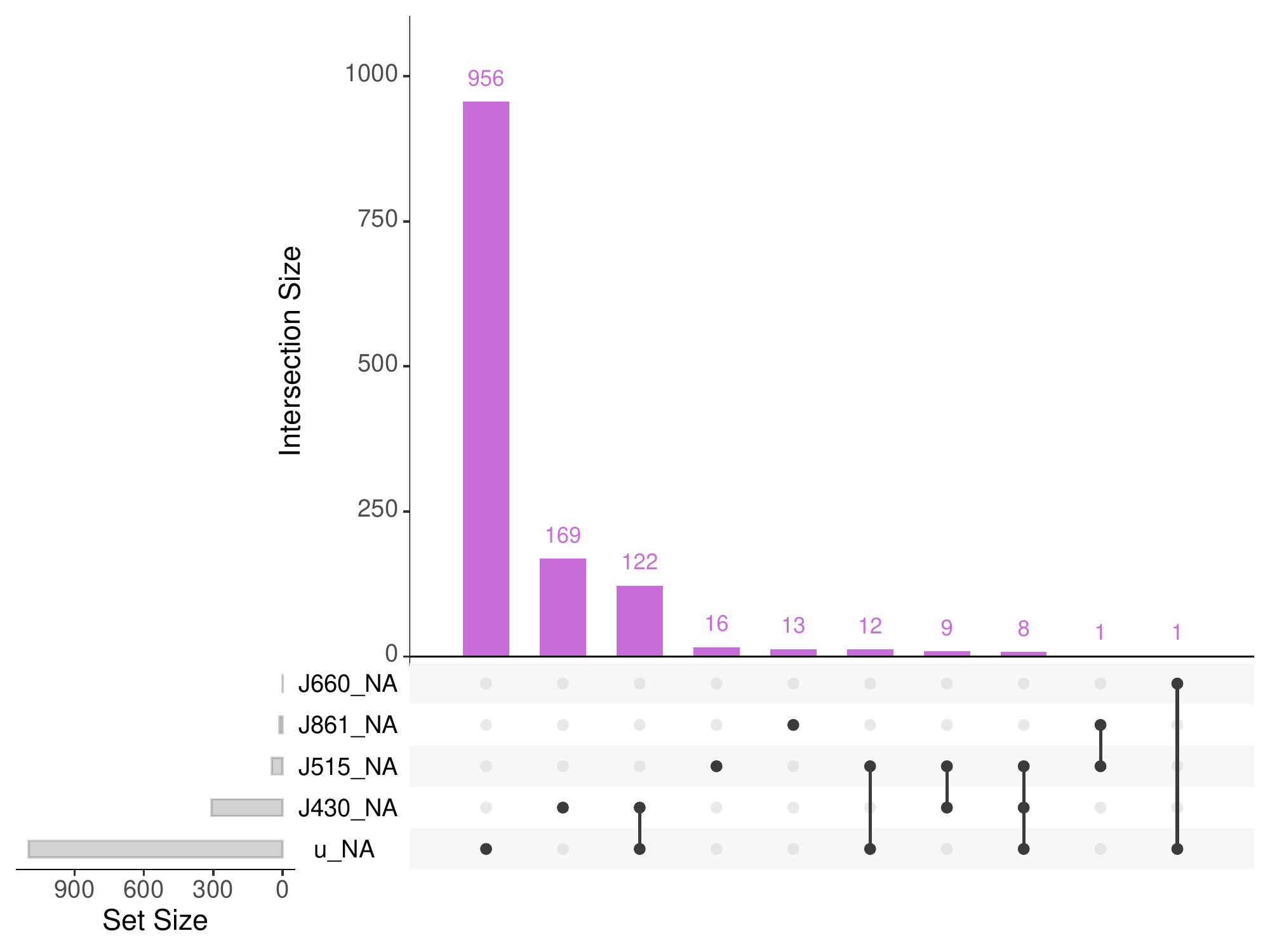}
 \caption{Missing data pattern in the labelled J-PLUS point source catalogue. Grey bars are the number of missing bands, the connected black dots indicate combinations of missing bands, and the violet histograms indicate the number of instances these combinations are missing.}
 \label{fig:NA_pattern}
\end{figure}

When constructing a catalogue, missing data is likely to occur, and here is no different. \autoref{fig:NA_pattern} shows the missing pattern of our data, from which 81.5 per cent have complete information, 13.3 per cent are missing the u band, and 1.7 per cent are missing u and J430 bands together. Hence, a naive removal of rows presenting missing information would throw away a non-negligible amount of data. The imputation here is not meant to provide "true" values for the missing bands but to marginalise them, thus enabling to use of the entire dataset. We employ a {\it Multiple Copula Imputation} (MCI). The method decomposes joint probability distributions into their marginal distributions and a copula function that couples them \citep{Nelsen10}.
Recently, \cite{Kuhn2021} used the method of MCI to aid the construction of the SPICY catalogue of young stellar objects in the Galactic midplane. Noteworthy applications of copula to astronomy include \cite{Sato2011}, and \cite{Lin2016}, who constructed likelihood functions for weak lensing analysis, and \cite{Andreani2018} that inferred bivariate luminosity and mass functions of galaxies.
Previous tests suggest that this method outperforms other popular approaches, such as multiple imputations via chained equations \citep{Stef2011} and Amelia \citep{Amelia2011}, in terms of bias and coverage, especially for non-Gaussian distributed variables  \citep{hoff2007}.  
The underlying idea of MCI is to derive conditional density functions of the missing variables given the observed ones through the corresponding conditional copulas and then impute missing values by drawing observations from them. Although the method employs a  Bayesian marginalisation under the hood, a critical difference is an assumption regarding the joint distribution of the data. The typical approach usually marginalises missing values under the assumption of a multivariate normal distribution. At the same time, the copula imputation relaxes such assumptions and computes the imputation on a transformed space (the copula space). 
In the present work we implement the MCI using the {\sc sbgcop} package \citep{sbgcop} within the {\sc r} language \citep{rcore19}. Copulas fit simultaneously to both training and target datasets. For a detailed discussion about Copula and applications, we refer the reader to \citet{Hofert2018}.

\subsection{Uncertainty aware Principal Component Analysis}

The last step of our analysis consists in projecting the 12 bands using \textit{Principal Component Analysis} (PCA). PCA is ubiquitous in data analysis because of some of its desired properties.  Generally speaking, PCA acts as a dimensionality reduction and variance modelling method. At each projection,  it minimises the information loss of the data by maximising the explained variance on each component \citep[e.g.][]{Jollife2016}. Because of its versatility, PCA is utilised in a broad range of astronomical studies \citep[e.g.,][]{Ishida2011,Ishida2011b,Ishida2013,DeSouza2014,wild2014, maltby2018,Yohana2021}. A standard approach to compute PCA is via the singular value decomposition (SVD) of the data matrix $\mathcal{X}$:
\begin{equation}
\mathcal{X} =  \mathcal{U}\Sigma \mathcal{V}^{\intercal}, 
\end{equation}
where $\mathcal{U}\Sigma$ gives the principal components, and the columns of $\mathcal{V}$  the corresponding coefficients of the linear combination of the original variables known as PC loadings.  The projection of a new point $x$ into the PCA space is then given by  $\hat{t}  = x\mathcal{V}$ so that $\hat{t}\mathcal{V}^\intercal$ is equal to $x$ in the original space.

Despite its versatility, the standard PCA has some drawbacks, it is not robust against outliers,  does not distinguish intrinsic variance from measurement errors, and does not perform well on data structures embedded in complex manifolds. This inspired the development of PCA extensions such as robust PCA (resilient to outliers), and kernel PCA (for non-linear structures). Here, we follow the framework proposed by \citet{WENTZELL1999, WENTZELL2009,Wentzell2012} and employ a PCA variant suitable to account for measurement errors. This uncertainty aware PCA finds a maximum likelihood projection of the data $\textbf{x}$ in a new subspace 
that considers a variance-covariance structure $\mathcal{Q} \equiv \mathcal{X}_{sd}^{-2}$, for the errors, this projection is given by
\begin{equation}
\hat{t} =  x \mathcal{Q}^{-1} \hat{\mathcal{V}} \left(\hat{\mathcal{V}}^{\intercal} \mathcal{Q}^{-1} \hat{\mathcal{V}}\right)^{-1}. 
\label{eq:MLPCA_proj}
\end{equation}

\noindent Therefore, we can approximate  $\textbf{x}$ in the original space by
\begin{equation}
\hat{\textbf{x}} =  \hat{\textbf{t}}\hat{\mathcal{V}}^\intercal = \textbf{x} \mathcal{Q}^{-1} \hat{\mathcal{V}} \left(\hat{\mathcal{V}}^{\intercal} \mathcal{Q}^{-1} \hat{\mathcal{V}}\right)^{-1} \hat{\mathcal{V}}^\intercal. 
\end{equation}

For the case in which the errors are all independent and identically normally distributed with fixed variance $\mathcal{Q} = \sigma^2\mathcal{I}$, where $\mathcal{I}$ is the identity matrix, the projection in (\ref{eq:MLPCA_proj}) recovers the one associated to the standard PCA:
\begin{equation}
\label{eq:pcaproj}
\hat{\textbf{t}} =  \textbf{x}  \hat{\mathcal{V}}. 
\end{equation}

Algorithm~\ref{alg:mlpca} shows a pseudocode describing the procedure,  which has been implemented in \texttt{R} language as the package \textit{RMLPCA} \citep{RMLPCA}, and in \texttt{python} \citep{Chen_2022}. In addition, codes snippets are available in Appendix~\ref{apend:mlpca} for both languages. 

\begin{algorithm}[ht]
    \caption{Maximum Likelihood PCA}\label{mlpca}
    \hspace*{\algorithmicindent} 
    \begin{algorithmic}[1]
        \Require\\
        \begin{itemize}
        \item Matrix $\mathcal{X}$;
        \item Error Matrix $\mathcal{X}_{sd}$;
        \end{itemize}
        \State Initialisation\\
        $\epsilon$ = \num{1e-10}; \Comment{Tolerance level};\\
        MaxIter = \num{1e5} \Comment{Max. Iterations};
        \State $n \gets ncol(X) \qquad m \gets nrow(X)$
         \State $i \gets 0$  \Comment{Loop counter};
         \State $\kappa$ $\gets$ $-$1 \Comment{Loop flag};
         \State $S_{old}$ $\gets$ 0 \Comment{Holds last value of objective function $S_{obj}$};
    \State Compute Singular Value Decomposition (SVD) 
      \State $\mathcal{X} = \mathcal{U}\Sigma\mathcal{V}^{\intercal}$
        \While{($\kappa$ $< 0$)} 
            \State $i \gets i+1$
            \State  $S_{obj}$ $\gets$ 0 
             \State \sbox0{$\vcenter{\hbox{$\begin{array}{|c|c|c|}
  \hline
  0.0 & \ldots & 0.0 \\ \hline
  \vdots & \ddots & \vdots \\ \hline
  0.0 & \ldots & 0.0 \\ \hline
\end{array}$}}$}%
$\mathcal{L_X}$ $\gets$ 
  $\underbrace{\vrule width0pt depth \dimexpr\dp0 + .3ex\relax\copy0}_{n}%
  \left.\kern-\nulldelimiterspace
    \vphantom{\copy0}
  \right\rbrace \scriptstyle m$
            \For{$j \in 1:n$}\\
      \State $\mathcal{Q}$ $\gets$ diag($\mathcal{X}^{-2}_{sd}$[, j]) 
       \State $\mathcal{F}$ $\gets$ $(\mathcal{U}^\intercal \mathcal{Q}\mathcal{U})^{-1}$ \
       \State $\mathcal{L_X}[, j]$ $\gets$ $\mathcal{U}\left( \mathcal{F} \left(\mathcal{U}^{\intercal}(\mathcal{Q}\mathcal{X}[, j])\right)\right)$
     \State  $\mathcal{D_{X}}$ $\gets$  $\mathcal{X}[, j] - \mathcal{L_X}[, j]$  \Comment{Residuals}
     \State  $S_{obj}$ $\gets$ $S_{obj}$ + $\mathcal{D^{\intercal}_{X}}\mathcal{Q}\mathcal{D_{X}}$
                \EndFor
  
  \If{$i \mod 2 = 1$}\Comment{Convergence check}
  \State $\epsilon^{\prime}$ $\gets$ $\left\|\frac{S_{old} - S_{obj}}{S_{obj}}\right\|$
  \If{$\epsilon^{\prime}$ < $\epsilon$}
  \State $\kappa$ $\gets$ 0
  \EndIf
  \If{$i > {\rm MaxIter}$}
  \State $\kappa$ $\gets$ 1 \Comment{Max Iterations exceeded}
  \EndIf
 \EndIf
 \If{$\kappa < 0$}{
 \State $S_{old}$ $\gets$ $S_{obj}$
 \State Compute SVD
 \State  $\mathcal{L_X} = \mathcal{U^{\prime}}\Sigma^{\prime}\mathcal{V^{\prime}}^{\intercal}$,   \qquad        
 $\mathcal{U} \gets \mathcal{U^{\prime}}$, \qquad
 $\Sigma \gets \Sigma^{\prime}$, \qquad
\State $\mathcal{V} \gets \mathcal{V^{\prime}}$,  \qquad
 $\mathcal{X} \gets  \mathcal{X}^{\intercal}$, \qquad
 $\mathcal{X}_{sd} \gets \mathcal{X}^{\intercal}_{sd}$, 
 \State  $n \gets ncol(\mathcal{X})$,  \qquad
 $\mathcal{U} \gets \mathcal{V}$
 \EndIf
 }
\EndWhile
\Ensure
 \State Compute final SVD
 \State  $\mathcal{\hat{L}_X} = \hat{\mathcal{U}}\hat{\Sigma}\mathcal{\hat{V}^{\intercal}}$
 \State Compute Matrix Deconvolution
 \State $\hat{\mathcal{X}} \gets \hat{\mathcal{U}}\hat{\Sigma}\mathcal{\hat{V}^{\intercal}}$
\State Compute standard PCA on $\hat{\mathcal{X}}$
    \end{algorithmic}
\label{alg:mlpca}
\end{algorithm}


\subsection{Flagging GC candidates}
\label{sec:mlpca}
\begin{figure}
\centering
 \includegraphics[width=\columnwidth]{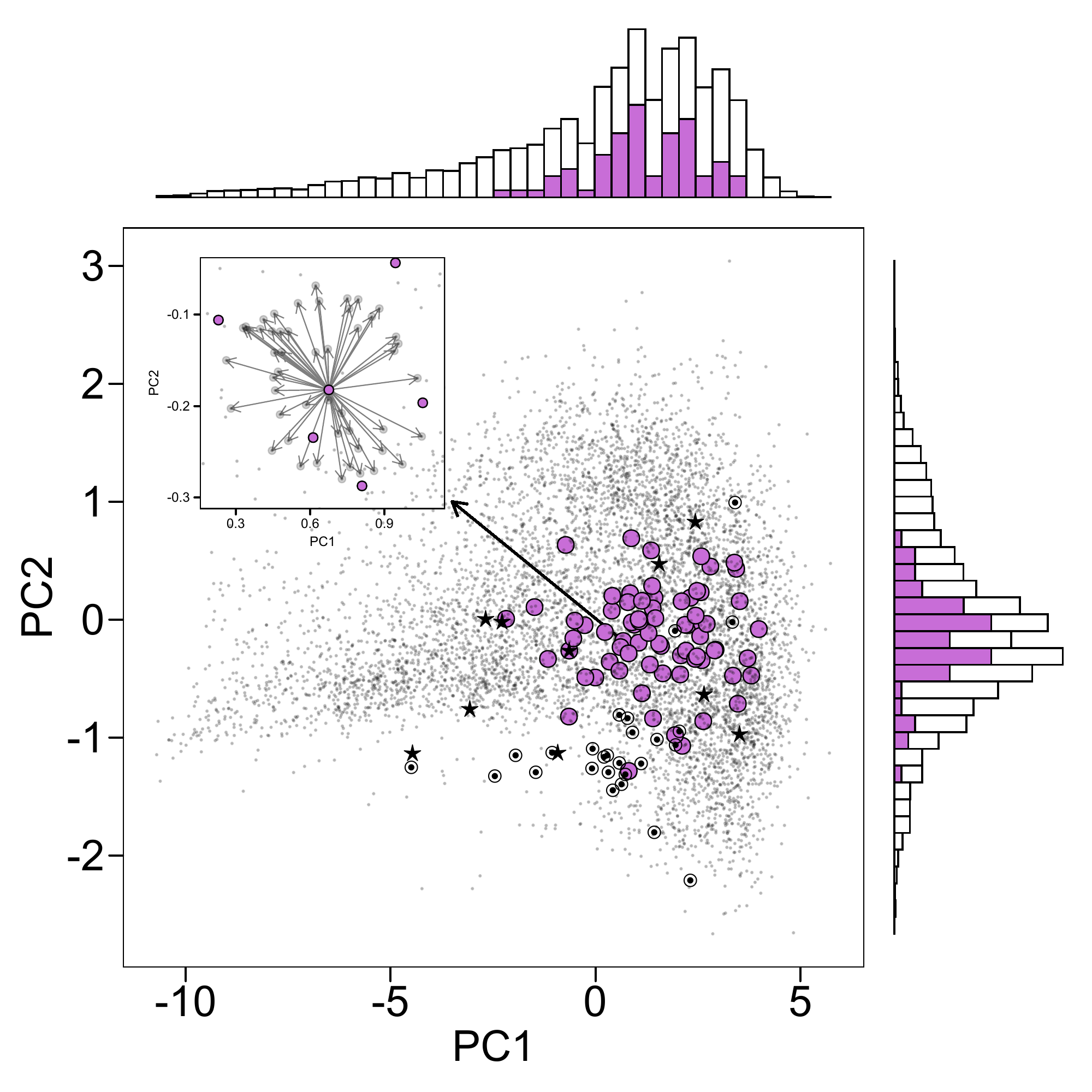}
 \caption{The PC1 vs. PC2 projection for the point sources and spectroscopically confirmed GCs, according to the legend. Background galaxies (dot-filled circles), and known Milky-Way Stars (black stars) are over-plotted as reference.
 The side histograms indicate the  PC2 and PC2 distributions for spectroscopically confirmed GCs (filled violet histogram) and point sources (open histogram). The inset panel illustrates the PSM heuristics using one GC as an example and the 50 closest points connected by grey arrows.}
 \label{fig:pca1}
\end{figure}

We search for GC candidates in the projected PCA space by matching confirmed GCs with their closest counterparts. \autoref{fig:pca1} displays the list of known GCs projected in two principal components, together with all available point sources. We also show the known galactic stars and background galaxies as a reference. A visual inspection suggests that most of the known GCs occupy a well-defined locus, with only a low to moderate contamination. The match relies on a non-parametric approach known as propensity score matching \citep[PSM;][]{HoImaKin07,Austin2011}. As stated in \autoref{eq:pcaproj}, the projection of each object into the PCA space is given by  $\textbf{x}\mathcal{V}$. After computing the rotation matrix $\mathcal{V}$, we can then express each component PC1 and PC2 in terms of the standardised\footnote{Subtracted by mean and divided by standard deviation.}  J-PLUS filters:

\begin{align}
&PC1 = 0.27u +  0.28J378 + 0.29J395 + 0.29J410 + 0.29J430  + \\ 
&0.30g + 0.30J515 + 0.30r + 0.29J660 + 0.28i + 0.28J861 + 0.28z,  \nonumber \\
&PC2 = 0.41u +  0.37J378 + 0.33J395 + 0.22J410 + 0.19J430  + \nonumber\\ 
&0.02g - 0.01J515 - 0.17r - 0.18J660 - 0.33i - 0.40J861 - 0.40z. \nonumber
\end{align}

\begin{figure}
\centering
 \includegraphics[width=\columnwidth]{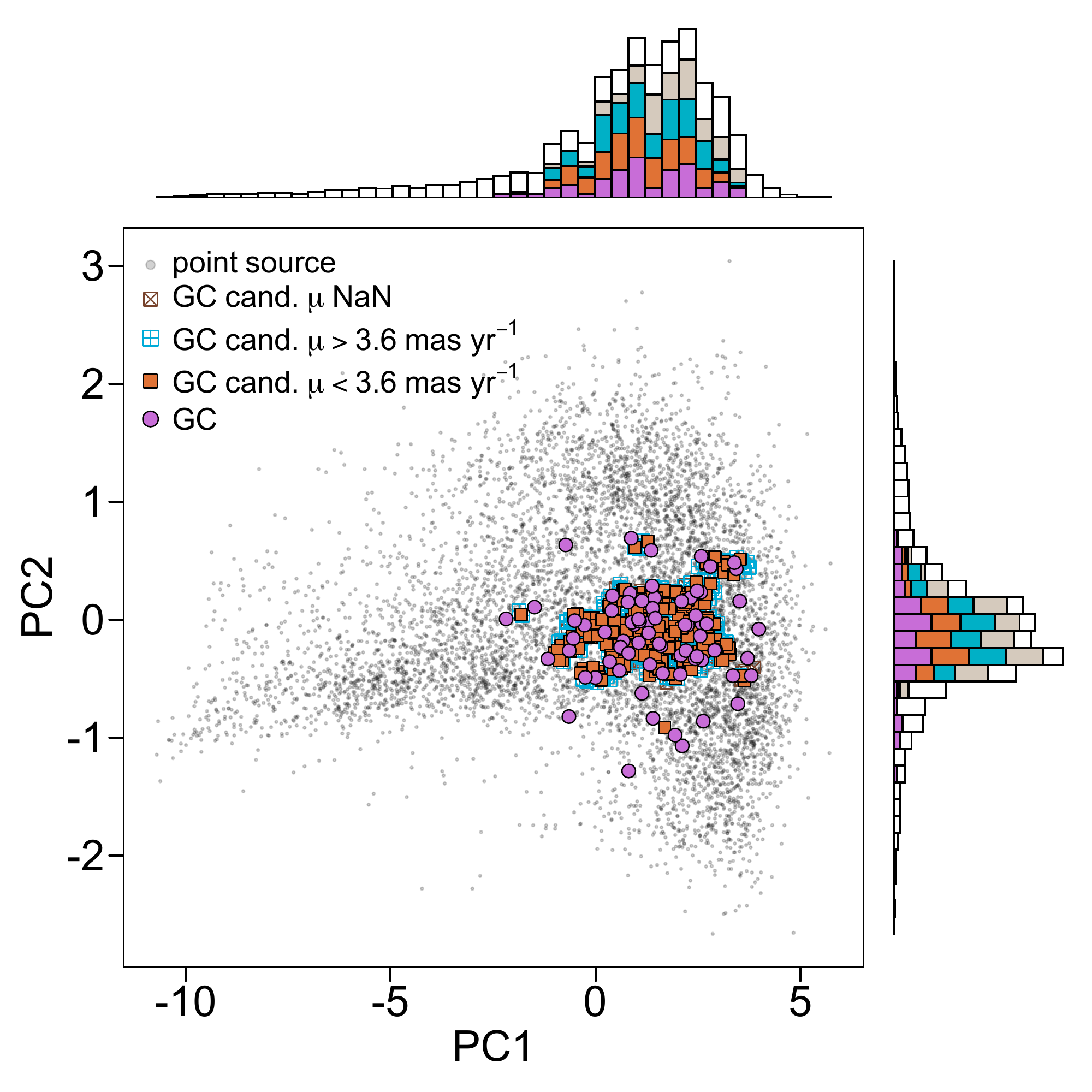}
 \caption{The PC1 vs. PC2 projection and distributions as in Fig.~\ref{fig:pca1} for point sources, spectroscopically confirmed GCs and GC candidates according to the legend.}
 \label{fig:psm_pca}
\end{figure}

For each confirmed GC, the method searches for the 50 closest candidates with replacement, i.e., a given point source can be matched with more than one confirmed GC -- the inset panel at \autoref{fig:pca1} illustrates the heuristics. We then disregard any point source that failed to match any given GC at least once. The choice of 50 ensures the stability of the search. Increasing this number does not affect the list of candidates. This process leads to a list of 642 candidates, which we flag accordingly to Gaia proper motions, i.e., if  $\mu$ is higher or lower than $3.6\,{\rm mas\,yr^{-1}}$ (see \autoref{fig:pm_hist}), which represents the 95 percent quantiles of the current distribution of known GCs or not available. We note that if we assume a typical GC tangential velocity of 250 km/s, a more strict cut of $\sim 0.014\, {\rm mas\,yr^{-1}}$ (assuming a distance of 3.6 Mpc) should take place. However, the existence of spectroscopically confirmed GCs with measured proper motions  $>  10\, {\rm mas\,yr^{-1}}$ suggests that a non-negligible measurement scatter around sources with nearly zero proper motion. Hence, this criterion does not represent a hard cut for the candidate's credibility, but it conveys a simple rule to tag the most probable candidates. \autoref{fig:psm_pca} shows the candidates projected in two principal components and conveys the intuition behind our approach; the method is conservative in the sense that it only flags within the coverage of the known GCs and in the densest regions. 
Despite their limited size, the samples of galactic stars and background galaxies described in Section\,\ref{sec:anc} convey some complementary information to test our procedure. Firstly, none of the Galactic stars and just one of the background galaxies were included in our selection of candidates to GCs. Besides, more than 60 per cent of the galactic stars have proper motion larger than $3.6\,{\rm mas\,yr^{-1}}$.
Thus, in \autoref{sec:res} we use this flag in the plots analysing the GC candidates colours and spatial distribution.

\section{Analysis and Results}
\label{sec:res}

Here we present the analysis of some basic properties of the catalogue of GC candidates we selected through the methodology outlined in \autoref{sec:met}. We study the spatial distribution, the colours and metallicities and show a few example J-PLUS spectral energy distributions (SEDs) of the candidates. 

\subsection{Spatial Distribution}
\begin{figure*}
\centering
 \includegraphics[width=.45\linewidth]{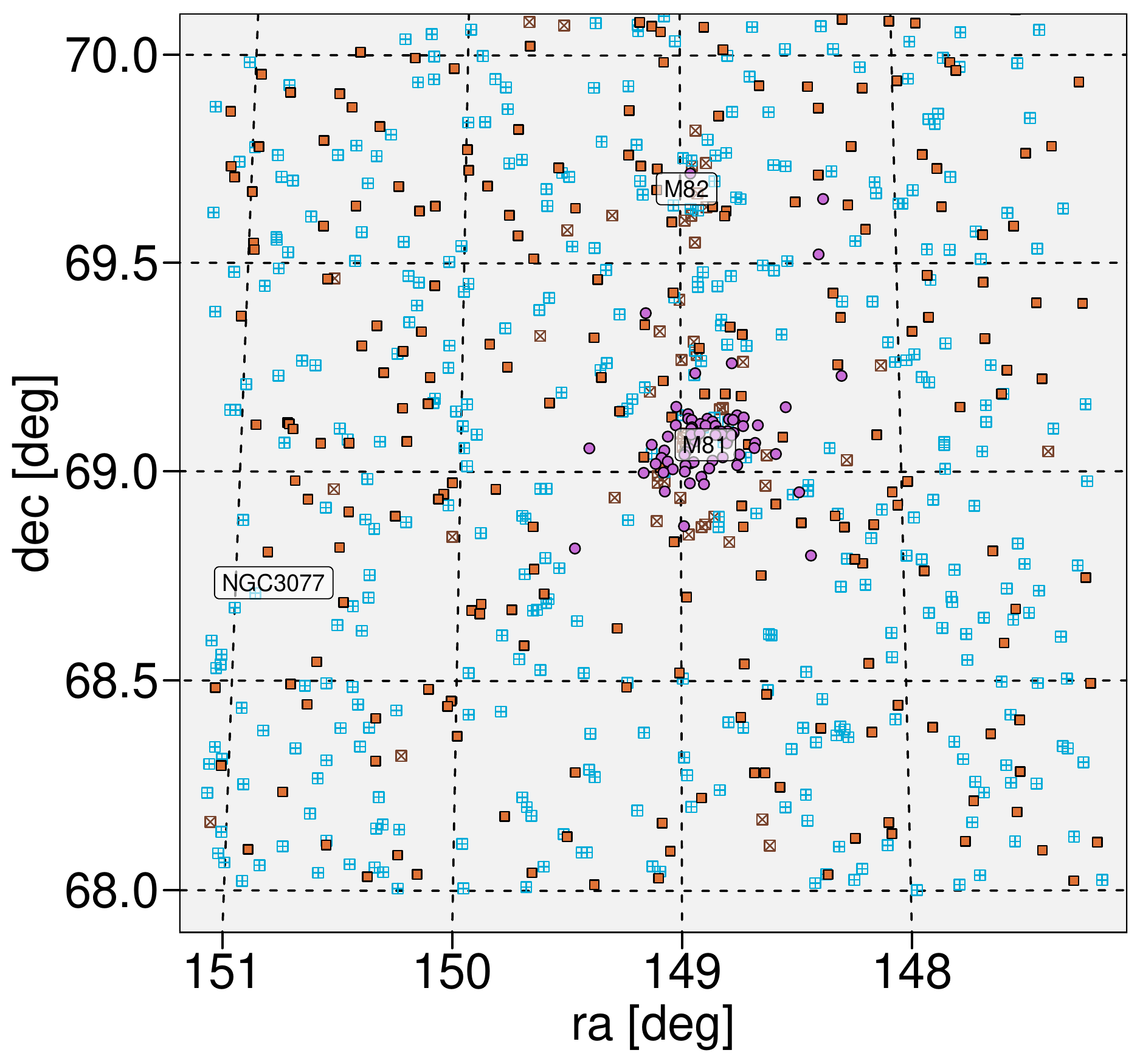}
 \includegraphics[width=.45\linewidth]{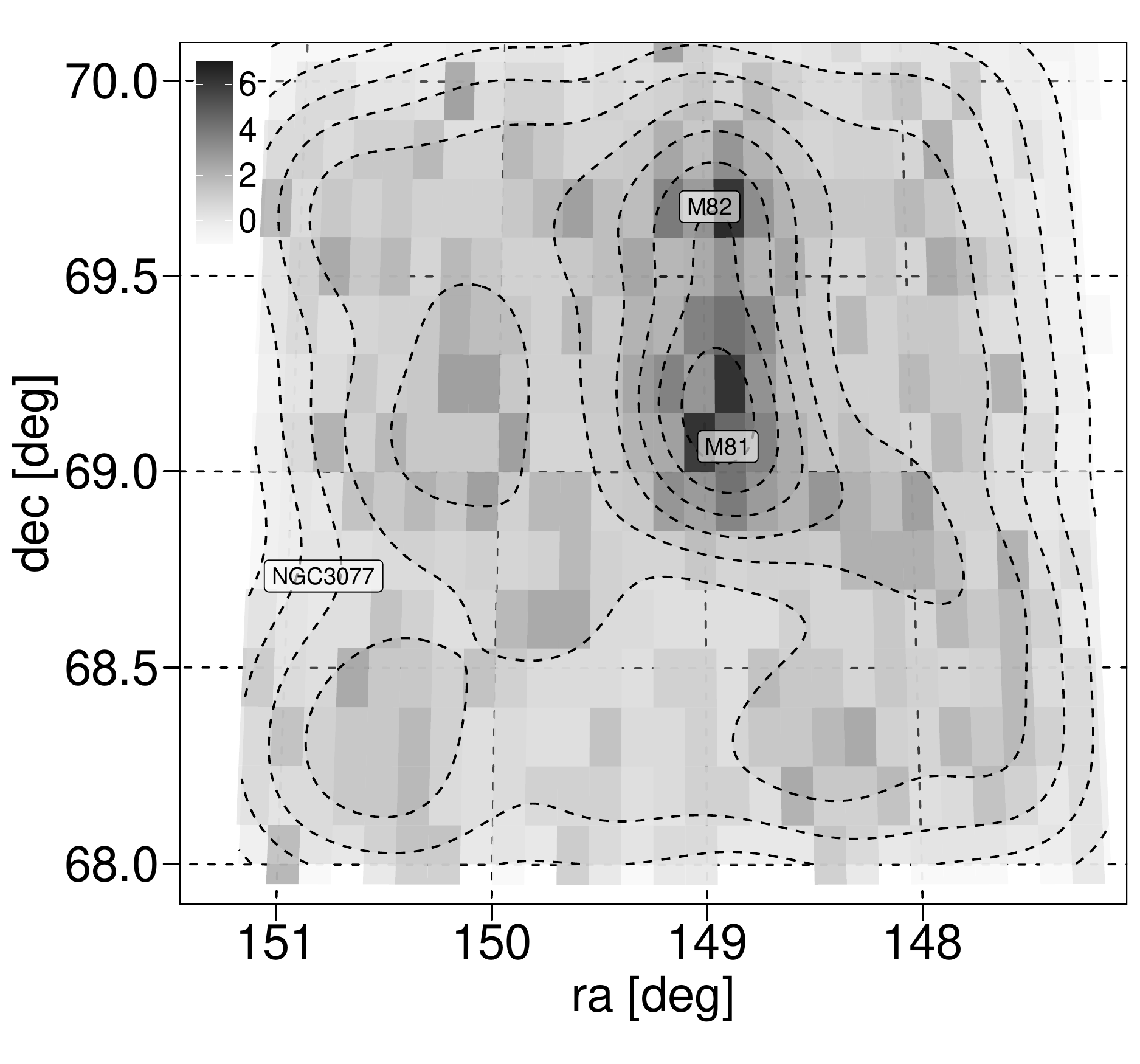}
 \caption{
 \textit{Left panel:} Spatial distribution in Aitoff projection of spectroscopically confirmed GCs and GC candidates shape-coded as in Fig.~\ref{fig:psm_pca}. \textit{Right panel:} Count distribution of GC candidates from low (light grey) to high (dark grey) counts with over-plotted density contours. A GC count excess between M\,81 and M\,82 provides a tantalising indication of a potential GC bridge.}
\label{fig:spatial}
\end{figure*}

In \autoref{fig:spatial} we show the spatial distribution of the candidates and previously known GCs in an Aitoff projection.
While most of the previously known GCs are close to M\,81, the new GC candidates appear to be distributed across the entire region.
The count distribution of GC candidates with over-plotted density contours shown in the right panel of \autoref{fig:spatial} reveals an overdensity of GC candidates extending from M\,81 in the direction of M\,82.
Such count excess provides tantalising evidence of a GC bridge.
Although a bridge of H\,{\sc i} gas between M\,81 and M\,82 is evident from \cite{deb18}, see \autoref{fig:triplet}, a bridge of GCs between these galaxies has not yet been reported. 
A bridge of GCs has been found to exist between the SMC and LMC \citep{bica2015}. If we go further out in the local group, towards Andromeda, the density of GCs around M\,31 seems higher in the direction of M33, along the major axis of the galaxy, if compared to the opposite direction (see Fig. 1 of \citealt{huxor2009}).
In Appendix~\ref{apend:spatial} we further discuss the potential bridge taking into account the different proper motion samples (\autoref{fig:spatial_split}).
Another interesting feature of \autoref{fig:spatial} is the absence of cluster candidates towards the south of M\,81, opposite to the location of M\,82.
Moreover, we do not see many GC-like objects close to NGC\,3077. This could be due to the fact that the stellar cluster population of this galaxy is dominated by younger clusters, whose SEDs differ from that of old globulars. Therefore, our methodology does not extract such objects from the data.

\begin{figure}
\centering
 \includegraphics[width=\columnwidth]{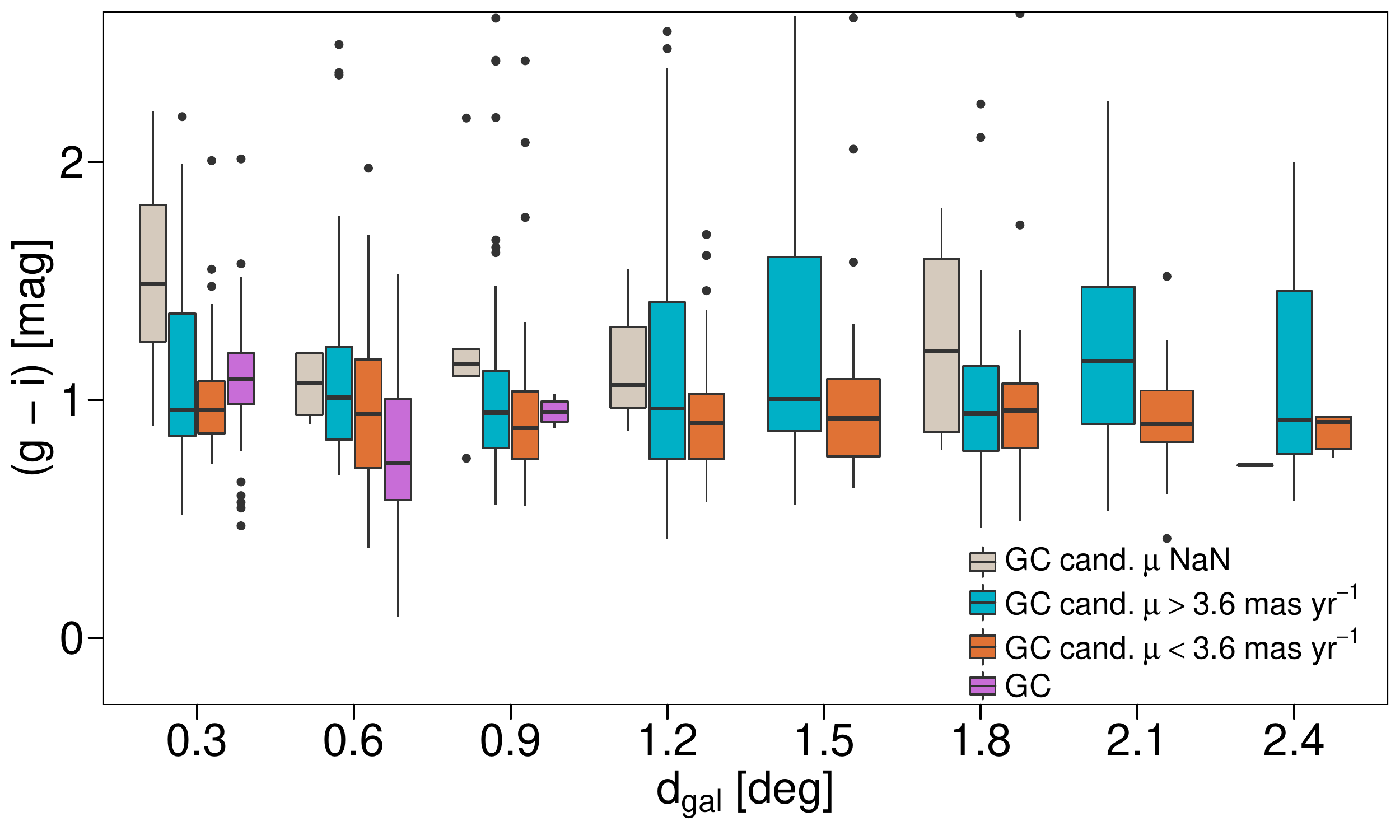}
 \caption{$(g-i)$ colour as a function of the projected distance, $d_{gal}$,  to the nearest galaxy in the triplet,  for GC candidates and spectroscopically confirmed GCs according to the legend. Confirmed GCs from the literature are in the range of intermediate to blue, and the candidates with lower proper motions (orange boxplots) follow a similar trend. This trend is compatible with halo and intracluster GCs.}
 \label{fig:gi_dgal}
\end{figure}

The $(g-i)$ colour of the candidates as a function of their minimum projected distance to the pair M\,81/M\,82 are indicated in \autoref{fig:gi_dgal}.
As expected, there is a marginal increase in the population of redder GCs closer to the galaxy,
indicating a metallicity gradient.
Although the direct separation of two colour modes into two metallicity modes should be taken carefully \citep{chi12, blakeslee12, pow17, lee19, fahrion2020},
the colours in optical bands have largely been used to separate GCs in bluer and redder GCs (`more metal-poor' and `more metal-rich', respectively), that present distinct behaviours in their spatial distribution \citep{bassino06,esc15} and kinematics \citep{sch10,pot13}.
This has been suggested as a consequence of the processes ruling the build-up of the GC system \citep{for11, choksi2018, kruijssen19}. 
From \autoref{fig:gi_dgal} it is clear that the GC systems in the triplet fit with the picture of redder GCs being more concentrated towards the centre of the hosts while the bluer ones being more extended to larger distances from the centre of the host and dominating at larger radii \citep{bassino06,brodie06,beasley20}. 

\subsection{Colours}

\begin{figure*}
\centering
 \includegraphics[width=1\linewidth]{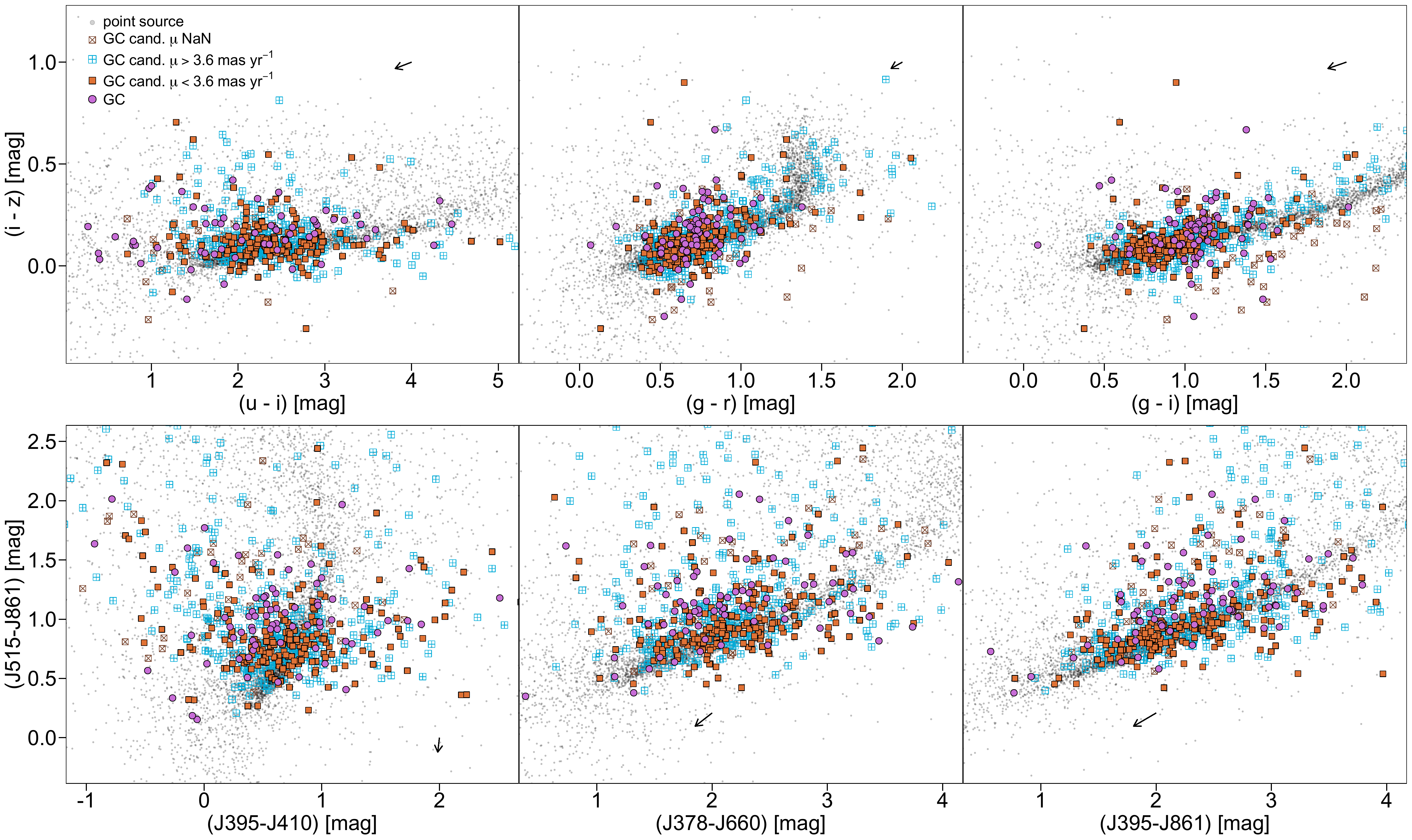}
 \caption{Example colour-colour diagrams for the point sources, spectroscopically confirmed GCs and GC candidates according to the legend. Overall, the candidates occupy a similar locus as the confirmed GC counterparts in different projections. The black arrows indicate the reddening vectors.}
 \label{fig:CCD_GCs}
\end{figure*}

The colour-colour diagrams and colour distributions of the candidates provide extra insight into how our method selects the GCs. \autoref{fig:CCD_GCs} show 6 representative colour-colour diagrams for the selected sources in the J-PLUS narrow and broad-band filters. The distribution of the sources seem plausible since most of the candidates are within the coverage of the known GCs. However, the candidates with higher proper motions ($\mu > 3.6~{\rm mas~yr^{-1}}$) are overall more spread than their lower proper motion counterparts. To illustrate the foreground extinction and to allow for easy comparison with other GC systems, we show a reddening vector calculated assuming $E(B-V)=0.068\pm0.012$ (\citealt{Schlafly_Finkbeiner2011}) towards the direction of the centre of the field analysed in this work.

\begin{figure*}
\centering
 \includegraphics[width=0.95\linewidth]{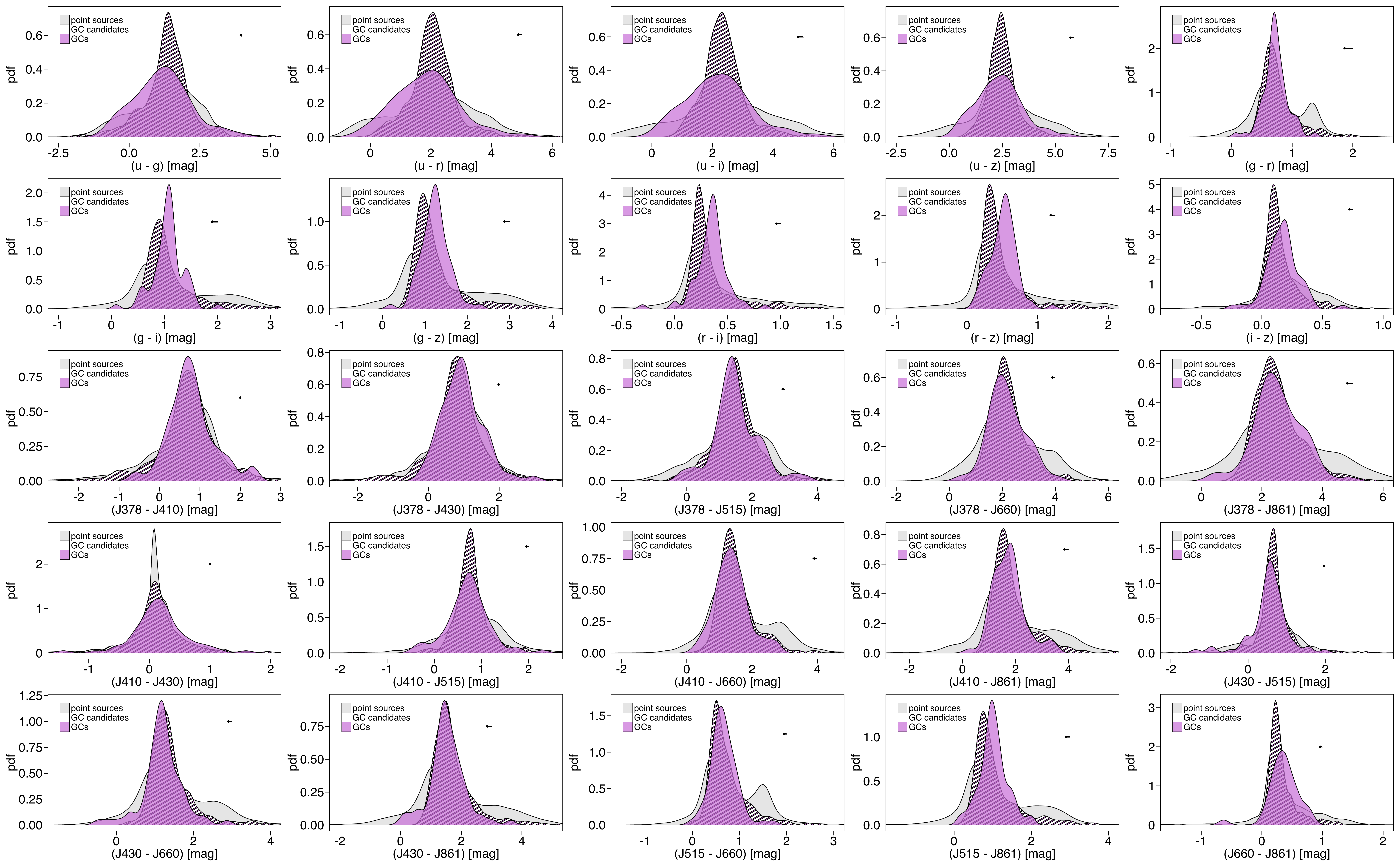}
 \caption{Narrow and broad-band colour distributions for point sources, spectroscopically confirmed GCs and GC candidates according to the legend. Our methodology selects GC candidates roughly consistent with the spectroscopically confirmed GCs in most colours, and excludes very red objects (more likely to be background galaxies) in some of them, as e.g. $(g-r)$. The black arrows indicate the reddening vectors.}
 \label{fig:coldist}
\end{figure*}

In \autoref{fig:coldist} we show narrow and broad-band colour distributions for point sources, spectroscopically confirmed GCs and GC candidates. 
From the distributions, it is clear that our methodology selects GC candidates roughly consistent with the spectroscopically confirmed GCs in most colours and excludes very red objects (see, e.g. the $(g-r)$ distributions) that are consistent with background galaxies.
In commonly used optical colours such as $(g-i)$ (\citealt{harris2016}) and $(g-z)$ (\citealt{peng06}, \citealt{beasley18}) our sample of GC candidates peaks at slightly bluer colours, if compared to the spectroscopically confirmed GCs. This goes in hand with what is seen in \autoref{fig:gi_dgal} and is not surprising, given that blue GCs are expected to dominate at large galactocentric radii.
Looking at the $(g-i)$ distribution of candidates, we find that  85 per cent of the low proper motion candidates fall in the range $0.6<(g-i)<1.6$\,mag, which is in agreement  with the typical colours of extragalactic GCs in this photometric system \citep[e.g.][]{sinnott10,chies11a,faifer11,cas19,enn20} under the assumption of foreground reddening $E_{(g-i)}\sim 0.13$\,mag \citep{Schlafly_Finkbeiner2011}. Moreover, there is a tail of candidates presenting $(g-i)>1.6$\,mag, they are mainly fainter than $r=19$\,mag and $\sim 2/3$ of them have higher proper motion values. 

\subsection{Metallicities}
\label{sec:feh}

\begin{figure*}
\centering
 \includegraphics[width=\linewidth]{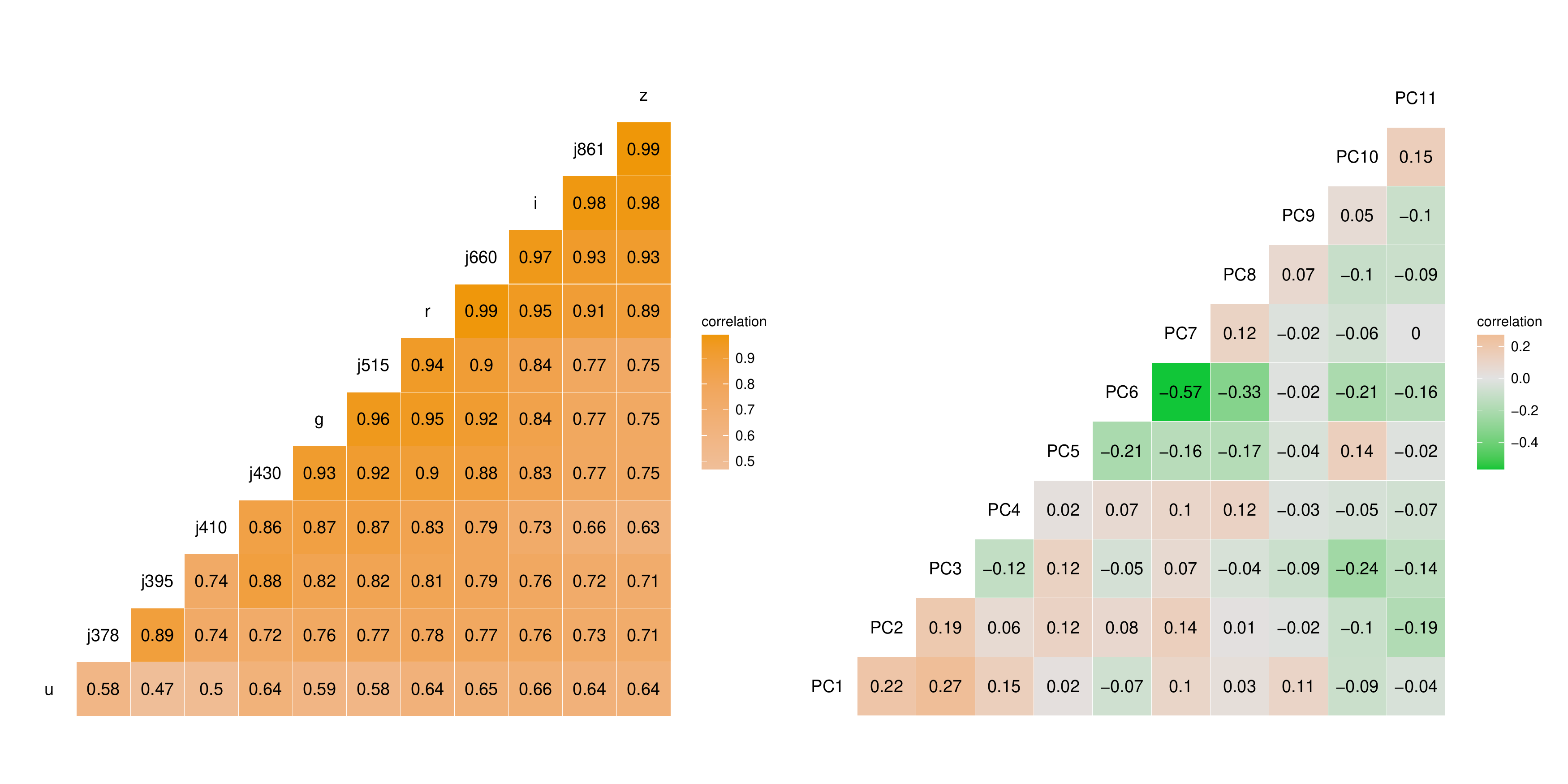}
 \caption{Pair-wise Pearson correlation of the J-PLUS photometric bands (left panel) and the principal components projections (right panel). The red and blue colours represent positive and negative correlation. The PCA projection heavily decreases correlation among the original vectors.}
 \label{fig:corr}
\end{figure*}

\begin{figure*}
\centering
 \includegraphics[width=\linewidth]{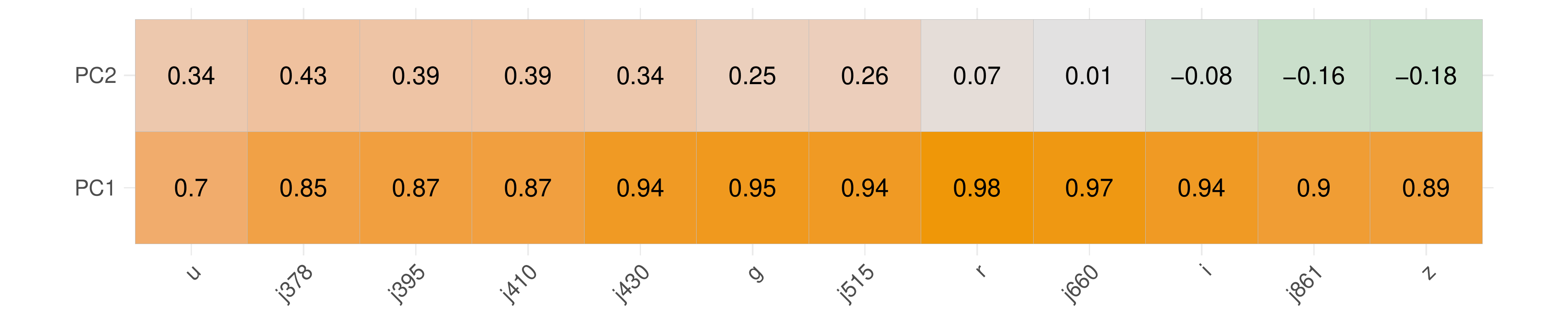}
 \caption{Pair-wise Pearson correlation between the J-PLUS photometric bands and the two first principal components projections.}
 \label{fig:corr_pca}
\end{figure*}

\begin{figure*}
\centering
\includegraphics[width=0.45\linewidth]{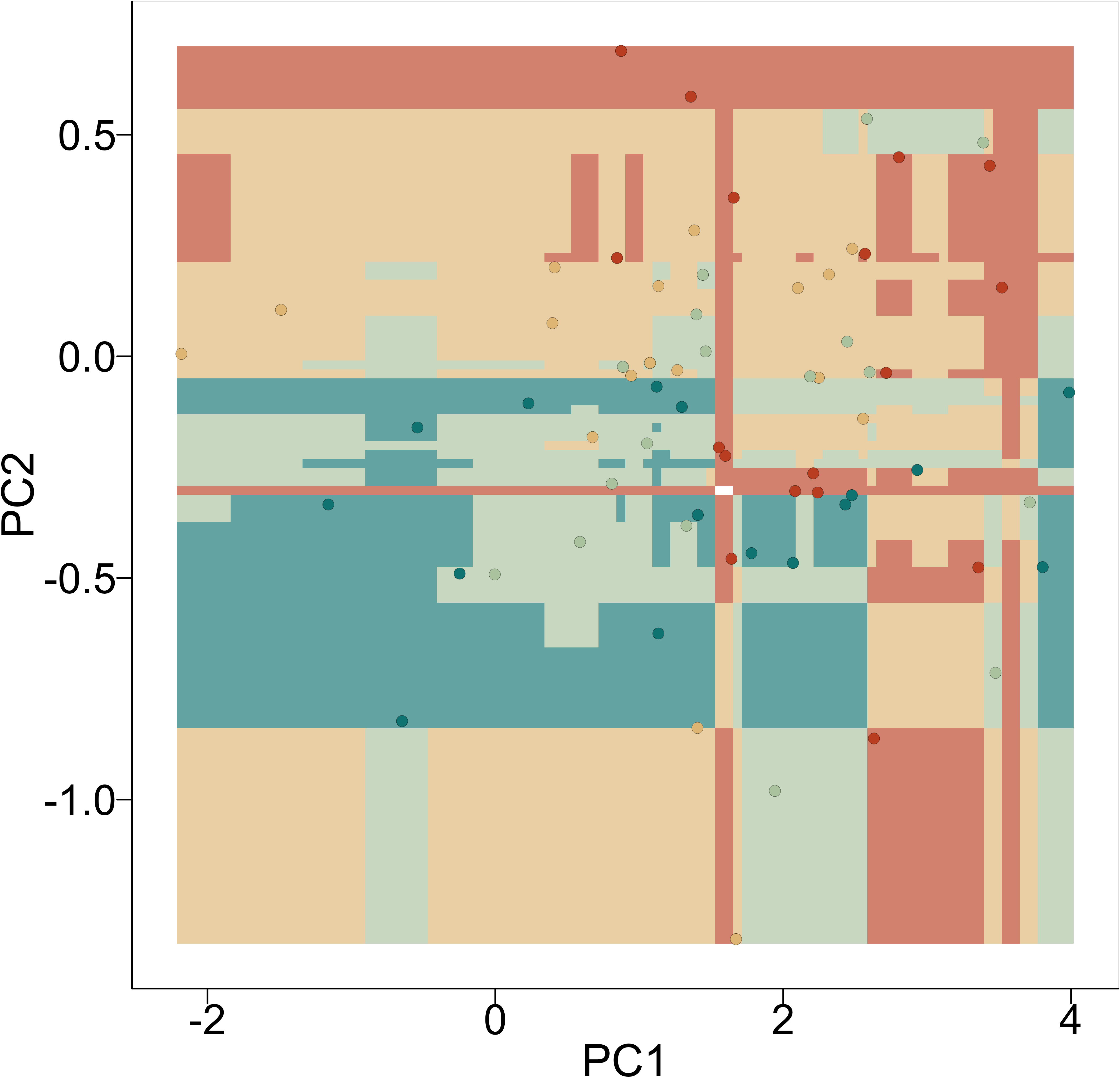}
\includegraphics[width=0.45\linewidth]{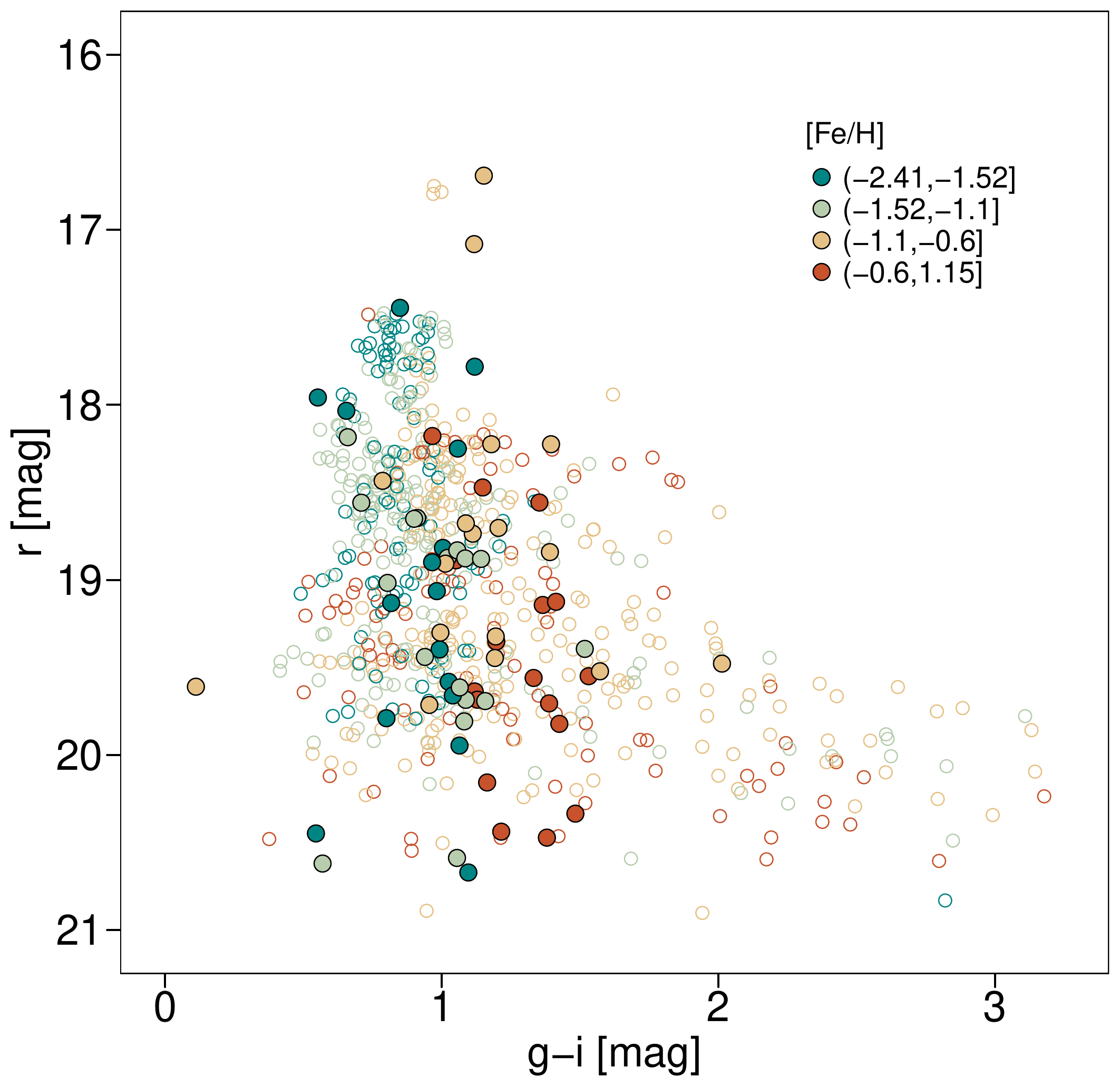}
 \caption{Left: Two-dimensional regression map via \texttt{XGBoost} of the predicted GC metallicity as function of the first two principal components vectors. Right: colour-magnitude distribution of the GC and GC candidates colour-coded in bins of metallicity.}
 \label{fig:fit_Fe}
\end{figure*}

Based on the sample of 73 spectroscopically confirmed GCs with derived metallicities (\citealt{nantais2010_spec}), we build a function to map the metallicity distribution of confirmed GCs to their respective photometric bands in order to infer the metallicity distribution function (MDF) of the GC candidates. More specifically, we model the GC metallicity as a function of  J-PLUS photometric bands with \texttt{XGBoost} \citep{Chen2016}, a scalable regression-tree-based model that outperforms deep-learning-based approaches when it is comes to model tabular data \citep{SHWARTZZIV2021}. In a nutshell, the intuition behind different regression models relies on how we approximate the underlying relationship between a given response variable, here defined by the metallicity, and a set of covariates, herein represented by the first two PCs. The choice for only two PCs is motivated by both being able to explain more than 97 percent of the data variance.
In a hypothetical scenario where the relationship could be amenable by a linear regression model, the linear relation would be written as:
\begin{align}
[Fe/H] = \beta_0 + \beta_1PC1 + \beta_2PC2. 
\end{align}
Where $\beta_0$ is the intercept and $\beta_1$ and $\beta_2$ are the respective slopes of each covariate. On the other hand, is a preconceived  parametric regression does not exists or is unknown, a non-parametric approach is needed. Different solutions have been proposed in the literature, notable examples include additive models, kernel-based models, neural networks and regression-tree models. While a comprehensive discussion about non-parametric regression is beyond the scope of this paper, the underlying intuition behind tree-based models is to approximate the unknown relation by a series of additive functions:
\begin{align}
[Fe/H] = \sum_{k=1}^{K}f_k(PC1,PC2).
\end{align}
Each $f_k$ corresponds to an independent tree structure $g(\boldsymbol{x},T_h)$, where $g$ is a step function and $T_h$ is the $h$-$th$ tree. The model then partitions the space of
covariates and fits a series of trees in each of them.
We choose \texttt{XGBoost} because empirical results have shown that it enables us to uncover complex relations without fine-tuning. Furthermore, we perform a PCA regression instead of using all twelve J-PLUS filters to mitigate co-linearity and over-fitting.   \autoref{fig:corr} displays a pair-wise correlation matrix of the filters and PCs. The filters are highly correlated, a feature known to cause regression models instability. The PCs, on the other hand, show weak to moderate correlation -- a {\it desiderata} for any regression analysis. \autoref{fig:corr_pca} displays the correlation between the first two PCs and the J-PLUS filters. The first PC correlates strongly with most of the colours, except for the $u$-band, which is expected since most colours are highly correlated. To convey intuition about the model solution, the left panel of \autoref{fig:fit_Fe} depicts the regression plane between the first two PCs and the [Fe/H] abundance. The colours represent different bins of metallicity. The points depict the metallicities of the training sample, while the contours the predicted value by the \texttt{XGBoost}. The right panel shows the predicted metallicities as open circles and the metallicities of known GCs as solid circles.

\begin{figure}
\centering
\includegraphics[width=\columnwidth]{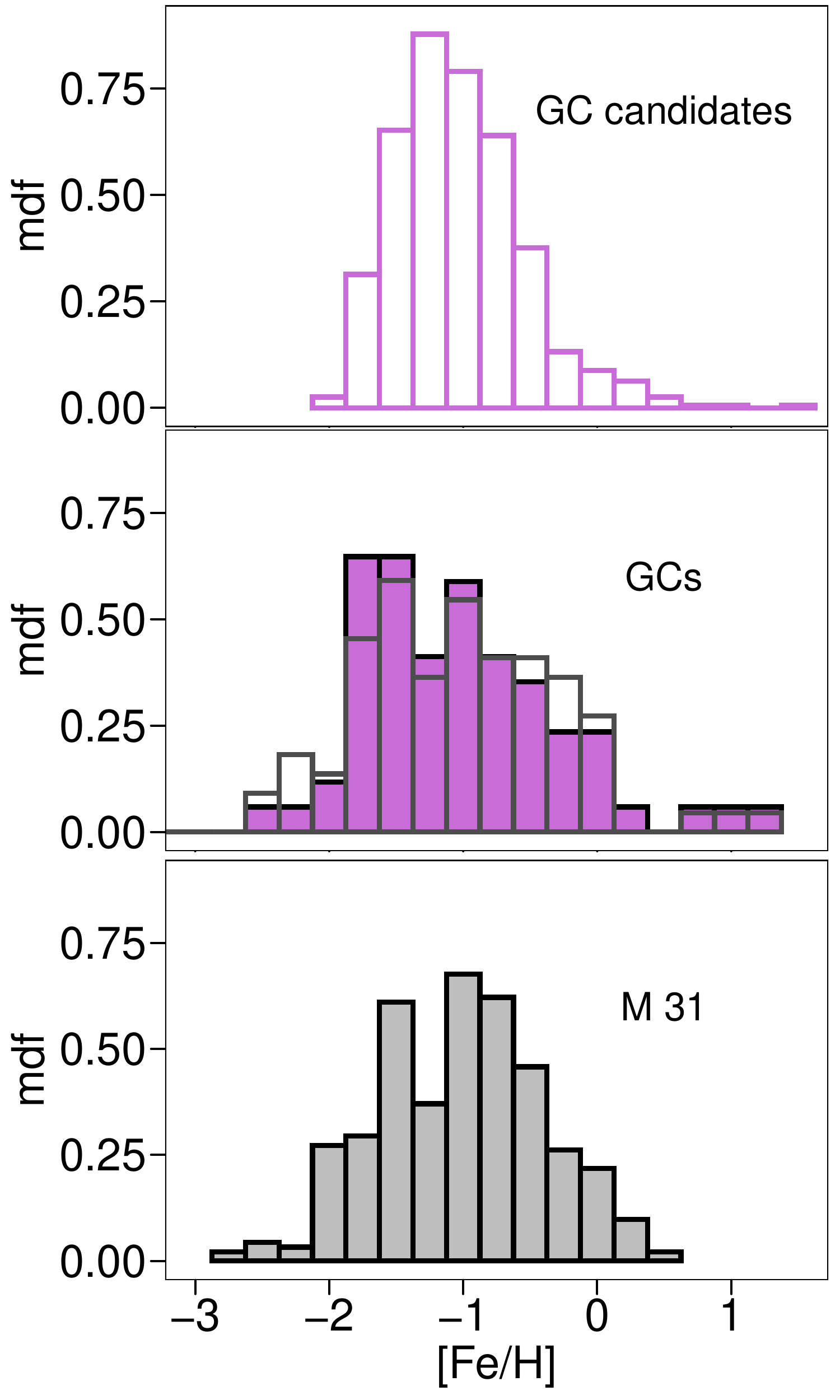}
 \caption{The metallicity distribution function of GC candidates (top panel), spectroscopically confirmed GCs (middle panel), and M 31 GCs (bottom panel, \citealt{caldwell16}). The solid grey histogram on the middle panel shows both the training sample we employ on this paper and the complete sample of GCs with metallicity available in the region (see \autoref{tab:litGCs}), for which we don't have J-PLUS counterparts.}
 \label{fig:Fe}
\end{figure}

In \autoref{fig:Fe} we show the MDF of the GC candidates and the spectroscopically confirmed GCs with metallicities available (including our training sample). We stress that we are using a small and biased training sample, which is far from ideal. Nevertheless, it can give us a grasp of our sample GC candidates' MDF shape. 
The top panel of \autoref{fig:Fe} shows that the MDF of the GC candidates is broad, -2.5 $\lesssim$ [Fe/H] $\lesssim$ 0.5, but has an important tail towards the metal-rich end (-0.5 $\lesssim$ [Fe/H] $\lesssim$ 0.7). 
A comparison to the work of \cite{Caldwell2011} and the catalogue published in \cite{caldwell16} for the M\,31 GC system (bottom panel of \autoref{fig:Fe}) shows similarities, including such metal-rich tail at similar values. While the Milky Way GC system shows two clear peaks, this is not seen for either M\,31 or the M\,81/M\,82 systems. 
A visual inspection of  \autoref{fig:Fe} indicates that our method is conservative and does not select GC candidates towards the tails of the MDF of the spectroscopically confirmed GCs. 
Thus, we caution the reader that we are not complete towards the extremes of metallicity distribution. 

\subsection{Spectral energy distributions}
\begin{figure*}
\centering
\includegraphics[width=\linewidth]{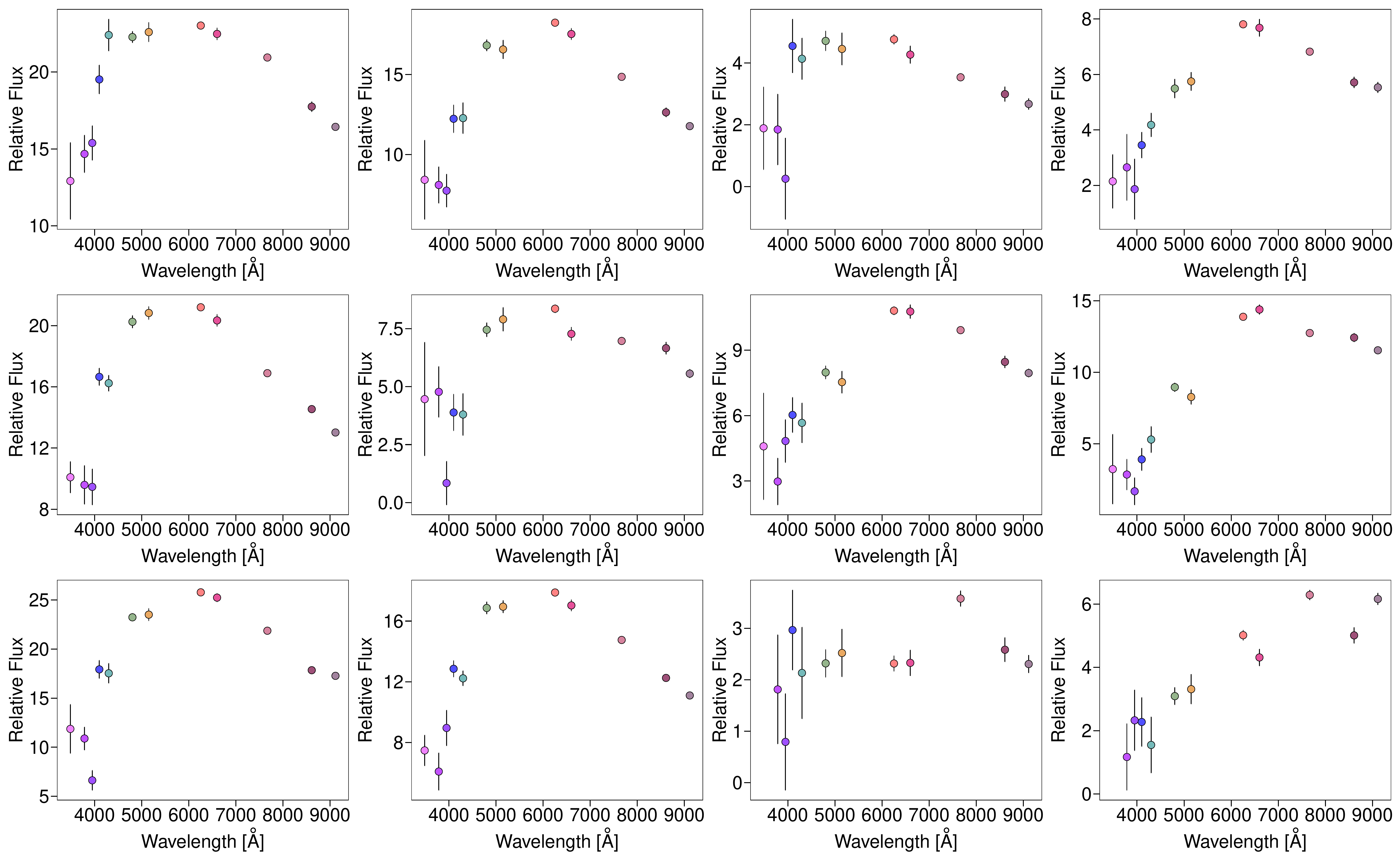}
 \caption{SED examples of the GC candidates. From left to right panels are divided in bins of metallicity -2.41 < [Fe/H] <= -1.155;  -1.155 < [Fe/H] < -1.095; -1.095 < [Fe/H] < -0.6 and 
 [Fe/H] > -0.6). From top to bottom are GC candidates with proper motion < 3.6, > 3.6 and no measured proper motion.}
 \label{fig:SEDs}
\end{figure*}

With the aim of illustrating the shapes of our GC candidate SEDs we show a few example SEDs of GCs of different metallicities and proper motion values in \autoref{fig:SEDs}. 
We divide the sample of GC candidates in bins of metallicity, according to the analysis presented in Section.~\ref{sec:feh}.
From left to right, the  panels are divided in bins of metallicity: -1.91 $<$ [Fe/H] $\leq$ -1.37;  -1.37 $<$ [Fe/H] $\leq$ -1.09; -1.09 $<$ [Fe/H] $\leq$ -0.76 and [Fe/H] $>$ -0.76). From top to bottom we show GC candidates with $\mu$ $<$  $3.6\,{\rm mas\,yr^{-1}}$, $\mu$ $>$ $3.6\,{\rm mas\,yr^{-1}}$ and no measured proper motion. A visual inspection suggests that the SEDs resemble significantly the SEDs of confirmed GC (\autoref{fig:gc_spectra}), specially the more metal-poor GC candidates (left panels).

\section{Summary and Concluding Remarks}
\label{sec:conc}

To build a large and homogeneous catalogue of GC candidates around the M\,81/M\,82/NGC\,3077 triplet with 12 J-PLUS broad and narrow-bands in 3 pointings, we develop a tailored statistical model. Our model accounts for missing data and small training sets and uses uncertainty aware PCA to flag GC candidates from a sample of point sources, starting from a training set of 73 spectroscopically confirmed GCs that we recover in J-PLUS.
GCs are  proxies of the formation assemblies of their host galaxies \citep{brodie06, beasley20} and of their environments \citep{lee2010, huxor2014,AM17,bortoli22}. With the lack of wide-field studies targeting GC systems of spiral galaxies \citep{kruijssen19}, such a catalogue is timely. 
This work showcases the power of principled statistical techniques allied with multi-band surveys in finding extragalactic GCs in and beyond the outskirts of their host galaxy in the local Universe. Our list increases the population of GC candidates around the triplet by 3-fold and provides a testbed for further studies of GC spatial distribution around spirals galaxies.

We study the spatial distribution of the candidate clusters and report an over-density of GC candidates, forming a potential bridge connecting M\,81 and M\,82. 
Interestingly, we do not recover a significant population of GC-like objects around NGC\,3077, probably due to the fact that the star cluster population of such galaxy is not made of old GCs, but of younger objects, whose SEDs are not part of our training sample.
As expected, blue GCs dominate at more considerable distances from M\,81 and M\,82.

The power of our method is further tested against colour-colour diagrams and colour distributions of the point sources,  the confirmed GCs and GC candidates. 
The bulk of our candidates tends to have bluer colours in typical $(g-i)$ and $(g-z)$ colours than the confirmed GCs. Our method generally excludes very red objects, typical of background galaxies.
We further map the metallicity distribution of the spectroscopically confirmed GCs into the metallicity distribution of the sample candidates.
Furthermore, find that the MDF of GC candidates is in the range of -2.5 $\lesssim$ [Fe/H] $\lesssim$ 0.5, but has a tail towards the metal-rich end (-0.5 $\lesssim$ [Fe/H] $\lesssim$ 0.7), similar to what is seen in the GC system of M\,31 (\citealt{Caldwell2011}). We present a few SEDs for GC candidates and discuss them in light of what is expected for GC systems.

The method presented here can be straightforwardly applied to other nearby systems in J-PLUS and S-PLUS surveys and more remote systems with the upcoming J-PAS survey. 
For the specific system studied in this work M\,81/M\,82/NGC\,3077, future developments include follow-up observations of a randomly selected sample of our candidate clusters in order to better quantify the contamination rate, SED fitting to derive stellar population parameters of a cleaner sample of GC candidates and a dedicated study to characterise the properties of the bridge of GCs connecting M\,81 and M\,82.

\section*{Acknowledgements}
We thank the anonymous referee for suggestions that improved the paper.
ACS acknowledges funding from the Conselho Nacional de Desenvolvimento
Científico e Tecnológico (CNPq) and the Rio Grande do Sul Research Foundation (FAPERGS) through grants CNPq-403580/2016-1, CNPq-11153/2018-6, PqG/FAPERGS-17/2551-0001, FAPERGS/CAPES 19/2551-0000696-9, L'Or\'eal UNESCO ABC \emph{Para Mulheres na Ci\^encia} and the Chinese Academy of Sciences (CAS) President's International Fellowship Initiative (PIFI) through grant 2021VMC0005.
R.S.S.\ was supported by the National Natural Science Foundation of China project 1201101284.
This work was funded with grants from Consejo Nacional de Investigaciones Cient\'{\i}ficas y T\'ecnicas de la Rep\'ublica Argentina, Agencia Nacional de Promoci\'on Cient\'{\i}fica y Tecnol\'ogica, and Universidad Nacional de La Plata, Argentina. 
We acknowledge Tamara Civera from the technical member of UPAD for her invaluable work and Alvaro Alvarez-Candal and Juan Antonio Fernández Ontiveros for comments that helped improve the paper.
J. V. acknowledges the technical members of the UPAD for their invaluable work: Juan Castillo, Tamara Civera, Javier Hernández, Ángel López, Alberto Moreno, and David Muniesa.
R.A.D. acknowledges support from CNPq through BP grant 308105/2018-4,
and the Financiadora de Estudos e Projetos – FINEP grants REF. 1217/13 –
01.13.0279.00 and REF 0859/10 – 01.10.0663.00 for hardware support for the
J-PLUS project through the National Observatory of Brazil.
Based on observations made with the JAST80 telescope at the  Observatorio Astrofísico de Javalambre (OAJ), in Teruel, owned, managed, and operated by the Centro de Estudios de Física del
Cosmos de Aragón (CEFCA). We acknowledge the OAJ Data Processing and Archiving Unit (UPAD) for reducing and calibrating the OAJ data used in this work. Funding for the J-PLUS Project has been provided by the Governments of Spain and Aragón through the Fondo de Inversiones de Teruel; the Aragón Government through the Research Groups E96, E103, E16\_17R, and E16\_20R; the Spanish Ministry of Science, Innovation and  Universities (MCIU/AEI/FEDER, UE) with grants PGC2018-097585-B-C21 and PGC2018-097585-B-C22; the Spanish  Ministry  of  Economy  and  Competitiveness (MINECO) under  AYA2015-66211-C2-1-P, AYA2015-66211-C2-2, AYA2012-30789, and ICTS-2009-14; and European  FEDER funding 
(FCDD10-4E-867, FCDD13-4E-2685). The Brazilian agencies FINEP, FAPESP, and the National 
Observatory of Brazil have also contributed to this project.

\section*{Data Availability}
The raw data underlying this article is available from J-PLUS DR2 \url{http://www.j-plus.es/datareleases/data_release_dr2}. The catalogue of GC candidates will be made available at Vizier and  \url{https://www.j-plus.es/ancillarydata/dr2_M81_triplet_GC_candidates}.
 

\newpage
\bibliographystyle{mnras}
\bibliography{ref.bib} 

\begin{thebibliography}{}
\makeatletter
\relax
\def\mn@urlcharsother{\let\do\@makeother \do\$\do\&\do\#\do\^\do\_\do\%\do\~}
\def\mn@doi{\begingroup\mn@urlcharsother \@ifnextchar [ {\mn@doi@}
  {\mn@doi@[]}}
\def\mn@doi@[#1]#2{\def\@tempa{#1}\ifx\@tempa\@empty \href
  {http://dx.doi.org/#2} {doi:#2}\else \href {http://dx.doi.org/#2} {#1}\fi
  \endgroup}
\def\mn@eprint#1#2{\mn@eprint@#1:#2::\@nil}
\def\mn@eprint@arXiv#1{\href {http://arxiv.org/abs/#1} {{\tt arXiv:#1}}}
\def\mn@eprint@dblp#1{\href {http://dblp.uni-trier.de/rec/bibtex/#1.xml}
  {dblp:#1}}
\def\mn@eprint@#1:#2:#3:#4\@nil{\def\@tempa {#1}\def\@tempb {#2}\def\@tempc
  {#3}\ifx \@tempc \@empty \let \@tempc \@tempb \let \@tempb \@tempa \fi \ifx
  \@tempb \@empty \def\@tempb {arXiv}\fi \@ifundefined
  {mn@eprint@\@tempb}{\@tempb:\@tempc}{\expandafter \expandafter \csname
  mn@eprint@\@tempb\endcsname \expandafter{\@tempc}}}

\bibitem[\protect\citeauthoryear{{Adebahr}, {Krause}, {Klein}, {Heald}  \&
  {Dettmar}}{{Adebahr} et~al.}{2017}]{adebahr2017}
{Adebahr} B.,  {Krause} M.,  {Klein} U.,  {Heald} G.,   {Dettmar} R.~J.,  2017,
  \mn@doi [\aap] {10.1051/0004-6361/201629616}, \href
  {https://ui.adsabs.harvard.edu/abs/2017A&A...608A..29A} {608, A29}

\bibitem[\protect\citeauthoryear{{Ahumada} et~al.,}{{Ahumada}
  et~al.}{2020}]{ahu20}
{Ahumada} R.,  et~al., 2020, \mn@doi [\apjs] {10.3847/1538-4365/ab929e}, \href
  {https://ui.adsabs.harvard.edu/abs/2020ApJS..249....3A} {249, 3}

\bibitem[\protect\citeauthoryear{{Alabi} et~al.,}{{Alabi}
  et~al.}{2017}]{alabi17}
{Alabi} A.~B.,  et~al., 2017, \mn@doi [\mnras] {10.1093/mnras/stx678}, \href
  {https://ui.adsabs.harvard.edu/abs/2017MNRAS.468.3949A} {468, 3949}

\bibitem[\protect\citeauthoryear{{Alamo-Mart{\'{\i}}nez} \&
  {Blakeslee}}{{Alamo-Mart{\'{\i}}nez} \& {Blakeslee}}{2017}]{AM17}
{Alamo-Mart{\'{\i}}nez} K.~A.,  {Blakeslee} J.~P.,  2017, \mn@doi [\apj]
  {10.3847/1538-4357/aa8f44}, \href
  {http://adsabs.harvard.edu/abs/2017ApJ...849....6A} {849, 6}

\bibitem[\protect\citeauthoryear{{Andreani}, {Boselli}, {Ciesla}, {Vio},
  {Cortese}, {Buat}  \& {Miyamoto}}{{Andreani} et~al.}{2018}]{Andreani2018}
{Andreani} P.,  {Boselli} A.,  {Ciesla} L.,  {Vio} R.,  {Cortese} L.,  {Buat}
  V.,   {Miyamoto} Y.,  2018, \mn@doi [\aap] {10.1051/0004-6361/201832873},
  \href {https://ui.adsabs.harvard.edu/abs/2018A&A...617A..33A} {617, A33}

\bibitem[\protect\citeauthoryear{Austin}{Austin}{2011}]{Austin2011}
Austin P.~C.,  2011, \mn@doi [Multivariate Behavioral Research]
  {10.1080/00273171.2011.568786}, 46, 399

\bibitem[\protect\citeauthoryear{{Bassino}, {Cellone}, {Forte}  \&
  {Dirsch}}{{Bassino} et~al.}{2003}]{bassino03}
{Bassino} L.~P.,  {Cellone} S.~A.,  {Forte} J.~C.,   {Dirsch} B.,  2003,
  \mn@doi [\aap] {10.1051/0004-6361:20021810}, \href
  {https://ui.adsabs.harvard.edu/abs/2003A&A...399..489B} {399, 489}

\bibitem[\protect\citeauthoryear{{Bassino}, {Faifer}, {Forte}, {Dirsch},
  {Richtler}, {Geisler}  \& {Schuberth}}{{Bassino} et~al.}{2006}]{bassino06}
{Bassino} L.~P.,  {Faifer} F.~R.,  {Forte} J.~C.,  {Dirsch} B.,  {Richtler} T.,
   {Geisler} D.,   {Schuberth} Y.,  2006, \mn@doi [\aap]
  {10.1051/0004-6361:20054563}, \href
  {https://ui.adsabs.harvard.edu/abs/2006A&A...451..789B} {451, 789}

\bibitem[\protect\citeauthoryear{{Beasley}}{{Beasley}}{2020}]{beasley20}
{Beasley} M.~A.,  2020, {Globular Cluster Systems and Galaxy Formation}.
Springer Nature Switzerland, pp 245--277, \mn@doi{10.1007/978-3-030-38509-5_9}

\bibitem[\protect\citeauthoryear{{Beasley}, {Trujillo}, {Leaman}  \&
  {Montes}}{{Beasley} et~al.}{2018}]{beasley18}
{Beasley} M.~A.,  {Trujillo} I.,  {Leaman} R.,   {Montes} M.,  2018, \mn@doi
  [\nat] {10.1038/nature25756}, \href
  {http://adsabs.harvard.edu/abs/2018Natur.555..483B} {555, 483}

\bibitem[\protect\citeauthoryear{{Bell}, {Monachesi}, {Harmsen}, {de Jong},
  {Bailin}, {Radburn-Smith}, {D'Souza}  \& {Holwerda}}{{Bell}
  et~al.}{2017}]{bell2017}
{Bell} E.~F.,  {Monachesi} A.,  {Harmsen} B.,  {de Jong} R.~S.,  {Bailin} J.,
  {Radburn-Smith} D.~J.,  {D'Souza} R.,   {Holwerda} B.~W.,  2017, \mn@doi
  [\apjl] {10.3847/2041-8213/aa6158}, \href
  {https://ui.adsabs.harvard.edu/abs/2017ApJ...837L...8B} {837, L8}

\bibitem[\protect\citeauthoryear{{Bertin} \& {Arnouts}}{{Bertin} \&
  {Arnouts}}{1996a}]{ber96}
{Bertin} E.,  {Arnouts} S.,  1996a, A\&AS, \href
  {http://adsabs.harvard.edu/abs/1996A%26AS..117..393B} {117, 393}

\bibitem[\protect\citeauthoryear{{Bertin} \& {Arnouts}}{{Bertin} \&
  {Arnouts}}{1996b}]{sex}
{Bertin} E.,  {Arnouts} S.,  1996b, \mn@doi [\aaps] {10.1051/aas:1996164},
  \href {https://ui.adsabs.harvard.edu/abs/1996A&AS..117..393B} {117, 393}

\bibitem[\protect\citeauthoryear{{Bica}, {Santiago}, {Bonatto}, {Garcia-Dias},
  {Kerber}, {Dias}, {Barbuy}  \& {Balbinot}}{{Bica} et~al.}{2015}]{bica2015}
{Bica} E.,  {Santiago} B.,  {Bonatto} C.,  {Garcia-Dias} R.,  {Kerber} L.,
  {Dias} B.,  {Barbuy} B.,   {Balbinot} E.,  2015, \mn@doi [\mnras]
  {10.1093/mnras/stv1720}, \href
  {https://ui.adsabs.harvard.edu/abs/2015MNRAS.453.3190B} {453, 3190}

\bibitem[\protect\citeauthoryear{{Blakeslee}}{{Blakeslee}}{1999}]{Blakeslee1999}
{Blakeslee} J.~P.,  1999, \mn@doi [\aj] {10.1086/301052}, \href
  {https://ui.adsabs.harvard.edu/abs/1999AJ....118.1506B} {118, 1506}

\bibitem[\protect\citeauthoryear{{Blakeslee}, {Cho}, {Peng}, {Ferrarese},
  {Jord{\'a}n}  \& {Martel}}{{Blakeslee} et~al.}{2012}]{blakeslee12}
{Blakeslee} J.~P.,  {Cho} H.,  {Peng} E.~W.,  {Ferrarese} L.,  {Jord{\'a}n} A.,
    {Martel} A.~R.,  2012, \mn@doi [\apj] {10.1088/0004-637X/746/1/88}, \href
  {http://adsabs.harvard.edu/abs/2012ApJ...746...88B} {746, 88}

\bibitem[\protect\citeauthoryear{{Bressan}, {Marigo}, {Girardi}, {Salasnich},
  {Dal Cero}, {Rubele}  \& {Nanni}}{{Bressan} et~al.}{2012}]{bre12}
{Bressan} A.,  {Marigo} P.,  {Girardi} L.,  {Salasnich} B.,  {Dal Cero} C.,
  {Rubele} S.,   {Nanni} A.,  2012, \mn@doi [\mnras]
  {10.1111/j.1365-2966.2012.21948.x}, \href
  {https://ui.adsabs.harvard.edu/abs/2012MNRAS.427..127B} {427, 127}

\bibitem[\protect\citeauthoryear{{Brito-Silva} et~al.,}{{Brito-Silva}
  et~al.}{2021}]{brito2022}
{Brito-Silva} D.,  et~al., 2021, arXiv e-prints, \href
  {https://ui.adsabs.harvard.edu/abs/2021arXiv211004423B} {p. arXiv:2110.04423}

\bibitem[\protect\citeauthoryear{{Brodie} \& {Strader}}{{Brodie} \&
  {Strader}}{2006}]{brodie06}
{Brodie} J.~P.,  {Strader} J.,  2006, \mn@doi [\araa]
  {10.1146/annurev.astro.44.051905.092441}, \href
  {http://adsabs.harvard.edu/abs/2006ARA%26A..44..193B} {44, 193}

\bibitem[\protect\citeauthoryear{{Brodie}, {Romanowsky}, {Strader}  \&
  {Forbes}}{{Brodie} et~al.}{2011}]{brodie11}
{Brodie} J.~P.,  {Romanowsky} A.~J.,  {Strader} J.,   {Forbes} D.~A.,  2011,
  \mn@doi [\aj] {10.1088/0004-6256/142/6/199}, \href
  {https://ui.adsabs.harvard.edu/abs/2011AJ....142..199B} {142, 199}

\bibitem[\protect\citeauthoryear{{Br{\"u}ns} \& {Kroupa}}{{Br{\"u}ns} \&
  {Kroupa}}{2012}]{bru12}
{Br{\"u}ns} R.~C.,  {Kroupa} P.,  2012, \mn@doi [\aap]
  {10.1051/0004-6361/201219693}, \href
  {https://ui.adsabs.harvard.edu/abs/2012A&A...547A..65B} {547, A65}

\bibitem[\protect\citeauthoryear{Buuren \& Groothuis-Oudshoorn}{Buuren \&
  Groothuis-Oudshoorn}{2011}]{Stef2011}
Buuren S.~V.,  Groothuis-Oudshoorn K.,  2011, \mn@doi [Journal of Statistical
  Software, Articles] {10.18637/jss.v045.i03}, 45, 1

\bibitem[\protect\citeauthoryear{{Buzzo} et~al.,}{{Buzzo}
  et~al.}{2022}]{buzzo2022}
{Buzzo} M.~L.,  et~al., 2022, \mn@doi [\mnras] {10.1093/mnras/stab3489}, \href
  {https://ui.adsabs.harvard.edu/abs/2022MNRAS.510.1383B} {510, 1383}

\bibitem[\protect\citeauthoryear{{Caldwell} \& {Romanowsky}}{{Caldwell} \&
  {Romanowsky}}{2016}]{caldwell16}
{Caldwell} N.,  {Romanowsky} A.~J.,  2016, VizieR Online Data Catalog, \href
  {https://ui.adsabs.harvard.edu/abs/2016yCat..18240042C} {p. J/ApJ/824/42}

\bibitem[\protect\citeauthoryear{{Caldwell}, {Schiavon}, {Morrison}, {Rose}  \&
  {Harding}}{{Caldwell} et~al.}{2011}]{Caldwell2011}
{Caldwell} N.,  {Schiavon} R.,  {Morrison} H.,  {Rose} J.~A.,   {Harding} P.,
  2011, \mn@doi [\aj] {10.1088/0004-6256/141/2/61}, \href
  {https://ui.adsabs.harvard.edu/abs/2011AJ....141...61C} {141, 61}

\bibitem[\protect\citeauthoryear{{Caso}, {Bassino}, {Richtler}, {Calder{\'o}n}
  \& {Smith Castelli}}{{Caso} et~al.}{2014}]{cas14}
{Caso} J.~P.,  {Bassino} L.~P.,  {Richtler} T.,  {Calder{\'o}n} J.~P.,   {Smith
  Castelli} A.~V.,  2014, \mn@doi [\mnras] {10.1093/mnras/stu876}, \href
  {https://ui.adsabs.harvard.edu/abs/2014MNRAS.442..891C} {442, 891}

\bibitem[\protect\citeauthoryear{{Caso}, {Bassino}  \& {G{\'o}mez}}{{Caso}
  et~al.}{2017}]{caso17}
{Caso} J.~P.,  {Bassino} L.~P.,   {G{\'o}mez} M.,  2017, \mn@doi [MNRAS]
  {10.1093/mnras/stx1393}, \href
  {http://adsabs.harvard.edu/abs/2017MNRAS.470.3227C} {470, 3227}

\bibitem[\protect\citeauthoryear{{Caso}, {Bassino}, {Richtler}  \&
  {Salinas}}{{Caso} et~al.}{2019}]{cas19}
{Caso} J.~P.,  {Bassino} L.~P.,  {Richtler} T.,   {Salinas} R.,  2019, \mn@doi
  [\mnras] {10.1093/mnras/sty3370}, \href
  {https://ui.adsabs.harvard.edu/abs/2019MNRAS.483.4371C} {483, 4371}

\bibitem[\protect\citeauthoryear{{Cenarro} et~al.,}{{Cenarro}
  et~al.}{2014}]{oaj}
{Cenarro} A.~J.,  et~al., 2014, in Observatory Operations: Strategies,
  Processes, and Systems V. p. 91491I, \mn@doi{10.1117/12.2055455}

\bibitem[\protect\citeauthoryear{{Cenarro} et~al.,}{{Cenarro}
  et~al.}{2019}]{cen19}
{Cenarro} A.~J.,  et~al., 2019, \mn@doi [\aap] {10.1051/0004-6361/201833036},
  \href {https://ui.adsabs.harvard.edu/abs/2019A&A...622A.176C} {622, A176}

\bibitem[\protect\citeauthoryear{Chen \& Guestrin}{Chen \&
  Guestrin}{2016}]{Chen2016}
Chen T.,  Guestrin C.,  2016, in Proceedings of the 22nd ACM SIGKDD
  International Conference on Knowledge Discovery and Data Mining. KDD '16.
Association for Computing Machinery, New York, NY, USA, p. 785–794,
  \mn@doi{10.1145/2939672.2939785}, \url
  {https://doi.org/10.1145/2939672.2939785}

\bibitem[\protect\citeauthoryear{Chen \& de Souza}{Chen \&
  de~Souza}{2022}]{Chen_2022}
Chen P.,  de Souza R.~S.,  2022, \mn@doi [Research Notes of the {AAS}]
  {10.3847/2515-5172/ac5c57}, 6, 51

\bibitem[\protect\citeauthoryear{{Chies-Santos}, {Larsen}, {Wehner},
  {Kuntschner}, {Strader}  \& {Brodie}}{{Chies-Santos}
  et~al.}{2011a}]{chies11a}
{Chies-Santos} A.~L.,  {Larsen} S.~S.,  {Wehner} E.~M.,  {Kuntschner} H.,
  {Strader} J.,   {Brodie} J.~P.,  2011a, \mn@doi [\aap]
  {10.1051/0004-6361/201015681}, \href
  {https://ui.adsabs.harvard.edu/abs/2011A&A...525A..19C} {525, A19}

\bibitem[\protect\citeauthoryear{{Chies-Santos}, {Larsen}, {Kuntschner},
  {Anders}, {Wehner}, {Strader}, {Brodie}  \& {Santos}}{{Chies-Santos}
  et~al.}{2011b}]{chies11}
{Chies-Santos} A.~L.,  {Larsen} S.~S.,  {Kuntschner} H.,  {Anders} P.,
  {Wehner} E.~M.,  {Strader} J.,  {Brodie} J.~P.,   {Santos} J.~F.~C.,  2011b,
  \mn@doi [\aap] {10.1051/0004-6361/201015683}, \href
  {https://ui.adsabs.harvard.edu/abs/2011A&A...525A..20C} {525, A20}

\bibitem[\protect\citeauthoryear{{Chies-Santos}, {Larsen}, {Cantiello},
  {Strader}, {Kuntschner}, {Wehner}  \& {Brodie}}{{Chies-Santos}
  et~al.}{2012}]{chi12}
{Chies-Santos} A.~L.,  {Larsen} S.~S.,  {Cantiello} M.,  {Strader} J.,
  {Kuntschner} H.,  {Wehner} E.~M.,   {Brodie} J.~P.,  2012, \mn@doi [\aap]
  {10.1051/0004-6361/201117169}, \href
  {https://ui.adsabs.harvard.edu/abs/2012A&A...539A..54C} {539, A54}

\bibitem[\protect\citeauthoryear{{Choksi} \& {Gnedin}}{{Choksi} \&
  {Gnedin}}{2019}]{choksi19}
{Choksi} N.,  {Gnedin} O.~Y.,  2019, \mn@doi [\mnras] {10.1093/mnras/stz2097},
  \href {https://ui.adsabs.harvard.edu/abs/2019MNRAS.488.5409C} {488, 5409}

\bibitem[\protect\citeauthoryear{{Choksi}, {Gnedin}  \& {Li}}{{Choksi}
  et~al.}{2018}]{choksi2018}
{Choksi} N.,  {Gnedin} O.~Y.,   {Li} H.,  2018, \mn@doi [\mnras]
  {10.1093/mnras/sty1952}, \href
  {https://ui.adsabs.harvard.edu/abs/2018MNRAS.480.2343C} {480, 2343}

\bibitem[\protect\citeauthoryear{{Cuevas-Otahola}, {Mayya}, {Puerari}  \&
  {Rosa-Gonz{\'a}lez}}{{Cuevas-Otahola} et~al.}{2021}]{cuevas21}
{Cuevas-Otahola} B.,  {Mayya} Y.~D.,  {Puerari} I.,   {Rosa-Gonz{\'a}lez} D.,
  2021, \mn@doi [\mnras] {10.1093/mnras/staa3513}, \href
  {https://ui.adsabs.harvard.edu/abs/2021MNRAS.500.4422C} {500, 4422}

\bibitem[\protect\citeauthoryear{{Davidge}}{{Davidge}}{2004}]{davidge2004}
{Davidge} T.~J.,  2004, \mn@doi [\aj] {10.1086/382096}, \href
  {https://ui.adsabs.harvard.edu/abs/2004AJ....127.1460D} {127, 1460}

\bibitem[\protect\citeauthoryear{{De B{\'o}rtoli}, {Caso}, {Ennis}  \&
  {Bassino}}{{De B{\'o}rtoli} et~al.}{2022}]{bortoli22}
{De B{\'o}rtoli} B.~J.,  {Caso} J.~P.,  {Ennis} A.~I.,   {Bassino} L.~P.,
  2022, \mn@doi [\mnras] {10.1093/mnras/stac010}, \href
  {https://ui.adsabs.harvard.edu/abs/2022MNRAS.510.5725D} {510, 5725}

\bibitem[\protect\citeauthoryear{De~Souza, Maio, Biffi  \& Ciardi}{De~Souza
  et~al.}{2014}]{DeSouza2014}
De~Souza R.~S.,  Maio U.,  Biffi V.,   Ciardi B.,  2014, \mn@doi [\mnras]
  {10.1093/mnras/stu274}, 440, 240

\bibitem[\protect\citeauthoryear{{Di Tullio Zinn} \& {Zinn}}{{Di Tullio Zinn}
  \& {Zinn}}{2015a}]{zinn2015}
{Di Tullio Zinn} G.,  {Zinn} R.,  2015a, \mn@doi [\aj]
  {10.1088/0004-6256/149/4/139}, \href
  {https://ui.adsabs.harvard.edu/abs/2015AJ....149..139D} {149, 139}

\bibitem[\protect\citeauthoryear{{Di Tullio Zinn} \& {Zinn}}{{Di Tullio Zinn}
  \& {Zinn}}{2015b}]{ditul15}
{Di Tullio Zinn} G.,  {Zinn} R.,  2015b, \mn@doi [\aj]
  {10.1088/0004-6256/149/4/139}, \href
  {https://ui.adsabs.harvard.edu/abs/2015AJ....149..139D} {149, 139}

\bibitem[\protect\citeauthoryear{{Doppel}, {Sales}, {Navarro}, {Abadi}, {Peng},
  {Toloba}  \& {Ramos-Almendares}}{{Doppel} et~al.}{2021}]{doppel2021}
{Doppel} J.~E.,  {Sales} L.~V.,  {Navarro} J.~F.,  {Abadi} M.~G.,  {Peng}
  E.~W.,  {Toloba} E.,   {Ramos-Almendares} F.,  2021, \mn@doi [\mnras]
  {10.1093/mnras/staa3915}, \href
  {https://ui.adsabs.harvard.edu/abs/2021MNRAS.502.1661D} {502, 1661}

\bibitem[\protect\citeauthoryear{{El-Badry}, {Quataert}, {Weisz}, {Choksi}  \&
  {Boylan-Kolchin}}{{El-Badry} et~al.}{2019}]{elbadry19}
{El-Badry} K.,  {Quataert} E.,  {Weisz} D.~R.,  {Choksi} N.,   {Boylan-Kolchin}
  M.,  2019, \mn@doi [\mnras] {10.1093/mnras/sty3007}, \href
  {https://ui.adsabs.harvard.edu/abs/2019MNRAS.482.4528E} {482, 4528}

\bibitem[\protect\citeauthoryear{{Ennis}, {Caso}, {Bassino}, {Salinas}  \&
  {G{\'o}mez}}{{Ennis} et~al.}{2020}]{enn20}
{Ennis} A.~I.,  {Caso} J.~P.,  {Bassino} L.~P.,  {Salinas} R.,   {G{\'o}mez}
  M.,  2020, \mn@doi [\mnras] {10.1093/mnras/staa2967}, \href
  {https://ui.adsabs.harvard.edu/abs/2020MNRAS.499.2554E} {499, 2554}

\bibitem[\protect\citeauthoryear{{Escudero}, {Faifer}, {Bassino},
  {Calder{\'o}n}  \& {Caso}}{{Escudero} et~al.}{2015}]{esc15}
{Escudero} C.~G.,  {Faifer} F.~R.,  {Bassino} L.~P.,  {Calder{\'o}n} J.~P.,
  {Caso} J.~P.,  2015, \mn@doi [\mnras] {10.1093/mnras/stv283}, \href
  {https://ui.adsabs.harvard.edu/abs/2015MNRAS.449..612E} {449, 612}

\bibitem[\protect\citeauthoryear{{Evans} et~al.,}{{Evans} et~al.}{2018}]{eva18}
{Evans} D.~W.,  et~al., 2018, \mn@doi [\aap] {10.1051/0004-6361/201832756},
  \href {https://ui.adsabs.harvard.edu/abs/2018A&A...616A...4E} {616, A4}

\bibitem[\protect\citeauthoryear{{Fabricius} et~al.,}{{Fabricius}
  et~al.}{2021}]{fab20}
{Fabricius} C.,  et~al., 2021, \mn@doi [\aap] {10.1051/0004-6361/202039834},
  \href {https://ui.adsabs.harvard.edu/abs/2021A&A...649A...5F} {649, A5}

\bibitem[\protect\citeauthoryear{{Fahrion} et~al.,}{{Fahrion}
  et~al.}{2020}]{fahrion2020}
{Fahrion} K.,  et~al., 2020, \mn@doi [\aap] {10.1051/0004-6361/202037686},
  \href {https://ui.adsabs.harvard.edu/abs/2020A&A...637A..27F} {637, A27}

\bibitem[\protect\citeauthoryear{{Faifer} et~al.,}{{Faifer}
  et~al.}{2011}]{faifer11}
{Faifer} F.~R.,  et~al., 2011, \mn@doi [\mnras]
  {10.1111/j.1365-2966.2011.19018.x}, \href
  {http://adsabs.harvard.edu/abs/2011MNRAS.416..155F} {416, 155}

\bibitem[\protect\citeauthoryear{{Fensch} et~al.,}{{Fensch}
  et~al.}{2020}]{fensch2020}
{Fensch} J.,  et~al., 2020, \mn@doi [\aap] {10.1051/0004-6361/202038550}, \href
  {https://ui.adsabs.harvard.edu/abs/2020A&A...644A.164F} {644, A164}

\bibitem[\protect\citeauthoryear{{Fisher} \& {Drory}}{{Fisher} \&
  {Drory}}{2008}]{fisher2008}
{Fisher} D.~B.,  {Drory} N.,  2008, \mn@doi [\aj]
  {10.1088/0004-6256/136/2/773}, \href
  {https://ui.adsabs.harvard.edu/abs/2008AJ....136..773F} {136, 773}

\bibitem[\protect\citeauthoryear{{Forbes}, {Spitler}, {Strader}, {Romanowsky},
  {Brodie}  \& {Foster}}{{Forbes} et~al.}{2011a}]{forbes11}
{Forbes} D.~A.,  {Spitler} L.~R.,  {Strader} J.,  {Romanowsky} A.~J.,  {Brodie}
  J.~P.,   {Foster} C.,  2011a, \mn@doi [\mnras]
  {10.1111/j.1365-2966.2011.18373.x}, \href
  {http://adsabs.harvard.edu/abs/2011MNRAS.413.2943F} {413, 2943}

\bibitem[\protect\citeauthoryear{{Forbes}, {Spitler}, {Strader}, {Romanowsky},
  {Brodie}  \& {Foster}}{{Forbes} et~al.}{2011b}]{for11}
{Forbes} D.~A.,  {Spitler} L.~R.,  {Strader} J.,  {Romanowsky} A.~J.,  {Brodie}
  J.~P.,   {Foster} C.,  2011b, \mn@doi [\mnras]
  {10.1111/j.1365-2966.2011.18373.x}, \href
  {https://ui.adsabs.harvard.edu/abs/2011MNRAS.413.2943F} {413, 2943}

\bibitem[\protect\citeauthoryear{{Forte} et~al.,}{{Forte}
  et~al.}{2019}]{forte19}
{Forte} J.~C.,  et~al., 2019, \mn@doi [\mnras] {10.1093/mnras/sty2746}, \href
  {https://ui.adsabs.harvard.edu/abs/2019MNRAS.482..950F} {482, 950}

\bibitem[\protect\citeauthoryear{{Gaia Collaboration} et~al.,}{{Gaia
  Collaboration} et~al.}{2016}]{pru16}
{Gaia Collaboration} et~al., 2016, \mn@doi [\aap]
  {10.1051/0004-6361/201629272}, \href
  {https://ui.adsabs.harvard.edu/abs/2016A&A...595A...1G} {595, A1}

\bibitem[\protect\citeauthoryear{{Gaia Collaboration} et~al.,}{{Gaia
  Collaboration} et~al.}{2021}]{gaia20}
{Gaia Collaboration} et~al., 2021, \mn@doi [\aap]
  {10.1051/0004-6361/202039657}, \href
  {https://ui.adsabs.harvard.edu/abs/2021A&A...649A...1G} {649, A1}

\bibitem[\protect\citeauthoryear{{Gonz{\'a}lez-L{\'o}pezlira}
  et~al.,}{{Gonz{\'a}lez-L{\'o}pezlira} et~al.}{2017}]{gonzalez2016}
{Gonz{\'a}lez-L{\'o}pezlira} R.~A.,  et~al., 2017, \mn@doi [\apj]
  {10.3847/1538-4357/835/2/184}, \href
  {https://ui.adsabs.harvard.edu/abs/2017ApJ...835..184G} {835, 184}

\bibitem[\protect\citeauthoryear{{Gonz{\'a}lez-L{\'o}pezlira}
  et~al.,}{{Gonz{\'a}lez-L{\'o}pezlira} et~al.}{2019}]{gonzalez2019}
{Gonz{\'a}lez-L{\'o}pezlira} R.~A.,  et~al., 2019, \mn@doi [\apj]
  {10.3847/1538-4357/ab113a}, \href
  {https://ui.adsabs.harvard.edu/abs/2019ApJ...876...39G} {876, 39}

\bibitem[\protect\citeauthoryear{{Harris}}{{Harris}}{1996}]{har96}
{Harris} W.~E.,  1996, \mn@doi [\aj] {10.1086/118116}, \href
  {https://ui.adsabs.harvard.edu/abs/1996AJ....112.1487H} {112, 1487}

\bibitem[\protect\citeauthoryear{{Harris}, {Calzetti}, {Gallagher}, {Smith}  \&
  {Conselice}}{{Harris} et~al.}{2004}]{harris2004}
{Harris} J.,  {Calzetti} D.,  {Gallagher} John~S. I.,  {Smith} D.~A.,
  {Conselice} C.~J.,  2004, \mn@doi [\apj] {10.1086/381669}, \href
  {https://ui.adsabs.harvard.edu/abs/2004ApJ...603..503H} {603, 503}

\bibitem[\protect\citeauthoryear{{Harris}, {Harris}  \& {Hudson}}{{Harris}
  et~al.}{2015}]{harris15}
{Harris} W.~E.,  {Harris} G.~L.,   {Hudson} M.~J.,  2015, \mn@doi [\apj]
  {10.1088/0004-637X/806/1/36}, \href
  {https://ui.adsabs.harvard.edu/abs/2015ApJ...806...36H} {806, 36}

\bibitem[\protect\citeauthoryear{{Harris}, {Blakeslee}, {Whitmore}, {Gnedin},
  {Geisler}  \& {Rothberg}}{{Harris} et~al.}{2016}]{harris2016}
{Harris} W.~E.,  {Blakeslee} J.~P.,  {Whitmore} B.~C.,  {Gnedin} O.~Y.,
  {Geisler} D.,   {Rothberg} B.,  2016, \mn@doi [\apj]
  {10.3847/0004-637X/817/1/58}, \href
  {https://ui.adsabs.harvard.edu/abs/2016ApJ...817...58H} {817, 58}

\bibitem[\protect\citeauthoryear{{Harris}, {Blakeslee}  \& {Harris}}{{Harris}
  et~al.}{2017}]{Harris2017}
{Harris} W.~E.,  {Blakeslee} J.~P.,   {Harris} G. L.~H.,  2017, \mn@doi [\apj]
  {10.3847/1538-4357/836/1/67}, \href
  {https://ui.adsabs.harvard.edu/abs/2017ApJ...836...67H} {836, 67}

\bibitem[\protect\citeauthoryear{{Harris} et~al.,}{{Harris}
  et~al.}{2020}]{harris2020}
{Harris} W.~E.,  et~al., 2020, \mn@doi [\apj] {10.3847/1538-4357/ab6992}, \href
  {https://ui.adsabs.harvard.edu/abs/2020ApJ...890..105H} {890, 105}

\bibitem[\protect\citeauthoryear{{Hernitschek} et~al.,}{{Hernitschek}
  et~al.}{2019}]{her19}
{Hernitschek} N.,  et~al., 2019, \mn@doi [\apj] {10.3847/1538-4357/aaf388},
  \href {https://ui.adsabs.harvard.edu/abs/2019ApJ...871...49H} {871, 49}

\bibitem[\protect\citeauthoryear{Ho, Imai, King  \& Stuart}{Ho
  et~al.}{2007}]{HoImaKin07}
Ho D.,  Imai K.,  King G.,   Stuart E.,  2007, Political Analysis, 15,
  199{\textendash}236

\bibitem[\protect\citeauthoryear{Hofert, Kojadinovic, Maechler  \& Yan}{Hofert
  et~al.}{2018}]{Hofert2018}
Hofert M.,  Kojadinovic I.,  Maechler M.,   Yan J.,  2018, {E}lements of
  {C}opula {M}odeling with \textsf{R}.
Springer Use R! Series, \url {http://www.springer.com/de/book/9783319896342}

\bibitem[\protect\citeauthoryear{{Hoff}}{{Hoff}}{2007}]{hoff2007}
{Hoff} P.~D.,  2007, \mn@doi [Ann. Appl. Stat.] {10.1214/07-AOAS107}, 1, 265

\bibitem[\protect\citeauthoryear{Hoff}{Hoff}{2018}]{sbgcop}
Hoff P.,  2018, sbgcop: Semiparametric Bayesian Gaussian Copula Estimation and
  Imputation.
\url {https://CRAN.R-project.org/package=sbgcop}

\bibitem[\protect\citeauthoryear{Honaker, King  \& Blackwell}{Honaker
  et~al.}{2011}]{Amelia2011}
Honaker J.,  King G.,   Blackwell M.,  2011, Journal of Statistical Software,
  45, 1

\bibitem[\protect\citeauthoryear{{Hudson}, {Harris}  \& {Harris}}{{Hudson}
  et~al.}{2014}]{Hudson2014}
{Hudson} M.~J.,  {Harris} G.~L.,   {Harris} W.~E.,  2014, \mn@doi [\apjl]
  {10.1088/2041-8205/787/1/L5}, \href
  {https://ui.adsabs.harvard.edu/abs/2014ApJ...787L...5H} {787, L5}

\bibitem[\protect\citeauthoryear{{Huxor}, {Ferguson}, {Barker}, {Tanvir},
  {Irwin}, {Chapman}, {Ibata}  \& {Lewis}}{{Huxor} et~al.}{2009}]{huxor2009}
{Huxor} A.,  {Ferguson} A.~M.~N.,  {Barker} M.~K.,  {Tanvir} N.~R.,  {Irwin}
  M.~J.,  {Chapman} S.~C.,  {Ibata} R.,   {Lewis} G.,  2009, \mn@doi [\apjl]
  {10.1088/0004-637X/698/2/L77}, \href
  {https://ui.adsabs.harvard.edu/abs/2009ApJ...698L..77H} {698, L77}

\bibitem[\protect\citeauthoryear{{Huxor} et~al.,}{{Huxor}
  et~al.}{2014}]{huxor2014}
{Huxor} A.~P.,  et~al., 2014, \mn@doi [\mnras] {10.1093/mnras/stu771}, \href
  {https://ui.adsabs.harvard.edu/abs/2014MNRAS.442.2165H} {442, 2165}

\bibitem[\protect\citeauthoryear{{Ishida} \& {de Souza}}{{Ishida} \& {de
  Souza}}{2011}]{Ishida2011}
{Ishida} E.~E.~O.,  {de Souza} R.~S.,  2011, \mn@doi [\aap]
  {10.1051/0004-6361/201015281}, \href
  {https://ui.adsabs.harvard.edu/abs/2011A&A...527A..49I} {527, A49}

\bibitem[\protect\citeauthoryear{{Ishida} \& {de Souza}}{{Ishida} \& {de
  Souza}}{2013}]{Ishida2013}
{Ishida} E.~E.~O.,  {de Souza} R.~S.,  2013, \mn@doi [\mnras]
  {10.1093/mnras/sts650}, \href
  {https://ui.adsabs.harvard.edu/abs/2013MNRAS.430..509I} {430, 509}

\bibitem[\protect\citeauthoryear{{Ishida}, {de Souza}  \& {Ferrara}}{{Ishida}
  et~al.}{2011}]{Ishida2011b}
{Ishida} E.~E.~O.,  {de Souza} R.~S.,   {Ferrara} A.,  2011, \mn@doi [\mnras]
  {10.1111/j.1365-2966.2011.19501.x}, \href
  {https://ui.adsabs.harvard.edu/abs/2011MNRAS.418..500I} {418, 500}

\bibitem[\protect\citeauthoryear{{Jang}, {Lim}, {Park}  \& {Lee}}{{Jang}
  et~al.}{2012}]{jan12}
{Jang} I.~S.,  {Lim} S.,  {Park} H.~S.,   {Lee} M.~G.,  2012, \mn@doi [\apjl]
  {10.1088/2041-8205/751/1/L19}, \href
  {https://ui.adsabs.harvard.edu/abs/2012ApJ...751L..19J} {751, L19}

\bibitem[\protect\citeauthoryear{Jolliffe \& Cadima}{Jolliffe \&
  Cadima}{2016}]{Jollife2016}
Jolliffe I.~T.,  Cadima J.,  2016, \mn@doi [Philos. Trans. R. Soc. A Math.
  Phys. Eng. Sci.] {10.1098/rsta.2015.0202}, 374, 20150202

\bibitem[\protect\citeauthoryear{{Karachentsev} \& {Kudrya}}{{Karachentsev} \&
  {Kudrya}}{2014}]{kk2014}
{Karachentsev} I.~D.,  {Kudrya} Y.~N.,  2014, \mn@doi [\aj]
  {10.1088/0004-6256/148/3/50}, \href
  {https://ui.adsabs.harvard.edu/abs/2014AJ....148...50K} {148, 50}

\bibitem[\protect\citeauthoryear{{Kruijssen}, {Pfeffer}, {Reina-Campos},
  {Crain}  \& {Bastian}}{{Kruijssen} et~al.}{2019}]{kruijssen19}
{Kruijssen} J.~M.~D.,  {Pfeffer} J.~L.,  {Reina-Campos} M.,  {Crain} R.~A.,
  {Bastian} N.,  2019, \mn@doi [\mnras] {10.1093/mnras/sty1609}, \href
  {https://ui.adsabs.harvard.edu/abs/2019MNRAS.486.3180K} {486, 3180}

\bibitem[\protect\citeauthoryear{{Kuhn}, {de Souza}, {Krone-Martins},
  {Castro-Ginard}, {Ishida}, {Povich}, {Hillenbrand}  \& {COIN
  Collaboration}}{{Kuhn} et~al.}{2021}]{Kuhn2021}
{Kuhn} M.~A.,  {de Souza} R.~S.,  {Krone-Martins} A.,  {Castro-Ginard} A.,
  {Ishida} E. E.~O.,  {Povich} M.~S.,  {Hillenbrand} L.~A.,   {COIN
  Collaboration} 2021, \mn@doi [\apjs] {10.3847/1538-4365/abe465}, \href
  {https://ui.adsabs.harvard.edu/abs/2021ApJS..254...33K} {254, 33}

\bibitem[\protect\citeauthoryear{{Laevens} et~al.,}{{Laevens}
  et~al.}{2014}]{laevens14}
{Laevens} B. P.~M.,  et~al., 2014, \mn@doi [\apjl]
  {10.1088/2041-8205/786/1/L3}, \href
  {https://ui.adsabs.harvard.edu/abs/2014ApJ...786L...3L} {786, L3}

\bibitem[\protect\citeauthoryear{{Larsen}}{{Larsen}}{1999}]{lar99}
{Larsen} S.~S.,  1999, \mn@doi [A\&AS] {10.1051/aas:1999509}, \href
  {http://adsabs.harvard.edu/abs/1999A%26AS..139..393L} {139, 393}

\bibitem[\protect\citeauthoryear{{Lee}, {Park}  \& {Hwang}}{{Lee}
  et~al.}{2010}]{lee2010}
{Lee} M.~G.,  {Park} H.~S.,   {Hwang} H.~S.,  2010, \mn@doi [Science]
  {10.1126/science.1186496}, \href
  {https://ui.adsabs.harvard.edu/abs/2010Sci...328..334L} {328, 334}

\bibitem[\protect\citeauthoryear{{Lee}, {Chung}  \& {Yoon}}{{Lee}
  et~al.}{2019}]{lee19}
{Lee} S.-Y.,  {Chung} C.,   {Yoon} S.-J.,  2019, \mn@doi [\apjs]
  {10.3847/1538-4365/aaecd4}, \href
  {https://ui.adsabs.harvard.edu/abs/2019ApJS..240....2L} {240, 2}

\bibitem[\protect\citeauthoryear{{Li} \& {Gnedin}}{{Li} \&
  {Gnedin}}{2019}]{li2019}
{Li} H.,  {Gnedin} O.~Y.,  2019, \mn@doi [\mnras] {10.1093/mnras/stz1114},
  \href {https://ui.adsabs.harvard.edu/abs/2019MNRAS.486.4030L} {486, 4030}

\bibitem[\protect\citeauthoryear{{Lim}, {Hwang}  \& {Lee}}{{Lim}
  et~al.}{2013}]{lim2013}
{Lim} S.,  {Hwang} N.,   {Lee} M.~G.,  2013, \mn@doi [\apj]
  {10.1088/0004-637X/766/1/20}, \href
  {https://ui.adsabs.harvard.edu/abs/2013ApJ...766...20L} {766, 20}

\bibitem[\protect\citeauthoryear{{Lin}, {Kilbinger}  \& {Pires}}{{Lin}
  et~al.}{2016}]{Lin2016}
{Lin} C.-A.,  {Kilbinger} M.,   {Pires} S.,  2016, \mn@doi [\aap]
  {10.1051/0004-6361/201628565}, \href
  {https://ui.adsabs.harvard.edu/abs/2016A&A...593A..88L} {593, A88}

\bibitem[\protect\citeauthoryear{{Longobardi} et~al.,}{{Longobardi}
  et~al.}{2018}]{longo18}
{Longobardi} A.,  et~al., 2018, \mn@doi [\apj] {10.3847/1538-4357/aad3d2},
  \href {https://ui.adsabs.harvard.edu/abs/2018ApJ...864...36L} {864, 36}

\bibitem[\protect\citeauthoryear{{L{\'o}pez-Sanjuan}
  et~al.,}{{L{\'o}pez-Sanjuan} et~al.}{2019}]{lop19}
{L{\'o}pez-Sanjuan} C.,  et~al., 2019, \mn@doi [\aap]
  {10.1051/0004-6361/201936405}, \href
  {https://ui.adsabs.harvard.edu/abs/2019A&A...631A.119L} {631, A119}

\bibitem[\protect\citeauthoryear{{Lupton}, {Blanton}, {Fekete}, {Hogg},
  {O'Mullane}, {Szalay}  \& {Wherry}}{{Lupton} et~al.}{2004}]{Lupton2004}
{Lupton} R.,  {Blanton} M.~R.,  {Fekete} G.,  {Hogg} D.~W.,  {O'Mullane} W.,
  {Szalay} A.,   {Wherry} N.,  2004, \mn@doi [\pasp] {10.1086/382245}, \href
  {https://ui.adsabs.harvard.edu/abs/2004PASP..116..133L} {116, 133}

\bibitem[\protect\citeauthoryear{{Ma} et~al.,}{{Ma} et~al.}{2017}]{ma2017}
{Ma} J.,  et~al., 2017, \mn@doi [\mnras] {10.1093/mnras/stx761}, \href
  {https://ui.adsabs.harvard.edu/abs/2017MNRAS.468.4513M} {468, 4513}

\bibitem[\protect\citeauthoryear{{Maltby}, {Almaini}, {Wild}, {Hatch},
  {Hartley}, {Simpson}, {Rowlands}  \& {Socolovsky}}{{Maltby}
  et~al.}{2018}]{maltby2018}
{Maltby} D.~T.,  {Almaini} O.,  {Wild} V.,  {Hatch} N.~A.,  {Hartley} W.~G.,
  {Simpson} C.,  {Rowlands} K.,   {Socolovsky} M.,  2018, \mn@doi [\mnras]
  {10.1093/mnras/sty1794}, \href
  {https://ui.adsabs.harvard.edu/abs/2018MNRAS.480..381M} {480, 381}

\bibitem[\protect\citeauthoryear{{Marchi-Lasch} et~al.,}{{Marchi-Lasch}
  et~al.}{2019}]{marchi19}
{Marchi-Lasch} S.,  et~al., 2019, \mn@doi [\apj] {10.3847/1538-4357/ab089c},
  \href {https://ui.adsabs.harvard.edu/abs/2019ApJ...874...29M} {874, 29}

\bibitem[\protect\citeauthoryear{{Mar{\'{\i}}n-Franch}, {Taylor}, {Cenarro},
  {Cristobal-Hornillos}  \& {Moles}}{{Mar{\'{\i}}n-Franch}
  et~al.}{2015}]{t80cam}
{Mar{\'{\i}}n-Franch} A.,  {Taylor} K.,  {Cenarro} J.,  {Cristobal-Hornillos}
  D.,   {Moles} M.,  2015, in IAU General Assembly. p. 2257381

\bibitem[\protect\citeauthoryear{{Mendes de Oliveira} et~al.,}{{Mendes de
  Oliveira} et~al.}{2019}]{mendes19}
{Mendes de Oliveira} C.,  et~al., 2019, \mn@doi [\mnras]
  {10.1093/mnras/stz1985}, \href
  {https://ui.adsabs.harvard.edu/abs/2019MNRAS.489..241M} {489, 241}

\bibitem[\protect\citeauthoryear{{Monachesi} et~al.,}{{Monachesi}
  et~al.}{2013}]{monachesi2013}
{Monachesi} A.,  et~al., 2013, \mn@doi [\apj] {10.1088/0004-637X/766/2/106},
  \href {https://ui.adsabs.harvard.edu/abs/2013ApJ...766..106M} {766, 106}

\bibitem[\protect\citeauthoryear{{Nantais} \& {Huchra}}{{Nantais} \&
  {Huchra}}{2010}]{nantais2010_spec}
{Nantais} J.~B.,  {Huchra} J.~P.,  2010, \mn@doi [\aj]
  {10.1088/0004-6256/139/6/2620}, \href
  {https://ui.adsabs.harvard.edu/abs/2010AJ....139.2620N} {139, 2620}

\bibitem[\protect\citeauthoryear{{Nantais}, {Huchra}, {McLeod}, {Strader}  \&
  {Brodie}}{{Nantais} et~al.}{2010}]{nantais2010_phot_ext}
{Nantais} J.~B.,  {Huchra} J.~P.,  {McLeod} B.,  {Strader} J.,   {Brodie}
  J.~P.,  2010, \mn@doi [\aj] {10.1088/0004-6256/139/4/1413}, \href
  {https://ui.adsabs.harvard.edu/abs/2010AJ....139.1413N} {139, 1413}

\bibitem[\protect\citeauthoryear{{Nantais}, {Huchra}, {Zezas}, {Gazeas}  \&
  {Strader}}{{Nantais} et~al.}{2011}]{nantais2011_photGCs}
{Nantais} J.~B.,  {Huchra} J.~P.,  {Zezas} A.,  {Gazeas} K.,   {Strader} J.,
  2011, \mn@doi [\aj] {10.1088/0004-6256/142/6/183}, \href
  {https://ui.adsabs.harvard.edu/abs/2011AJ....142..183N} {142, 183}

\bibitem[\protect\citeauthoryear{Nelsen}{Nelsen}{2010}]{Nelsen10}
Nelsen R.~B.,  2010, An Introduction to Copulas.
Springer Publishing Company, Incorporated

\bibitem[\protect\citeauthoryear{{Norris}, {van de Ven}, {Kannappan},
  {Schinnerer}  \& {Leaman}}{{Norris} et~al.}{2019}]{norris19}
{Norris} M.~A.,  {van de Ven} G.,  {Kannappan} S.~J.,  {Schinnerer} E.,
  {Leaman} R.,  2019, \mn@doi [\mnras] {10.1093/mnras/stz2096}, \href
  {https://ui.adsabs.harvard.edu/abs/2019MNRAS.488.5400N} {488, 5400}

\bibitem[\protect\citeauthoryear{{Oehm}, {Thies}  \& {Kroupa}}{{Oehm}
  et~al.}{2017}]{oehm2017}
{Oehm} W.,  {Thies} I.,   {Kroupa} P.,  2017, \mn@doi [\mnras]
  {10.1093/mnras/stw3381}, \href
  {https://ui.adsabs.harvard.edu/abs/2017MNRAS.467..273O} {467, 273}

\bibitem[\protect\citeauthoryear{{Okamoto}, {Arimoto}, {Ferguson}, {Bernard},
  {Irwin}, {Yamada}  \& {Utsumi}}{{Okamoto} et~al.}{2015}]{oka15}
{Okamoto} S.,  {Arimoto} N.,  {Ferguson} A. M.~N.,  {Bernard} E.~J.,  {Irwin}
  M.~J.,  {Yamada} Y.,   {Utsumi} Y.,  2015, \mn@doi [\apjl]
  {10.1088/2041-8205/809/1/L1}, \href
  {https://ui.adsabs.harvard.edu/abs/2015ApJ...809L...1O} {809, L1}

\bibitem[\protect\citeauthoryear{{Peng} et~al.,}{{Peng} et~al.}{2006}]{peng06}
{Peng} E.~W.,  et~al., 2006, \mn@doi [\apj] {10.1086/498210}, \href
  {http://adsabs.harvard.edu/abs/2006ApJ...639...95P} {639, 95}

\bibitem[\protect\citeauthoryear{{Peng} et~al.,}{{Peng} et~al.}{2008}]{pen08}
{Peng} E.~W.,  et~al., 2008, \mn@doi [ApJ] {10.1086/587951}, \href
  {http://adsabs.harvard.edu/abs/2008ApJ...681..197P} {681, 197}

\bibitem[\protect\citeauthoryear{{Perelmuter}, {Brodie}  \&
  {Huchra}}{{Perelmuter} et~al.}{1995}]{perelmuter1995}
{Perelmuter} J.-M.,  {Brodie} J.~P.,   {Huchra} J.~P.,  1995, \mn@doi [\aj]
  {10.1086/117547}, \href
  {https://ui.adsabs.harvard.edu/abs/1995AJ....110..620P} {110, 620}

\bibitem[\protect\citeauthoryear{{Pota} et~al.,}{{Pota} et~al.}{2013}]{pot13}
{Pota} V.,  et~al., 2013, \mn@doi [\mnras] {10.1093/mnras/sts029}, \href
  {https://ui.adsabs.harvard.edu/abs/2013MNRAS.428..389P} {428, 389}

\bibitem[\protect\citeauthoryear{{Powalka} et~al.,}{{Powalka}
  et~al.}{2017}]{pow17}
{Powalka} M.,  et~al., 2017, \mn@doi [\apj] {10.3847/1538-4357/aa77b1}, \href
  {https://ui.adsabs.harvard.edu/abs/2017ApJ...844..104P} {844, 104}

\bibitem[\protect\citeauthoryear{{R Core Team}}{{R Core Team}}{2019}]{rcore19}
{R Core Team} 2019, R: A Language and Environment for Statistical Computing.
R Foundation for Statistical Computing, Vienna, Austria, \url
  {https://www.R-project.org/}

\bibitem[\protect\citeauthoryear{{Reina-Campos}, {Trujillo-Gomez}, {Deason},
  {Kruijssen}, {Pfeffer}, {Crain}, {Bastian}  \& {Hughes}}{{Reina-Campos}
  et~al.}{2021}]{reina2021}
{Reina-Campos} M.,  {Trujillo-Gomez} S.,  {Deason} A.~J.,  {Kruijssen}
  J.~M.~D.,  {Pfeffer} J.~L.,  {Crain} R.~A.,  {Bastian} N.,   {Hughes} M.~E.,
  2021, arXiv e-prints, \href
  {https://ui.adsabs.harvard.edu/abs/2021arXiv210607652R} {p. arXiv:2106.07652}

\bibitem[\protect\citeauthoryear{{Saito} et~al.,}{{Saito}
  et~al.}{2005}]{saito2005}
{Saito} Y.,  et~al., 2005, \mn@doi [\apj] {10.1086/427645}, \href
  {https://ui.adsabs.harvard.edu/abs/2005ApJ...621..750S} {621, 750}

\bibitem[\protect\citeauthoryear{{Santos Barbosa}}{{Santos
  Barbosa}}{2020}]{RMLPCA}
{Santos Barbosa} R.,  2020, RMLPCA: Maximum Likelihood Principal Component
  Analysis.
\url {https://CRAN.R-project.org/package=RMLPCA}

\bibitem[\protect\citeauthoryear{{Sato}, {Ichiki}  \& {Takeuchi}}{{Sato}
  et~al.}{2011}]{Sato2011}
{Sato} M.,  {Ichiki} K.,   {Takeuchi} T.~T.,  2011, \mn@doi [\prd]
  {10.1103/PhysRevD.83.023501}, \href
  {https://ui.adsabs.harvard.edu/abs/2011PhRvD..83b3501S} {83, 023501}

\bibitem[\protect\citeauthoryear{{Schlafly} \& {Finkbeiner}}{{Schlafly} \&
  {Finkbeiner}}{2011}]{Schlafly_Finkbeiner2011}
{Schlafly} E.~F.,  {Finkbeiner} D.~P.,  2011, \mn@doi [\apj]
  {10.1088/0004-637X/737/2/103}, \href
  {https://ui.adsabs.harvard.edu/abs/2011ApJ...737..103S} {737, 103}

\bibitem[\protect\citeauthoryear{{Schuberth}, {Richtler}, {Hilker}, {Dirsch},
  {Bassino}, {Romanowsky}  \& {Infante}}{{Schuberth} et~al.}{2010}]{sch10}
{Schuberth} Y.,  {Richtler} T.,  {Hilker} M.,  {Dirsch} B.,  {Bassino} L.~P.,
  {Romanowsky} A.~J.,   {Infante} L.,  2010, \mn@doi [\aap]
  {10.1051/0004-6361/200912482}, \href
  {https://ui.adsabs.harvard.edu/abs/2010A&A...513A..52S} {513, A52}

\bibitem[\protect\citeauthoryear{{Schuberth}, {Richtler}, {Hilker}, {Salinas},
  {Dirsch}  \& {Larsen}}{{Schuberth} et~al.}{2012}]{schuberth12}
{Schuberth} Y.,  {Richtler} T.,  {Hilker} M.,  {Salinas} R.,  {Dirsch} B.,
  {Larsen} S.~S.,  2012, \mn@doi [\aap] {10.1051/0004-6361/201015038}, \href
  {https://ui.adsabs.harvard.edu/abs/2012A&A...544A.115S} {544, A115}

\bibitem[\protect\citeauthoryear{{Sharina}, {Chandar}, {Puzia}, {Goudfrooij}
  \& {Davoust}}{{Sharina} et~al.}{2010}]{sha10}
{Sharina} M.~E.,  {Chandar} R.,  {Puzia} T.~H.,  {Goudfrooij} P.,   {Davoust}
  E.,  2010, \mn@doi [\mnras] {10.1111/j.1365-2966.2010.16510.x}, \href
  {https://ui.adsabs.harvard.edu/abs/2010MNRAS.405..839S} {405, 839}

\bibitem[\protect\citeauthoryear{Shwartz-Ziv \& Armon}{Shwartz-Ziv \&
  Armon}{2021}]{SHWARTZZIV2021}
Shwartz-Ziv R.,  Armon A.,  2021, \mn@doi [Information Fusion]
  {https://doi.org/10.1016/j.inffus.2021.11.011}

\bibitem[\protect\citeauthoryear{{Sinnott}, {Hou}, {Anderson}, {Harris}  \&
  {Woodley}}{{Sinnott} et~al.}{2010}]{sinnott10}
{Sinnott} B.,  {Hou} A.,  {Anderson} R.,  {Harris} W.~E.,   {Woodley} K.~A.,
  2010, \mn@doi [\aj] {10.1088/0004-6256/140/6/2101}, \href
  {https://ui.adsabs.harvard.edu/abs/2010AJ....140.2101S} {140, 2101}

\bibitem[\protect\citeauthoryear{{Smercina} et~al.,}{{Smercina}
  et~al.}{2020}]{sme20}
{Smercina} A.,  et~al., 2020, \mn@doi [\apj] {10.3847/1538-4357/abc485}, \href
  {https://ui.adsabs.harvard.edu/abs/2020ApJ...905...60S} {905, 60}

\bibitem[\protect\citeauthoryear{{Strader}, {Brodie}, {Cenarro}, {Beasley}  \&
  {Forbes}}{{Strader} et~al.}{2005}]{strader05}
{Strader} J.,  {Brodie} J.~P.,  {Cenarro} A.~J.,  {Beasley} M.~A.,   {Forbes}
  D.~A.,  2005, \mn@doi [\aj] {10.1086/432717}, \href
  {https://ui.adsabs.harvard.edu/abs/2005AJ....130.1315S} {130, 1315}

\bibitem[\protect\citeauthoryear{{Tully} et~al.,}{{Tully}
  et~al.}{2013}]{tully2013}
{Tully} R.~B.,  et~al., 2013, \mn@doi [\aj] {10.1088/0004-6256/146/4/86}, \href
  {https://ui.adsabs.harvard.edu/abs/2013AJ....146...86T} {146, 86}

\bibitem[\protect\citeauthoryear{{Villaume}, {Foreman-Mackey}, {Romanowsky},
  {Brodie}  \& {Strader}}{{Villaume} et~al.}{2020}]{villaume20}
{Villaume} A.,  {Foreman-Mackey} D.,  {Romanowsky} A.~J.,  {Brodie} J.,
  {Strader} J.,  2020, \mn@doi [\apj] {10.3847/1538-4357/aba616}, \href
  {https://ui.adsabs.harvard.edu/abs/2020ApJ...900...95V} {900, 95}

\bibitem[\protect\citeauthoryear{{Voggel}, {Seth}, {Sand}, {Hughes}, {Strader},
  {Crnojevic}  \& {Caldwell}}{{Voggel} et~al.}{2020}]{voggel2020}
{Voggel} K.~T.,  {Seth} A.~C.,  {Sand} D.~J.,  {Hughes} A.,  {Strader} J.,
  {Crnojevic} D.,   {Caldwell} N.,  2020, \mn@doi [\apj]
  {10.3847/1538-4357/ab6f69}, \href
  {https://ui.adsabs.harvard.edu/abs/2020ApJ...899..140V} {899, 140}

\bibitem[\protect\citeauthoryear{{Webb} \& {Carlberg}}{{Webb} \&
  {Carlberg}}{2021}]{webb21}
{Webb} J.~J.,  {Carlberg} R.~G.,  2021, \mn@doi [\mnras]
  {10.1093/mnras/stab353}, \href
  {https://ui.adsabs.harvard.edu/abs/2021MNRAS.502.4547W} {502, 4547}

\bibitem[\protect\citeauthoryear{Wentzell}{Wentzell}{2009}]{WENTZELL2009}
Wentzell P.,  2009, in Brown S.~D.,  Tauler R.,   Walczak B.,  eds, ,
  Comprehensive Chemometrics.
Elsevier, Oxford, pp 507--558, \mn@doi{10.1016/B978-044452701-1.00057-0}

\bibitem[\protect\citeauthoryear{Wentzell \& Hou}{Wentzell \&
  Hou}{2012}]{Wentzell2012}
Wentzell P.~D.,  Hou S.,  2012, \mn@doi [Journal of Chemometrics]
  {https://doi.org/10.1002/cem.2428}, 26, 264

\bibitem[\protect\citeauthoryear{Wentzell \& Lohnes}{Wentzell \&
  Lohnes}{1999}]{WENTZELL1999}
Wentzell P.~D.,  Lohnes M.~T.,  1999, \mn@doi [Chemometrics and Intelligent
  Laboratory Systems] {https://doi.org/10.1016/S0169-7439(98)00090-2}, 45, 65

\bibitem[\protect\citeauthoryear{{West}, {Cote}, {Jones}, {Forman}  \&
  {Marzke}}{{West} et~al.}{1995}]{west95}
{West} M.~J.,  {Cote} P.,  {Jones} C.,  {Forman} W.,   {Marzke} R.~O.,  1995,
  \mn@doi [\apjl] {10.1086/309748}, \href
  {https://ui.adsabs.harvard.edu/abs/1995ApJ...453L..77W} {453, L77}

\bibitem[\protect\citeauthoryear{{Wild} et~al.,}{{Wild}
  et~al.}{2014}]{wild2014}
{Wild} V.,  et~al., 2014, \mn@doi [\mnras] {10.1093/mnras/stu212}, \href
  {https://ui.adsabs.harvard.edu/abs/2014MNRAS.440.1880W} {440, 1880}

\bibitem[\protect\citeauthoryear{{Yohana}, {Ma}, {Li}, {Chen}  \&
  {Dai}}{{Yohana} et~al.}{2021}]{Yohana2021}
{Yohana} E.,  {Ma} Y.-Z.,  {Li} D.,  {Chen} X.,   {Dai} W.-M.,  2021, \mn@doi
  [\mnras] {10.1093/mnras/stab1197}, \href
  {https://ui.adsabs.harvard.edu/abs/2021MNRAS.504.5231Y} {504, 5231}

\bibitem[\protect\citeauthoryear{{de Blok} et~al.,}{{de Blok}
  et~al.}{2018}]{deb18}
{de Blok} W.~J.~G.,  et~al., 2018, \mn@doi [\apj] {10.3847/1538-4357/aad557},
  \href {https://ui.adsabs.harvard.edu/abs/2018ApJ...865...26D} {865, 26}

\makeatother
\end{thebibliography}



\appendix

\onecolumn
\section{Literature sources in the region of the M~81 triplet}
\label{apend:lit}

In Table \ref{tab:litGCs} we present the previously catalogued 105 confirmed globular clusters from \cite{perelmuter1995}, \cite{nantais2010_spec}, \cite{sha10}, \cite{jan12} and \cite{lim2013}. We detect 95 of these in J-PLUS. However, only 73 are detected in at least 11 of the 12 J-PLUS filters. As outlined in Sect.\ref{sec:met}, these are the ones used in our statistical analysis. The other 22 are either detected in less than 11 filters or do not pass the stellarity $>$ 0.5 cut, and are marked with $^\square$ in the table.

\begin{longtable}{lccccc}
\caption{\label{tab:litGCs}Spectroscopically confirmed GCs from the literature in increasing order of right ascension. Columns include Literature ID and references (\citealt{perelmuter1995}$^*$,\citealt{nantais2010_spec}$^\mathsection$, \citealt{sha10}$^{\diamond}$, \citealt{jan12}$^{\ddagger}$ and \citealt{lim2013}$^{\dag}$), equatorial coordinates,  r\_J-PLUS,  (g-i)\_J-PLUS, metallicity estimate.}\\ \hline\hline
\multicolumn{1}{c}{{\rm ID}} &\multicolumn{1}{c}{{\rm $\alpha$}} &\multicolumn{1}{c}{{\rm $\delta$}} &\multicolumn{1}{c}{r} &\multicolumn{1}{c}{(g-i)} &\multicolumn{1}{c}{{\rm [Fe/H]}}\\
\multicolumn{1}{c}{}&\multicolumn{1}{c}{{\rm (J2000)}}& \multicolumn{1}{c}{{\rm (J2000)}} &\multicolumn{1}{c}{{\rm [mag]}} &\multicolumn{1}{c}{{\rm [mag]}} &\multicolumn{1}{c}{}\\   
\hline
Id70349$^{*}$ & 09\,53\,03.2 & 69\,13\,47.4  & 19.818 & 0.880 & -2.41 \\
JM81GC-2$^{\ddagger}$ & 09\,53\,20.2 & 69\,39\,16.4 &  17.751 & 0.983 & -2.30 \\
JM81GC-1$^{\ddagger}$ & 09\,53\,26.2 & 69\,31\,17.5 &  18.559 & 0.916 &  -- \\
Is90103$^*$ & 09\,53\,39.8 & 68\,48\,00.7  & 17.958 & 0.552 & -2.20  \\
$^\square$Id70319$^*$ & 09\,53\,42.9 & 69\,13\,23.9  & 20.357 & 1.328 & -2.31 \\
Is80172$^{*}$ & 09\,53\,51.7 & 68\,57\,04.4  & 18.682 & 1.069 & -0.77 \\
Nan-4$^\mathsection$ & 09\,54\,04.9 & 69\,09\,18.8  & 19.550 & 1.532 & +1.10 \\
Id50357$^{*}$ & 09\,54\,11.2 & 69\,02\,06.6  & 19.199 & 1.262 & -3.62 \\
Is50394$^{*}$ & 09\,54\,16.5 & 69\,02\,34.4  & 19.017 & 0.804 & -1.50 \\
Is51027$^{*}$ & 09\,54\,20.0 & 69\,09\,11.1  & 19.151 & 0.630 & -2.47 \\
$^\square$Nan-16$^\mathsection$ & 09\,54\,25.1 & 69\,08\,04.1  & 19.876 & 1.380 & -0.72 \\
Nan-28$^\mathsection$ & 09\,54\,35.7 & 69\,06\,43.3  & 18.736 & 1.112 & -- \\
Nan-31$^\mathsection$ & 09\,54\,38.8 & 69\,04\,10.5  & 19.125 & 1.413 & +0.08 \\
Nan-32$^\mathsection$ & 09\,54\,39.7 & 69\,03\,26.8  & 19.685 & 1.087 & -1.51 \\
$^\square$Nan-45$^\mathsection$ & 09\,54\,46.5 & 69\,10\,54.3  & 20.333 & 1.087 & -1.63 \\
$^\square$Nan-51$^\mathsection$ & 09\,54\,50.3 & 69\,05\,08.7  & 20.448 & 0.857 & -2.10 \\
Nan-55$^\mathsection$ & 09\,54\,51.1 & 69\,07\,50.5  & 19.705 & 1.387 & -0.43 \\
Nan-71$^\mathsection$ & 09\,54\,56.1 & 69\,02\,31.4  & 19.323 & 1.195 & -0.64 \\
Nan-79$^\mathsection$ & 09\,54\,58.5 & 69\,08\,08.8  & 20.335 & 1.483 & -0.44 \\
Is50286$^*$ & 09\,54\,58.9 & 69\,00\,58.2  & -- & -- & -0.04 \\
Id50826$^*$ & 09\,54\,58.9 & 69\,00\,58.2  & 19.885 & 0.470 & -1.46 \\
$^\square$Nan-82$^\mathsection$ & 09\,54\,59.8 & 69\,09\,27.8  & 20.176 & 1.459 & -1.04 \\
Nan-90$^\mathsection$ & 09\,55\,02.2 & 69\,05\,38.1  & 17.866 & 1.037 & -- \\
Nan-96$^\mathsection$ & 09\,55\,02.8 & 69\,07\,29.8  & 19.352 & 1.198 & -0.29 \\
$^\square$Nan-97$^\mathsection$ & 09\,55\,03.3 & 69\,02\,24.0  & 20.466 & 0.762 & -1.69 \\
Is40165$^*$ & 09\,55\,03.8 & 69\,15\,37.8  & 18.034 & 0.656 & -1.57 \\
Nan-100$^\mathsection$ & 09\,55\,04.4 & 69\,05\,16.2  & 19.440 & 0.939 & -1.17 \\
Is50037$^*$ & 09\,55\,06.4 & 68\,56\,26.0  & -- & -- & -2.34 \\
Nan-109$^\mathsection$ & 09\,55\,07.3 & 69\,07\,34.6  & 20.472 & 1.380 & +1.15 \\
$^\square$Nan-114$^\mathsection$ & 09\,55\,08.4 & 69\,04\,11.5  & 19.975 & 1.473 & -0.81 \\
Nan-115$^\mathsection$ & 09\,55\,09.0 & 69\,05\,51.9  & 18.841 & 1.391 & -0.89 \\
Nan-116$^\mathsection$ & 09\,55\,09.1 & 69\,04\,28.7  & 18.877 & 1.085 & -1.44 \\
Nan-118$^\mathsection$ & 09\,55\,09.8 & 69\,04\,08.0  & 17.082 & 1.117 & -0.81 \\
Nan-129$^\mathsection$ & 09\,55\,14.3 & 69\,02\,06.6  & 19.808 & 1.082 & -1.48 \\
$^\square$Nan-130$^\mathsection$ & 09\,55\,15.2 & 69\,00\,26.1  & 19.657 & 1.038 & -0.57 \\
Nan-131$^\mathsection$ & 09\,55\,15.3 & 69\,05\,24.5  & 19.446 & 1.192 & -0.79 \\
Nan-136$^\mathsection$ & 09\,55\,15.6 & 69\,05\,48.2  & 19.684 & 1.129 & -0.54 \\
Nan-145$^\mathsection$ & 09\,55\,19.2 & 69\,05\,50.5  & 19.143 & 1.365 & -0.37 \\
$^\square$Nan-153$^\mathsection$ & 09\,55\,20.2 & 69\,05\,38.0  & 20.614 & 1.641 & -0.99 \\
Nan-158$^\mathsection$ & 09\,55\,21.4 & 69\,05\,32.1  & 18.583 & 1.057 & -- \\
Nan-160$^\mathsection$ & 09\,55\,21.9 & 69\,06\,38.0  & 16.691 & 1.151 & -0.86 \\
Nan-162$^\mathsection$ & 09\,55\,22.1 & 69\,05\,19.2  & 17.736 & 1.118 & -- \\
Nan-175$^\mathsection$ & 09\,55\,25.2 & 69\,07\,15.0  & 18.874 & 1.021 & -1.37 \\
Nan-179$^\mathsection$ & 09\,55\,25.7 & 69\,01\,40.2  & 17.180 & 1.484 & -1.26 \\
Nan-188$^\mathsection$ & 09\,55\,29.1 & 69\,00\,31.3  & 20.438 & 1.215 & -0.05 \\
Nan-190$^\mathsection$ & 09\,55\,29.7 & 69\,05\,12.1  & -- & -- & -0.82 \\
Nan-193$^\mathsection$ & 09\,55\,30.1 & 69\,06\,06.6  & -- & -- & -- \\
$^\square$Nan-194$^\mathsection$ & 09\,55\,30.1 & 69\,01\,59.8  & 20.227 & 1.019 & -- \\
Nan-199$^\mathsection$ & 09\,55\,30.8 & 69\,07\,39.1  & 17.782 & 1.119 & -- \\
Nan-209$^\mathsection$ & 09\,55\,32.9 & 69\,06\,40.1  & 18.471 & 1.040 & -- \\
$^\square$Nan-215$^\mathsection$ & 09\,55\,34.4 & 69\,06\,42.6  & 20.508 & 0.572 & -1.97 \\
Nan-218$^\mathsection$ & 09\,55\,34.9 & 68\,58\,15.0  & 18.703 & 1.205 & -0.80 \\
$^\square$Nan-227$^\mathsection$ & 09\,55\,37.2 & 69\,06\,35.9  & 18.639 & 1.639 & -1.86 \\
Nan-228$^\mathsection$ & 09\,55\,37.3 & 69\,02\,07.9  & -- & -- & -0.15 \\
Nan-231$^\mathsection$ & 09\,55\,37.8 & 68\,59\,17.9  & 19.822 & 1.425 & -0.40 \\
Nan-232$^\mathsection$ & 09\,55\,37.8 & 69\,03\,28.2  & -- & -- & -1.41 \\
Nan-236$^\mathsection$ & 09\,55\,38.5 & 69\,06\,55.4  & 19.478 & 2.013 & -0.95 \\
Nan-239$^\mathsection$ & 09\,55\,39.4 & 69\,05\,33.0  & 19.692 & 1.157 & -1.49 \\
Nan-244$^\mathsection$ & 09\,55\,40.0 & 69\,02\,29.9  & -- & -- & -0.88 \\
Nan-246$^\mathsection$ & 09\,55\,40.0 & 69\,04\,10.5  & -- & -- & -1.77 \\
Nan-247$^\mathsection$ & 09\,55\,40.5 & 69\,05\,25.1  & 19.394 & 1.517 & -1.38 \\
$^\square$Nan-253$^\mathsection$ & 09\,55\,41.9 & 68\,55\,00.9  & 19.557 & 1.379 & -0.12 \\
Nan-256$^\mathsection$ & 09\,55\,43.4 & 69\,03\,51.9  & -- & -- & -0.35 \\
Is40181$^{*}$ & 09\,55\,44.1 & 69\,14\,11.7  & 18.558 & 1.353 & +0.64 \\
Nan-258$^\mathsection$ & 09\,55\,44.2 & 69\,04\,24.5  & 19.522 & 1.572 & -1.08 \\
Nan-270$^\mathsection$ & 09\,55\,46.1 & 69\,01\,26.0  & 19.714 & 0.955 & -0.56 \\
Nan-275$^\mathsection$ & 09\,55\,47.7 & 69\,06\,25.6  & 18.831 & 1.056 & -1.42 \\
Nan-276$^\mathsection$ & 09\,55\,48.0 & 69\,07\,28.2  & 20.620 & 0.569 & -1.31 \\
Nan-277$^\mathsection$ & 09\,55\,48.0 & 69\,03\,52.3  & -- & -- & -0.40 \\
Nan-279$^\mathsection$ & 09\,55\,48.5 & 69\,06\,12.4  & 19.637 & 1.119 & -0.22 \\
Lim-523 & 09\,55\,48.6 & 69\,42\,58.4  & 18.403 & 0.934 & -- \\
Nan-280$^\mathsection$ & 09\,55\,48.8 & 69\,05\,22.6  & 19.660 & 1.040 & -1.69 \\
Nan-282$^\mathsection$ & 09\,55\,49.2 & 69\,01\,15.6  & 18.677 & 1.088 & -1.08 \\
Nan-288$^\mathsection$ & 09\,55\,50.2 & 68\,58\,23  & 19.132 & 0.817 & -1.85 \\
Nan-292$^\mathsection$ & 09\,55\,51.3 & 69\,03\,23.9  & 18.646 & 0.911 & -1.50 \\
Nan-293$^\mathsection$ & 09\,55\,51.9 & 69\,07\,39.9  & 18.472 & 1.147 & -0.57 \\
Nan-294$^\mathsection$ & 09\,55\,51.9 & 69\,08\,19.4  & 18.250 & 1.058 & -- \\
$^\square$Nan-295$^\mathsection$ & 09\,55\,52.1 & 69\,07\,10.9  & 19.491 & 1.366 & -- \\
Lim-617$^{\dag}$ & 09\,55\,53.0 & 69\,42\,11.9  & 19.737 & 1.448 & --  \\
Nan-301$^\mathsection$ & 09\,55\,54.5 & 69\,02\,52.9  & 18.225 & 1.179 & -0.90 \\
Nan-302$^\mathsection$ & 09\,55\,55.0 & 69\,00\,56.4  & 18.225 & 1.395 & -0.91 \\
$^\square$Nan-304$^\mathsection$ & 09\,55\,55.3 & 69\,03\,37.8  & 19.444 & 1.405 & +0.12 \\
Nan-307$^\mathsection$ & 09\,55\,55.7 & 69\,00\,03.5  & 18.899 & 0.965 & -- \\
Nan-309$^\mathsection$ & 09\,55\,56.2 & 69\,02\,28.8  & 20.448 & 0.546 & -1.81 \\
Is60045$^{*}$ & 09\,55\,56.9 & 68\,52\,13.4  & 18.434 & 0.786 & -1.03 \\
$^\square$Nan-315$^\mathsection$ & 09\,55\,57.7 & 69\,02\,23.5  & 19.005 & 1.229 & -- \\
$^\square$Nan-330$^\mathsection$ & 09\,56\,03.1 & 69\,07\,19.9  & 19.378 & 1.430 & -0.37 \\
Nan-337$^\mathsection$ & 09\,56\,05.0 & 69\,09\,21.7  & 19.301 & 0.995 & -0.61 \\
Nan-340$^\mathsection$ & 09\,56\,05.5 & 69\,06\,43.4  & 18.907 & 1.013 & -- \\
$^\square$Nan-353$^\mathsection$ & 09\,56\,08.7 & 69\,02\,24.8  & 20.090 & 1.154 & -1.10 \\
Nan-354$^\mathsection$ & 09\,56\,08.8 & 69\,00\,23.8  & 20.671 & 1.096 & -1.94 \\
Nan-365$^\mathsection$ & 09\,56\,14.1 & 69\,05\,05.7  & 20.588 & 1.055 & -1.26 \\
Nan-367$^\mathsection$ & 09\,56\,14.3 & 69\,01\,30.2  & 19.946 & 1.064 & -1.63 \\
Nan-377$^\mathsection$ & 09\,56\,17.5 & 68\,57\,12.3  & 19.614 & 1.065 & -1.22 \\
$^\square$Nan-378$^\mathsection$ & 09\,56\,17.8 & 68\,59\,18.9  & 19.652 & 0.963 & -1.08 \\
Nan-379$^\mathsection$ & 09\,56\,17.8 & 69\,03\,04.9  & 18.880 & 1.142 & -1.33 \\
Nan-385$^\mathsection$ & 09\,56\,18.9 & 68\,59\,55.6  & 19.790 & 0.800 & -1.79 \\
Nan-388$^\mathsection$ & 09\,56\,21.1 & 69\,02\,01.8  & 18.651 & 0.900 & -1.46 \\
Nan-398$^\mathsection$ & 09\,56\,27.5 & 69\,01\,10.1  & 17.446 & 0.849 & -1.53 \\
Nan-404$^\mathsection$ & 09\,56\,31.7 & 69\,03\,55.2  & 20.157 & 1.163 & -0.26 \\
$^\square$Nan-410$^\mathsection$ & 09\,56\,36.9 & 69\,01\,46.6  & 19.318 & 1.412 & -0.30 \\
Is40083$^{*}$ & 09\,56\,38.5 & 69\,22\,50.3  & 18.185 & 0.660 & -1.29 \\
Is50225$^{*}$ & 09\,56\,40.6 & 68\,59\,52.6  & 18.178 & 0.966 & -0.04 \\
HoIX-4-1038$^{\diamond}$ & 09\,57\,40.0 & 69\,03\,25.0  & 19.145 & 0.090 & -- \\
Id30244$^{*}$ & 09\,57\,54.9 & 68\,49\,00.4  & 19.582 & 1.025 & -1.76 \\
\hline
\end{longtable}

\twocolumn
\section{MLPCA}
\label{apend:mlpca}
In the following, we show a snippet code to run the algorithm   in \texttt{R} and \texttt{python}. The code uses as input a matrix of covariates X, a error matrix Xsd of same dimension, and the desired dimension of the projection, p < rank(X). 

\lstdefinestyle{RStyle} {
language=R,
keywordstyle= \color{blue},
identifierstyle=\ttfamily,
commentstyle=\color{orange!30!red},
stringstyle=\ttfamily\color{gray},
xleftmargin=1pt,
framexleftmargin=20pt,
showstringspaces=false,
basicstyle=\footnotesize,
numbersep=10pt,
tabsize=2,
breaklines=true,
prebreak = \raisebox{0ex}[0ex][0ex]{\ensuremath{\hookleftarrow}},
breakatwhitespace=false,
aboveskip={1.5\baselineskip},
columns=flexible,
extendedchars=true
}
\lstdefinestyle{pyStyle} {
language=python,
keywordstyle= \color{blue},
identifierstyle=\ttfamily,
commentstyle=\color{orange!30!red},
stringstyle=\ttfamily\color{gray},
xleftmargin=1pt,
framexleftmargin=20pt,
showstringspaces=false,
basicstyle=\footnotesize,
numbersep=10pt,
tabsize=2,
breaklines=true,
prebreak = \raisebox{0ex}[0ex][0ex]{\ensuremath{\hookleftarrow}},
breakatwhitespace=false,
aboveskip={1.5\baselineskip},
columns=flexible,
extendedchars=true
}
\lstset{language=R}
\lstset{language=python}

\begin{lstlisting}[caption={\texttt{R} script},captionpos=t,style=RStyle]
MLPCA <- function(X, Xsd,p,MaxIter = 1e5) {
  #Initialization
  epsilon <- 1e-10 # Convergence Limit
  MaxIter <- MaxIter # Maximum no. of iterations
  m <- nrow(X)
  n <-ncol(X)
  VarX <- Xsd^2 # Variance
  DecomX <- RSpectra::svds(X, p) #SVD
  U <- DecomX$u
  S <- diag(DecomX$d)
  V <- DecomX$v
  i <- 0 # Loop counter
  Sold <- 0 # Holds last value of objective function
  k <- -1 # Loop flag
  while (k < 0) {
    i <- i + 1 #Loop counter
    # Evaluate objective function 
    Sobj <- 0 # Initialize sum
    LX <- matrix(data = 0,nrow = nrow(X),ncol =ncol(X))
    for (j in 1:n){
      Q <- diag(1/VarX[, j])
      F <- solve(t(U) %*% Q %*% U)
      LX[,j]<- U %*% (F %*% (base::t(U) %*% (Q %*% X[, j])))
      Dx <- matrix(X[, j] - LX[, j]) # Residual Vector
      Sobj <- Sobj + base::t(Dx) %*% Q %*% Dx }
    # Convergence check
    if (i %% 2 == 1) {
      ConvCalc <- base::abs(Sold - Sobj)/Sobj
      if (ConvCalc < epsilon){
        k <- 0}
      if (i > MaxIter) {
        k <- 1
        stop("MaxIter exceeded")}}
    #Flip matrices
    if (k < 0) {
      Sold <- Sobj
      DecomLX <- RSpectra::svds(LX, p)
      U <- DecomLX$u
      S <- diag(DecomLX$d)
      V <- DecomLX$v
      X <- t(X)
      VarX <- t(VarX) 
      n <- ncol(X)
      U <- V}}
  DecomFinal <- RSpectra::svds(LX, p)
  U <- DecomFinal$u
  S <- diag(DecomFinal$d)
  V <- DecomFinal$v
  out <- list("U" = U, "S" = S,"V" = V)
  return(out)}
\end{lstlisting} 

\newpage
\begin{lstlisting}[caption={\texttt{Python} script},captionpos=t,style=pyStyle]
import datetime
import numpy as np
from numpy.linalg import inv
from sklearn.datasets import load_iris
import scipy.sparse.linalg as sp
def MLPCA(X, Xsd, p, MaxIter=1e5):
    epsilon = 1e-10
    MaxIter = MaxIter
    m = X.shape[0]
    n = X.shape[1]
    VarX = np.multiply(Xsd, Xsd)
    U, o, V = sp.svds(X,k=p)
    i = 0
    Sold = 0
    k = -1
    while (k < 0):
        i = i + 1
        Sobj = 0
        LX = np.mat(np.zeros((X.shape[0], X.shape[1])))
        for j in range(0, n):
            Q = np.diagflat(1 / VarX[:, j])
            F = inv(U.T @ Q @ U)
            LX[:, j] = U @ (F @ (U.T @ (Q @ X[:, j])))
            Dx = np.mat(X[:, j] - LX[:, j])
            Sobj = Sobj + Dx.T @ Q @ Dx
        if i % 2 == 1:
            ConvCalc = np.abs(Sold - Sobj) / Sobj
            if ConvCalc < epsilon:
                k = 0
            if i > MaxIter:
                k = 1
                exit("MaxIter exceeded")
        if k < 0:
            Sold = Sobj
            U, o, V = sp.svds(LX,k=p)
            V = V.T
            X = X.T
            VarX = VarX.T
            n = X.shape[1]
            U = V
    U, o, V = sp.svds(LX,k=p)
    S = np.mat(np.diag(o))
    V = V.T
    return U, S, V
\end{lstlisting} 
%


\section{Spatial Distribution}
\label{apend:spatial}
\begin{figure*}
\centering
 \includegraphics[width=\linewidth]{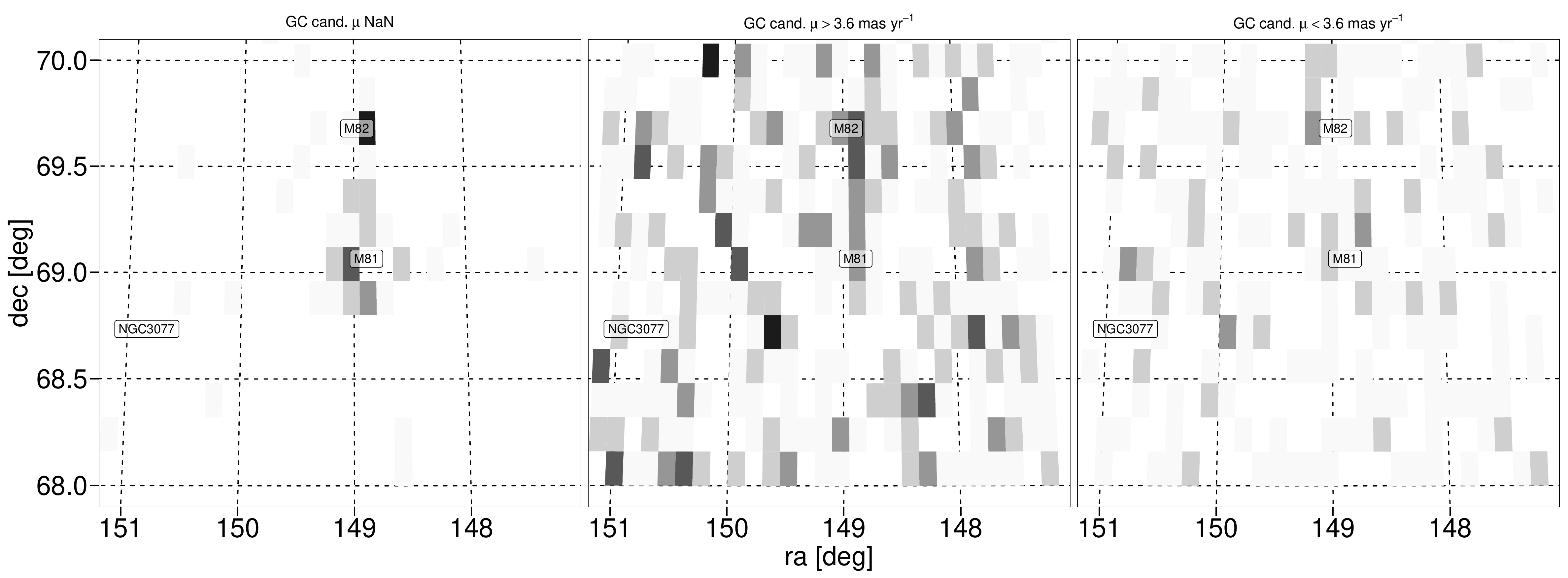}
 \caption{Spatial distribution in Aitoff projection of GC candidates divided by proper motion categories.}
 \label{fig:spatial_split}
\end{figure*}

Having information on proper motions, in \autoref{fig:spatial_split}, we show the spatial distribution for each of the GC candidates category colour-coded by counts of GC within each square bin of 0.15\,deg on each side. Despite the compelling evidence of GC bridge between M\,81 and M\,82, considerably contribution comes from GC without proper motion information. Thus, spectroscopic follow-up will be carried out around the region to confirm the potential bridge hypothesis. 

\section{Flux excess}
\label{apend:flux_ex}
\begin{figure}
 \includegraphics[width=\columnwidth]{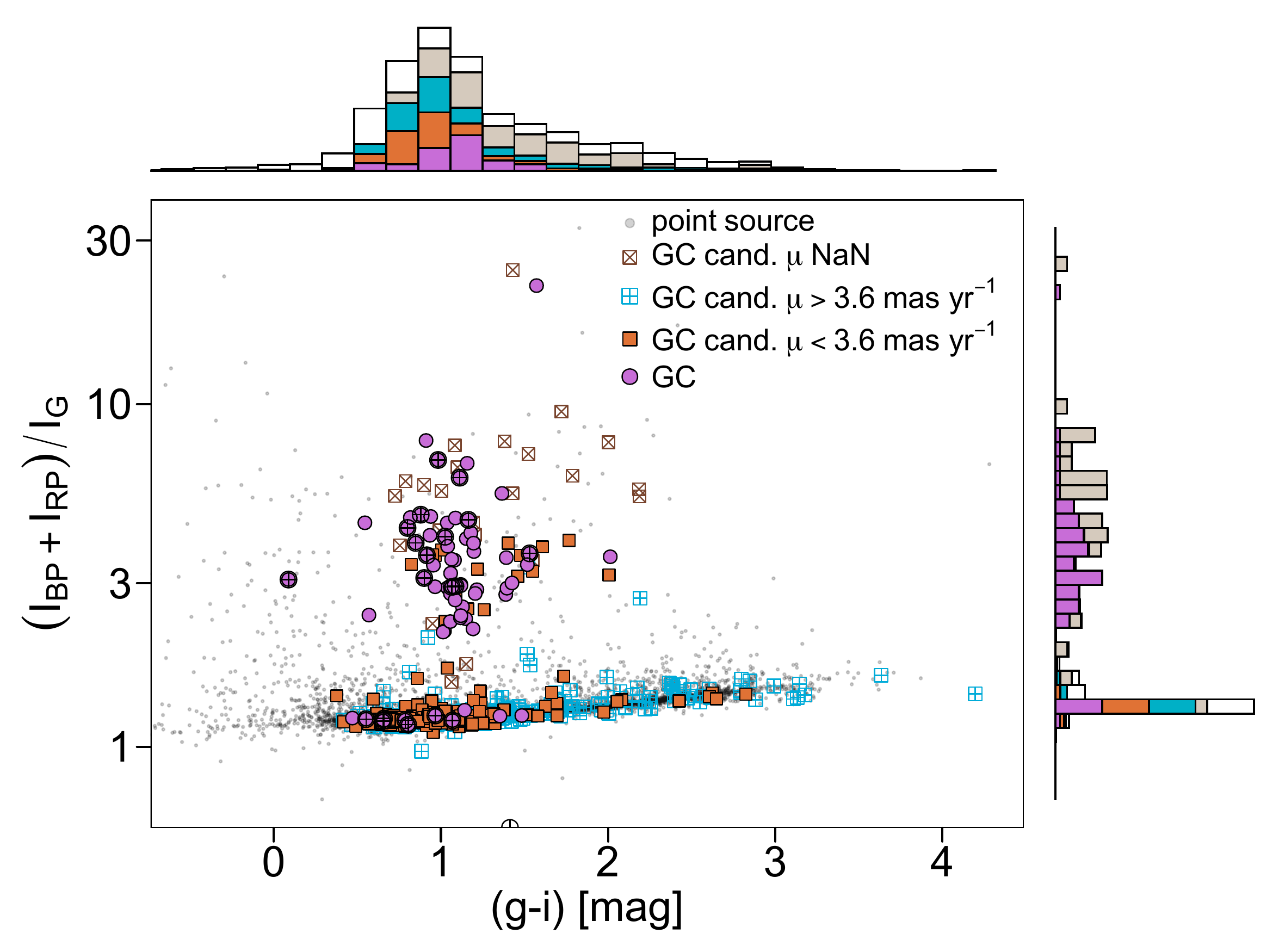}
 \caption{Flux ratio from the Gaia\,EDR3 passbands as a function of the $(g-i)$ colour for point sources, spectroscopically confirmed GCs and GC candidates according to the legend. Inner crosses highlight halo GC candidates.}
 \label{fig:exc1}
\end{figure}

Besides the astrometric information used in \autoref{sec:anc}, Gaia\,EDR3 provides photometry in three bands, $G$, $G_{BP}$ and $G_{RP}$. The first one covers the wavelength range from 330\,nm to 1050\,nm and results from the profile-fitting of the sources in the astrometric field. The latter ones are integrated over a rectangular aperture from the low-resolution spectra observed with two different prisms, and their joint range of wavelength matches with that of the $G$ band, with slightly different transmission curves \citep{eva18}. The similarity between these passbands leads to the definition of the flux ratio, $(I_{BP}+I_{RP})/I_G$, as a proxy of crowded regions. 
Although the flux ratio from the Gaia passbands, $(I_{BP}+I_{RP})/I_G$, is assumed as an indication of crowded regions, large values can also represent extended objects. This is particularly relevant in the case of GCs in nearby systems, like the M\,82/M\,82/NGC\,3077 triplet, for which the typical effective radii ($\sim 0.2$\,arcsec, from ${\rm r_{eff}\sim 3\,pc}$, e.g. \citealt{har96}, 2010 Edition, and \citealt{cas14}, and the distance assumed in this paper) is comparable to the pixel scale for the Gaia astrometric CCD \citep{pru16}. \autoref{fig:exc1} shows the flux excess for GC candidates as a function of the $(g-i)$ colour with blue squares. 
The majority of the GC candidates present flux excess close to unity, as expected from point sources (Figure\,\ref{fig:exc1} and \citealt{fab20}). Still, there are $\sim 20$ plausible candidates and a few less plausible ones with flux excess larger than 2. In contrast, almost all the flux excess for Galactic stars and background galaxies are below 2. However, we are aware that several confirmed GCs also present flux excess close to unity. This feature cannot be used to unequivocally separate GCs from Galactic stars in our sample (but see \citealt{voggel2020}).

\section{Comparison with foreground extinction estimates}
\label{apend:extinction}
To independently test the photometric calibration of the catalogue and to estimate the mean extinction of the GCs embedded in the M\,81 disk, 
we assumed the following approach
based on the spectroscopic metallicities available in the literature for a fraction of the confirmed GCs \citep[e.g.][]{perelmuter1995,nantais2010_spec}, that belong to our photometric catalogue. For such GCs, we calculated simulated magnitudes in the broad-bands from J-PLUS through the SSPs from the CMD\,$3.1$ web interface\footnote{\url{http://stev.oapd.inaf.it/cgi-bin/cmd_3.1}}, by means of the PARSEC evolutionary tracks \citep{bre12} and a Chabrier log-normal initial mass function, and a fiducial age of 10\,Gyrs. Then, the absorption is estimated as the difference between the simulated and real magnitudes, considering the absorption coefficients from López-Sanjuan et al. (2019) and $A_V=3.1 \times E(B-V).$ we estimate the magnitudes from the difference between observed and expected colours, and the transformations from \citet{lop19}. We note that the procedure is largely uncertain, due to the combination of the errors in the spectroscopic metallicities, the assumption of a fiducial age for all the GCs, systematic effects from the SSPs, and the photometric errors. However, it serves the purpose to estimate a mean $A_V$.
By restricting the sample to GCs at projected distances from the galaxy centre to be larger than 12\,arcmin, the mean absorptions from the broad-band colours are around $A_V\sim 0.19-0.22$\,mag. The estimated absorptions for these GCs 
should be ruled by Galactic foreground extinction, which is settled at $A_V\sim 0.22$\,mag from 
\cite{Schlafly_Finkbeiner2011}. The GCs at less than 12\,arcmin from M\,81 centre
are typically embedded in its disk, their mean absorption reach $A_V\sim 0.44-0.48$\,mag 
leading to a mean intrinsic absorption in the disk of M\,81 of $A_V\sim 0.22-0.26$\,mag. This is considerably lower than the mean value calculated by \citet{nantais2011_photGCs}, but it is also an estimation from optical data, and the samples in both analysis are not the same.


\bsp	
\label{lastpage}
\end{document}